\documentclass[aps,pra,reprint,superscriptaddress,floatfix]{revtex4-2}

\usepackage[T1]{fontenc}
\usepackage[utf8]{inputenc}
\usepackage[english]{babel}

\usepackage{amsmath,amssymb,mathtools}
\usepackage{graphicx}
\usepackage{dcolumn}
\usepackage{bm}

\usepackage{xcolor}

\usepackage{tikz}
\usetikzlibrary{hobby,backgrounds,calc,fit,trees,positioning,3d,intersections,shapes.geometric,shapes.misc,arrows,arrows.meta,decorations.pathreplacing}
\definecolor{ftcyan}{RGB}{42,185,209}
\definecolor{ftpurple}{RGB}{198,81,205}
\definecolor{ftgreen}{RGB}{59,133,93}
\definecolor{ncred}{RGB}{235,40,40}
\definecolor{ncgreen}{RGB}{0,165,65}
\definecolor{ncblue}{RGB}{35,80,235}
\tikzset{
 ft classical/.style={-{Stealth[length=1.6mm]},dashed,draw=black!70,line width=.55pt},
 ft note/.style={font=\scriptsize,align=center},
 ft block/.style={draw=black!65,rounded corners=1.2mm,align=center,
   inner sep=2mm,font=\small,line width=.5pt},
 ft flow/.style={-{Stealth[length=1.8mm]},line width=.6pt},
 nc error/.style={draw=ncred!85!black,fill=ncred,fill opacity=.32,
   draw opacity=1,text=black,text opacity=1,line width=.65pt},
 nc gate/.style={draw=ncblue!85!black,fill=ncblue,fill opacity=.32,
   draw opacity=1,text=black,text opacity=1,line width=.65pt},
 nc correction/.style={draw=ncgreen!85!black,fill=ncgreen,fill opacity=.32,
   draw opacity=1,text=black,text opacity=1,line width=.65pt},
 nc access/.style={nc gate,fill opacity=.18},
 nc records/.style={ft block,nc gate,fill opacity=.12},
 nc report/.style={ft block,nc correction,fill opacity=.24}
}

\tikzstyle{arrow} = [thick,->,>=stealth]
\usepackage{tikz-3dplot}
\usepackage{quantikz}
\usepackage{yquant}
\useyquantlanguage{groups}
\usepackage{tikzit}
\tikzstyle{gate}=[shape=rectangle, text height=1.5ex, text depth=0.25ex, yshift=0.5mm, fill=white, draw=black, minimum height=5mm, yshift=-0.5mm, minimum width=5mm, font={\small}, tikzit category=circuit]
\tikzstyle{big gate}=[shape=rectangle, text height=1.5ex, text depth=0.25ex, yshift=0.5mm, fill=white, draw=black, minimum height=10mm, yshift=-0.5mm, minimum width=5mm, font={\small}, tikzit category=circuit]
\tikzstyle{Z dot}=[inner sep=0mm, minimum size=2mm, shape=circle, draw=black, fill={rgb,255: red,221; green,255; blue,221}, tikzit category=zx]
\tikzstyle{Z phase dot}=[minimum size=5mm, font={\footnotesize\boldmath}, shape=rectangle, rounded corners=2mm, inner sep=0.2mm, outer sep=-2mm, scale=0.8, tikzit shape=rectangle, draw=black, fill={rgb,255: red,221; green,255; blue,221}, tikzit draw=blue, tikzit category=zx]
\tikzstyle{X dot}=[Z dot, shape=circle, draw=black, fill={rgb,255: red,255; green,136; blue,136}, tikzit category=zx]
\tikzstyle{X phase dot}=[Z phase dot, tikzit shape=rectangle, tikzit draw=blue, fill={rgb,255: red,255; green,136; blue,136}, font={\footnotesize\boldmath}, tikzit category=zx]
\tikzstyle{hadamard}=[fill=yellow, draw=black, shape=rectangle, inner sep=0.6mm, minimum height=1.5mm, minimum width=1.5mm, tikzit category=zx]
\tikzstyle{paulibox}=[fill={rgb,255: red,221; green,221; blue,255}, draw=black, shape=rectangle, inner sep=0.6mm, minimum height=5mm, minimum width=5mm, font={\footnotesize}, text height=1.5ex, text depth=0.25ex, tikzit category=zx]
\tikzstyle{vertex}=[inner sep=0mm, minimum size=1mm, shape=circle, draw=black, fill=black, tikzit category=misc]
\tikzstyle{vertex set}=[inner sep=0mm, minimum size=1mm, shape=circle, draw=black, fill=white, font={\footnotesize\boldmath}, tikzit category=misc]
\tikzstyle{small black dot}=[fill=black, draw=black, shape=circle, inner sep=0pt, minimum width=1.2mm, tikzit category=circuit]
\tikzstyle{cnot ctrl}=[fill=black, draw=black, shape=circle, inner sep=0pt, minimum width=1.2mm, tikzit category=circuit]
\tikzstyle{cnot targ}=[fill=white, draw=white, shape=circle, tikzit category=circuit, label={center:$\oplus$}, inner sep=0pt, minimum width=2.1mm, tikzit fill={rgb,255: red,102; green,204; blue,255}, tikzit draw=black]
\tikzstyle{ket}=[fill=white, draw=black, shape=regular polygon, regular polygon sides=3, regular polygon rotate=-30, scale=0.7, inner sep=1pt, tikzit category=circuit, tikzit shape=rectangle, tikzit fill=green]
\tikzstyle{bra}=[fill=white, draw=black, shape=regular polygon, regular polygon sides=3, regular polygon rotate=30, scale=0.7, inner sep=1pt, tikzit category=circuit, tikzit shape=rectangle, tikzit fill=red]
\tikzstyle{scalar}=[shape=rectangle, text height=1.5ex, text depth=0.25ex, yshift=0.5mm, fill=white, draw=black, minimum height=5mm, yshift=-0.5mm, minimum width=5mm, font={\small}]
\tikzstyle{clabel}=[fill=white, draw=none, shape=rectangle, tikzit fill={rgb,255: red,56; green,255; blue,242}, font={\footnotesize}, inner sep=1pt, tikzit category=labels]
\tikzstyle{empty diagram}=[draw={gray!40!white}, dashed, shape=rectangle, minimum width=1cm, minimum height=1cm, tikzit category=misc]
\tikzstyle{amap}=[fill=white, draw=black, shape=NEbox, tikzit category=asymmetric, tikzit fill=yellow, tikzit shape=rectangle]
\tikzstyle{amap conj}=[fill=white, draw=black, shape=NWbox, tikzit category=asymmetric, tikzit fill=green, tikzit shape=rectangle]
\tikzstyle{amap adj}=[fill=white, draw=black, shape=SEbox, tikzit category=asymmetric, tikzit fill=red, tikzit shape=rectangle]
\tikzstyle{amap trans}=[fill=white, draw=black, shape=SWbox, tikzit category=asymmetric, tikzit fill=orange, tikzit shape=rectangle]
\tikzstyle{astate}=[fill=white, draw=black, shape=NEtriangle, tikzit category=asymmetric, tikzit shape=circle, tikzit fill=yellow]
\tikzstyle{astate conj}=[fill=white, draw=black, shape=NWtriangle, tikzit category=asymmetric, tikzit shape=circle, tikzit fill=green]
\tikzstyle{astate adj}=[fill=white, draw=black, shape=SEtriangle, tikzit category=asymmetric, tikzit shape=circle, tikzit fill=red]
\tikzstyle{astate trans}=[fill=white, draw=black, shape=SWtriangle, tikzit category=asymmetric, tikzit shape=circle, tikzit fill=orange]

\tikzstyle{hadamard edge}=[-, dashed, dash pattern=on 2pt off 0.5pt, thick, draw={rgb,255: red,68; green,136; blue,255}]
\tikzstyle{star edge}=[-, dashed, dash pattern=on 2pt off 0.5pt, thick, draw={rgb,255: red,255; green,136; blue,68}]
\tikzstyle{box edge}=[-, dashed, dash pattern=on 2pt off 0.5pt, thick, draw={rgb,255: red,203; green,192; blue,225}]
\tikzstyle{brace edge}=[-, tikzit draw=blue, decorate, decoration={brace,amplitude=1mm,raise=-1mm}]
\tikzstyle{diredge}=[->]
\tikzstyle{double edge}=[-, double, shorten <=-1mm, shorten >=-1mm, double distance=2pt]
\tikzstyle{gray edge}=[-, {gray!60!white}]
\tikzstyle{pointer edge}=[->, very thick, gray]
\tikzstyle{boldedge}=[-, line width=1.6pt, shorten <=-0.17mm, shorten >=-0.17mm]
\tikzstyle{bidir edge}=[<->, very thick, draw={rgb,255: red,191; green,191; blue,191}]
\tikzstyle{separator edge}=[-, dashed, dash pattern=on 2pt off 0.5pt, thick, draw={rgb,255: red,153; green,153; blue,153}]

\usepackage{physics}
\usepackage{tensor}
\usepackage{cancel}
\usepackage{bbold}
\usepackage{makecell}
\usepackage{orcidlink}
\usepackage[normalem]{ulem} 

\usepackage{hyperref}
\newcommand{\floor}[1]{\lfloor #1 \rfloor}

\begin{document}

\title{Iteratively decoded magic state distillation}

\author{Kwok Ho Wan \orcidlink{0000-0002-1762-1001}}
\email{((initials))1496@((9.81))mail.com}
\affiliation{Universal Quantum Ltd, Gemini House, Haywards Heath, RH16 1XQ, United Kingdom}
\affiliation{Blackett Laboratory, Imperial College London, London SW7 2AZ, United Kingdom}
\affiliation{Mathematical Institute, University of Oxford, Oxford OX2 6GG, United Kingdom}
\date{\today}

\begin{abstract}
We present numerical simulation results for the 7-to-1 and 15-to-1 state distillation circuits, constructed using transversal CNOTs acting on multiple surface code patches. 
The distillation circuits are decoded iteratively using the method outlined in [arXiv:2407.20976]. We show that, with a re-configurable qubit architecture, 
we can perform fast magic state distillation in $\sim\mathcal{O}(1)$ code cycles. 
We confirm that both circuits suppress an injected input logical error rate $p$ to $\mathcal{O}(p^3)$ in the presence of additional circuit-level noise. 
This is done with two types of stabiliser proxies, distilling logical $\ket{-}$ and $\ket{Y}$ states, the latter is the
intended state of the 7-to-1 circuit while a stabiliser-proxy for the 15-to-1 circuit. We then also provide numerical evidences for actual $\ket{T}$ state distillation using the 15-to-1 circuit with a faulty-$T$ measurement, leveraging recent near-Clifford simulation tools.
Finally, we outline how ZX-calculus and Pauli webs can be used to benchmark stabiliser proxies for these distillation circuits.
\end{abstract}

\maketitle

\section{Introduction}
\label{section:intro}
The surface code \cite{Dennis_2002,Fowler_2012,fowler2019lowoverheadquantumcomputation} is a strong candidate for large-scale fault-tolerant quantum computation. Every Clifford gate ($H,S,\text{CNOT}$) on the surface code permits a transversal implementation given a re-configurable hardware architecture or flying qubits \cite{zhou2024algorithmicfaulttolerancefast}, while the non-Clifford $T = \begin{pmatrix} 1 & 0 \\ 0 & e^{i\frac{\pi}{4}} \end{pmatrix}$ gate cannot be implemented transversally \cite{eastin_knill_2009,Bravyi_Konig_2013}. One approach to bypass this restriction is to perform $T$ gates via gate teleportation. This is achieved by consuming the magic state: $\ket{T} \propto \ket{0}+e^{i\frac{\pi}{4}}\ket{1}$ in the process. 

Before the consumption of $\ket{T}$ state is possible, magic state injection techniques are needed to encode faulty logical $\ket{T}$ states from physical $\ket{T}$ states \cite{Li_2015,LaoInjection2022,PRXQuantum.5.010302,gidney2024magicstatecultivationgrowing,wan2026exactlogicalerrorrates}. Then, the 15-to-1 distillation protocol (also known as distillation factory) is used to output a higher quality logical $\ket{T}$ state \cite{PhysRevA.71.022316,Bravyi_2012,Haah2018codesprotocols}. Similarly, another resource state: $\ket{Y} \propto \ket{0} + i\ket{1}$ can also be injected, distilled using the 7-to-1 distillation protocol \cite{fowler2013bridgeloweroverheadquantum,Herr_2017,zhou2024algorithmicfaulttolerancefast} and consumed in a similar manner to perform the $S = \begin{pmatrix} 1 & 0 \\ 0 & i \end{pmatrix}$ gate. Lower overhead methods of initialising a high quality logical $\ket{Y}$ state has recently become available \cite{Gidney_2024}, this may make the 7-to-1 distillation protocol obsolete. Alternatively, the transversal logical $S$ gate can be implemented with physical $S, \ S^{\dagger}$ and $\text{CZ}$ gates on the surface code \cite{Moussa_Fold_2016,Breuckmann_2024,zhou2024algorithmicfaulttolerancefast}. Nevertheless, studying the 7-to-1 protocol can be insightful as the 7-to-1 and 15-to-1 protocols have many parallels. 

In general, state distillation aims to use multiple faulty resource states with input error $p$, and output a higher quality $\ket{T/Y}$ state with suppressed output error rate of $p_{\text{out}} = \mathcal{O}(p^3)$. To leading order in $p$, $p\xrightarrow[]{\text{15-to-1}}35p^3$ \cite{PhysRevA.71.022316,Bravyi_2012} and $p\xrightarrow[]{\text{7-to-1}}7p^3$ \cite{Herr_2017,fowler2013bridgeloweroverheadquantum} 

Historically, state distillation protocols are the leading contributor to the spacetime volume footprint of a fault-tolerant quantum computer \cite{Herr_2017,fowler2013bridgeloweroverheadquantum,fowler2019lowoverheadquantumcomputation}. Magic state distillation using lattice surgery based techniques requires a spacetime overhead of $\mathcal{O}(d^3)$ \textit{qubit-cycles} (physical data qubits $\times$ code cycles), where $d$ is the distance of the surface code. This spacetime overhead had drastically decreased from $\mathcal{O}(1500d^3)$ \cite{Herr_2017,fowler2013bridgeloweroverheadquantum} to $\mathcal{O}(624d^3)$ \cite{fowler2019lowoverheadquantumcomputation} or $\mathcal{O}(396d^3)$ for the $\ket{CCZ}$ state distillation factory \footnote{With a catalyst $\ket{T}$ state, the output state from a $\ket{CCZ}$ factory can be transformed into two $\ket{T}$ states \cite{Gidney2019efficientmagicstate}.} \cite{Gidney2019efficientmagicstate}. However, all of the above mentioned state distillation schemes will output one $\ket{T}$ state (2 $\ket{T}$ states for the $\ket{CCZ}$ factory) every $5.5d$ to $6.5d$ code cycles. For larger code distances, the time complexity cost will pose run-time challenges for trapped-ion and neutral-atom platforms, where the code cycle times are considerably slower, in the kHz regime \cite{postler2023demonstrationfaulttolerantsteanequantum,PhysRevX.11.041058,ryananderson2024highfidelityfaulttolerantteleportationlogical,Bluvstein2024} compared to MHz in superconducting platforms \cite{acharya2024quantumerrorcorrectionsurface}. 

To bridge this three orders of magnitude gap in code cycle rate \cite{Webber_2022,webber2020efficientqubitroutingglobally,Lekitsch_2017,Akhtar_2023, Bluvstein2024,cain2024correlated,zhou2024algorithmicfaulttolerancefast}, we use the iterative transversal CNOT decoding technique from \cite{wan2024iterativetransversalcnotdecoder} to collapse the time complexity of both the 15-to-1 and 7-to-1 distillation protocols to $\mathcal{O}(1)$, removing its dependence on $d$. We numerically confirm the 7-to-1 distillation simulation results from \cite{zhou2024algorithmicfaulttolerancefast}, without the need for a hyper-edge decoding graph or most-likely error (MLE) decoders \cite{cain2024correlated,zhou2024algorithmicfaulttolerancefast}. In our approach, each surface code patch is decoded separately before iteratively propagating Pauli-frames from each corresponding patch \cite{wan2024iterativetransversalcnotdecoder}. Furthermore, we extended our approach to numerically characterise the 15-to-1 distillation protocol. We confirm the $p \rightarrow \mathcal{O}(p^3)$ scaling in both distillation protocols using stabiliser proxies. 

{We also directly simulate non-Clifford $\ket{T}$ state distillation using faulty-$T$ measurements in the 15-to-1 circuit, combining the iterative transversal CNOT decoder with near-Clifford simulation (section~\ref{section:surface_code_non_clifford}).}

{With three rounds of syndrome extraction of padding between transversal CNOTs and initialisations, the 7-to-1 and 15-to-1 proxy circuits require $15$ and $18$ syndrome rounds respectively, including initialisation of the data patches. Their data-qubit costs are $8d^2\times15+7d^2\times3=141d^2$ and $16d^2\times18+15d^2\times3=333d^2$ qubit-cycles. Following the convention of Ref.~\cite{Litinski_2019_magic}, these counts include physical data qubits and exclude measurement ancillas, resource preparation (e.g. a short, two-round postselected injection stage~\cite{Li_2015}) and {output verification for estimating logical error rates}. {In our simulations, inplace $\ket{Y}$ preparation and readout, each use $\floor{d/2}+2$ additional syndrome rounds, are excluded from these costs.} {Noisy logical $\ket{Y}$ and $\ket{T}$ states can be prepared using two rounds of syndrome extraction per postselected injection attempt~\cite{Li_2015,LaoInjection2022,wan2025pauliwebyranglestate} (see section~\ref{section:surface_code_sim}).}} In our simulations, transversal CNOTs are instantaneous and noiseless. Hardware noise and scheduling costs will be explored in work in preparation~\cite{updating_impact}.

We benchmark both circuits with two types of stabiliser proxies: 
\begin{enumerate}
\item distilling logical $\ket{-}$ states, which is a proxy for both the 7-to-1 and 15-to-1 circuits, and 
\item distilling logical $\ket{Y}$ states (injected using the inplace method~\cite{Gidney_2024}), which is a genuine logical $\ket{Y}$ state distillation for the 7-to-1 circuit but remains a stabiliser proxy for the 15-to-1 circuit (intended state is $\ket{T}$).
\end{enumerate}

This paper is structured as follows. Firstly, we introduce the 7-to-1 and 15-to-1 distillation circuits in section \ref{section:state_distillation}. Then, we present our contributions in section \ref{section:surface_code_sim}, simulating each distillation circuit with multiple surface code patches entangled with transversal CNOTs. {Section~\ref{section:surface_code_non_clifford} extends this study to non-Clifford $\ket{T}$ state distillation.} Finally, we provide discussions and future directions in the last section.

\section{State distillation protocols}
\label{section:state_distillation}
In this section, we review the 7-to-1 and 15-to-1 distillation protocols and their underlying quantum error correcting codes: Steane and Reed-Muller code. We then discuss the post-selection protocol required in both distillation protocols to output an error suppressed distilled state.

The 15-to-1 $\ket{T}$ and 7-to-1 $\ket{Y}$ state distillation circuits are CNOT circuits followed by a resource state consumption stage, measurement and post-selection. They rely on the ability to perform specific gates via gate teleportation. This involves consuming many copies of faulty $\ket{T}$ ($\ket{Y}$) states and post-selecting measurement results to detect errors. One simple error is $Z\cdot T$ ($Z\cdot Y$) performed on the output state instead of $T$ ($Y$). 

The circuits in equation \ref{eq:Z_S_T_gate_by_teleportation} show consumption of $\ket{-}\propto \ket{0}-\ket{1}$, $\ket{Y}$ and $\ket{T}$ states to perform a $Z$, $S$ and $T$ gate respectively, up to measurement-conditioned Clifford byproducts \cite{Briegel_2009}.
\begin{equation}
\label{eq:Z_S_T_gate_by_teleportation}
\begin{split}
& \begin{tikzpicture}
\begin{yquantgroup}
\registers{
qubit {$\ket{\phi}$} q[+1];
qubit {$\ket{-}$} q[+1];  
}
\circuit{
cnot q[1] | q[0];       
measure q[1];   
output {$n$} q[1];   
output {$(-1)^{n} Z \ket{\phi}$} q[0];
}
\end{yquantgroup}
\end{tikzpicture}
\\
&\begin{tikzpicture}
\begin{yquantgroup}
\registers{
qubit {$\ket{\phi}$} q[+1];
qubit {$\ket{Y}$} q[+1];  
}
\circuit{
cnot q[1] | q[0];       
measure q[1];   
output {$n$} q[1];   
output {$Z^{n} S \ket{\phi}$} q[0];
}
\end{yquantgroup}
\end{tikzpicture}
\\
& \begin{tikzpicture}
\begin{yquantgroup}
\registers{
qubit {$\ket{\phi}$} q[+1];
qubit {$\ket{T}$} q[+1];
}
\circuit{
cnot q[1] | q[0];   
measure q[1];
output {$n$} q[1];
output {$(S^{\dagger})^{n} T \ket{\phi}$} q[0];
}
\end{yquantgroup}
\end{tikzpicture}
\end{split}
\end{equation}
The 7-to-1 $\ket{Y}$ and 15-to-1 $\ket{T}$ distillation protocols consume resource states needed to perform the $S$ and $T$ gates respectively. However, they also distill all resource states that perform integer powers of $S$ and $T$ respectively \cite{Haah2018codesprotocols,Herr_2017}. The tuples, $(\text{resource state}, \text{gate by teleportation})$, below in equation \ref{eq:tuple} show what can be distilled by each protocol:
\begin{equation}
\begin{split}
     \text{7-to-1 protocol: } & \Big\{ (\ket{-}, Z), \ (\ket{Y}, S) \Big\} \\ 
     \text{15-to-1 protocol: } & \Big\{ (\ket{-}, Z), \ (\ket{Y}, S), \ (\ket{T}, T) \Big\} \ .
\end{split}
\label{eq:tuple}
\end{equation}
In order to investigate the effectiveness of either distillation circuits, we looked at the stabiliser proxy of $\ket{-}/\ket{Y}$ state distillation using the 7-to-1 and 15-to-1 circuits in our numerical simulations. In practice, simple transversal implementation of the logical $Z$ gate exists for surface code and will never realistically be performed in this manner with gate teleportation. We assume numerical results from $\ket{-}/\ket{Y}$ state distillation provide an order-of-magnitude estimation to the actual performance of $\ket{T}$ state distillation with the 15-to-1 circuit. {Section~\ref{section:surface_code_non_clifford} presents our non-Clifford faulty-$T$-measurement simulations.}

We will now provide some background on the 7-to-1 and 15-to-1 distillation circuits along with the post-selection required to achieve the cubic error suppression. Unless otherwise specified, all the gates and measurements from this point onwards are understood to be in the surface code logical basis.

\subsection{7-to-1 distillation protocol}
The 7-to-1 distillation protocol relies on the $[[7,1,3]]$ Steane code's transversal $S$ gate, $\bar{S}_{\text{Steane}}$ \cite{eczoo_steane}. Figure \ref{fig:steane_code_labelling} shows the Tanner graph of the self-dual Steane code. The encoding qubits are arranged as circular nodes, lying on the corners, mid-points and centroid of an equilateral triangle. The rectangular nodes are stabilisers ($X$ or $Z$) of the code connected to qubits involved in each stabiliser check.

\begin{figure}[!h]
\centering
\resizebox{0.95\linewidth}{!}{
    \begin{tikzpicture}[scale=4,every node/.style={minimum size=1cm},on grid]
        \begin{scope}[xshift=0,every node/.append style={yslant=0,xslant=0},yslant=0,xslant=0,rotate=-30]
            \node[fill=white,shape=circle,draw=black] (NA) at (0,0) {1};
            \node[fill=white,shape=circle,draw=black] (NB) at (3,1.73205080757) {2};
            \node[fill=white,shape=circle,draw=black] (ND) at (0,3.46410161514) {4};

            \node[fill=white,shape=circle,draw=black] (NE) at ($(NA)!0.5!(ND)$) {5};
            \node[fill=white,shape=circle,draw=black] (NC) at ($(NA)!0.5!(NB)$) {3};
            \node[fill=white,shape=circle,draw=black] (NF) at ($(ND)!0.5!(NB)$) {6};

            \node[fill=white,shape=circle,draw=black] (NG) at ($(NC)!0.33!(ND)$) {7};

            \node[fill=white,shape=rectangle,draw=black] (A1) at ($(NB)!0.5!(NG)$) {$A_1$};

            \node[fill=white,shape=rectangle,draw=black] (A2) at ($(NA)!0.5!(NG)$) {$A_2$};

            \node[fill=white,shape=rectangle,draw=black] (A3) at ($(ND)!0.5!(NG)$) {$A_3$};

            \path [-,line width=0.1cm,black,opacity=1] (A1) edge node {} (NB);
            \path [-,line width=0.1cm,black,opacity=1] (A1) edge node {} (NC);
            \path [-,line width=0.1cm,black,opacity=1] (A1) edge node {} (NF);
            \path [-,line width=0.1cm,black,opacity=1] (A1) edge node {} (NG);

            \path [-,line width=0.1cm,black,opacity=1] (A2) edge node {} (NA);
            \path [-,line width=0.1cm,black,opacity=1] (A2) edge node {} (NC);
            \path [-,line width=0.1cm,black,opacity=1] (A2) edge node {} (NE);
            \path [-,line width=0.1cm,black,opacity=1] (A2) edge node {} (NG);

            \path [-,line width=0.1cm,black,opacity=1] (A3) edge node {} (ND);
            \path [-,line width=0.1cm,black,opacity=1] (A3) edge node {} (NE);
            \path [-,line width=0.1cm,black,opacity=1] (A3) edge node {} (NF);
            \path [-,line width=0.1cm,black,opacity=1] (A3) edge node {} (NG);

            \path [-,line width=0.05cm,black,opacity=.1] (NC) edge node {} (NG);
            \path [-,line width=0.05cm,black,opacity=.1] (NF) edge node {} (NG);
            \path [-,line width=0.05cm,black,opacity=.1] (NE) edge node {} (NG);
            \begin{pgfonlayer}{background}
                
            \path [-,line width=0.05cm,black,opacity=.1] (NA) edge node {} (ND);
            \path [-,line width=0.05cm,black,opacity=.1] (NA) edge node {} (NB);
            \path [-,line width=0.05cm,black,opacity=.1] (ND) edge node {} (NB);

            \coordinate [] (A) at (NA);
            \coordinate [] (B) at (NB);
            \coordinate [] (C) at (NC);
            \coordinate [] (D) at (ND);
            \coordinate [] (E) at (NE);
            \coordinate [] (F) at (NF);
            \coordinate [] (G) at (NG);
            
            \draw[fill=blue!30, opacity=.25,draw=none] (B)--(C)--(G)--(F)--cycle;

            \draw[fill=green!30, opacity=.25,draw=none] (C)--(A)--(E)--(G)--cycle;

            \draw[fill=red!30, opacity=.25,draw=none] (D)--(E)--(G)--(F)--cycle;

            \end{pgfonlayer}

        \end{scope}

    \end{tikzpicture}}

    \caption{Tanner graph for the Steane code, encoded using 7 qubits arranged in the usual geometric manner on a triangle, rectangular nodes: $\{A_1,A_2,A_3\}$ are the self-dual $X$ or $Z$ stabiliser checks, while circular nodes 1-7 are the qubit labels.}
    \label{fig:steane_code_labelling}
\end{figure}

We take the 7-to-1 distillation circuit (figure \ref{fig:y_state_distillation_circuit_Y_consumption_optimised}) from \cite{zhou2024algorithmicfaulttolerancefast} with the first and fourth redundant transversal CNOTs removed. There is a one-to-one mapping of qubit labels between this circuit and qubit labels on the triangle in figure \ref{fig:steane_code_labelling}. The additional $0^{\text{th}}$ qubit in the circuit is the output subsystem of this distillation circuit. 
\begin{figure}
    \centering
    \resizebox{0.99\linewidth}{!}{
    \begin{tikzpicture}
    \begin{yquant}
    qubit {0: $\ket{{+}}$} q[+1];
    qubit {} a[+1]; discard a[0];
    qubit {1: $\ket{{+}}$} q[+1];
    qubit {} a[+1]; discard a[1];
    qubit {2: $\ket{{+}}$} q[+1];
    qubit {} a[+1]; discard a[2];
    qubit {3: $\ket{{0}}$\hspace{0.1cm}} q[+1];
    qubit {} a[+1]; discard a[3];
    qubit {4: $\ket{{+}}$} q[+1];
    qubit {} a[+1]; discard a[4];
    qubit {5: $\ket{{0}}$\hspace{0.1cm}} q[+1];
    qubit {} a[+1]; discard a[5];
    qubit {6: $\ket{{0}}$\hspace{0.1cm}} q[+1];
    qubit {} a[+1]; discard a[6];
    qubit {7: $\ket{{0}}$\hspace{0.1cm}} q[+1];

    [blue, thick, label=0]
    barrier (-);
    
    cnot q[5] | q[1];
    cnot q[6] | q[2];

    [blue, thick, label=1]
    barrier (-);

    align q;
    
    cnot q[2] | q[0];
    cnot q[6] | q[4];
    
    align q;
    
    cnot q[3] | q[1];
    cnot q[7] | q[5];

    [blue, thick, label=2]
    barrier (-);

    align q;
    
    cnot q[1] | q[0];
    cnot q[3] | q[2];
    cnot q[5] | q[4];
    cnot q[7] | q[6];
    
    [blue, thick, label=3]
    barrier (-);

    align a;

    init {$\ket{{Y}}$} a[0];
    cnot a[0] | q[1];
    init {$\ket{{Y}}$} a[1];
    cnot a[1] | q[2];
    init {$\ket{{Y}}$} a[2];
    cnot a[2] | q[3];
    init {$\ket{{Y}}$} a[3];
    cnot a[3] | q[4];
    init {$\ket{{Y}}$} a[4];
    cnot a[4] | q[5];
    init {$\ket{{Y}}$} a[5];
    cnot a[5] | q[6];
    init {$\ket{{Y}}$} a[6];
    cnot a[6] | q[7];
    [blue, thick, label=4]
    barrier (-);
    
    hspace {-2mm} a;
    
    measure a;
    
    text {$n_1$} a[0];
    text {$n_2$} a[1];
    text {$n_3$} a[2];
    text {$n_4$} a[3];
    text {$n_5$} a[4];
    text {$n_6$} a[5];
    text {$n_7$} a[6];
    align a, q;
    
    discard a;
    hspace {-7mm} q;

    measure {\tiny{$M_{{X}}$}} q[1],q[2],q[3],q[4],q[5],q[6],q[7];

    text {$m_1$} q[1];
    text {$m_2$} q[2];
    text {$m_3$} q[3];
    text {$m_4$} q[4];
    text {$m_5$} q[5];
    text {$m_6$} q[6];
    text {$m_7$} q[7];
    
    text {$\sim\ket{{Y}}$} q[0];

    discard q;
    \end{yquant}
    \end{tikzpicture}
    }
    \caption{The 7-to-1 distillation circuit from \cite{zhou2024algorithmicfaulttolerancefast} with the first and fourth redundant CNOTs removed as the CNOTs were acting on either $\ket{{+},{+}}$ or $\ket{{0},{0}}$. 
    Three rounds of syndrome extraction are performed on every patch at each time step marked by a blue dotted line, independent of $d$.
    }
\label{fig:y_state_distillation_circuit_Y_consumption_optimised}
\end{figure}

The 7-to-1 distillation circuit consists of 3 parallel layers of transversal CNOTs (separated by blue dotted lines in figure \ref{fig:y_state_distillation_circuit_Y_consumption_optimised}) performed on surface code patches initialised in the $\ket{+}$ or $\ket{0}$ state, followed by $\ket{Y}$ state consumption that perform $S$ gates on patches 1 to 7, before measuring out the corresponding patches in the $X$ or $Z$ basis. Prior to the state consumption and readout stage, this circuit encodes the state stabilised by:

\begin{equation}
\Big\langle X_0 \overbrace{\prod_{j=1}^{7}X_j}^{\bar{X}_{\text{Steane}}}, \ Z_0 \underbrace{\prod_{j=1}^{7}Z_j}_{\bar{Z}_{\text{Steane}}} \Big\rangle \cup \mathcal{G}_{\text{Steane}} \ ,
\label{eq:entangled_0_steane}
\end{equation}

\begin{equation}
    \begin{split}
\mathcal{G}_{\text{Steane}} = \Big\langle & X_1 X_3 X_5 X_7, X_2 X_3 X_6 X_7, \\ 
& X_4 X_5 X_6 X_7, Z_1 Z_3 Z_5 Z_7, \\
& Z_2 Z_3 Z_6 Z_7, Z_4 Z_5 Z_6 Z_7\Big\rangle \ ,
    \end{split}
\label{eq:g_steane}
\end{equation}
where $\mathcal{G}_{\text{Steane}}$ consists of Steane code stabilisers and $\bar{X}_{\text{Steane}}, \ \bar{Z}_{\text{Steane}}$ are the $X$, $Z$ logical operators of the Steane code. Equation \ref{eq:entangled_0_steane} is equivalent to a Bell state of the $0^{\text{th}}$ surface code patch entangled with the Steane code encoded with patches 1 to 7. The state consumption stage uses 7 $\ket{Y}$ states to perform an $S$ gate on surface code patches 1 to 7 through measurements in the $Z$ basis (binary results $n_j$). This is equivalent to performing a $\bar{S}_{\text{Steane}}$ gate at the Steane code level. After the $Z$ measurements, the resulting state is stabilised by:
\begin{equation}
\begin{split}
\Big\langle   
& (-Y_0) \bar{X}_{\text{Steane}} \cdot (-1)^{\sum_{k=1}^{7}n_k}, \\
& X_1 X_3 X_5 X_7 (-1)^{n_1+n_3+n_5+n_7}, \\ 
& X_2 X_3 X_6 X_7 (-1)^{n_2+n_3+n_6+n_7}, \\ 
& X_4 X_5 X_6 X_7 (-1)^{n_4+n_5+n_6+n_7}\Big\rangle \\
& \cup \text{ $Z$ type operators} \ .
\end{split}
\label{eq:Y_state_consumption_G}
\end{equation}
Next, perform the $X$ basis measurements (binary results $m_j$), the output state is stabilised by:
\begin{equation}
\begin{split}
\Big\langle 
& Y_0 \cdot (-1)^{\big(1+\sum_{k=1}^{7}n_k+m_k\big)} \Big\rangle \ .
\end{split} 
\label{eq:Y_state_consumption_G_mx}
\end{equation}
Note that we have managed to force the $0^{\text{th}}$ output surface code patch to be stabilised by $Y_0$ up to a $\pm$ sign, depending on measurement results. 

However, the input resource states or the distillation circuit can be faulty. The measurements results in the $X$ and $Z$ basis ($m_j, n_j\in\{0,1\}$) are used to construct the Steane code stabilisers. These stabiliser measurement results will be used to post-select and identify erroneous output states.

See detailed discussion in \cite{Herr_2017} on the 7-to-1 post-selection procedure. In summary:
\begin{enumerate}
    \item Construct the three weight-4 $X$ type Steane code stabilisers: 
    \begin{equation}
    \begin{split}
            X_1 X_3 X_5 X_7 = & \prod_{l\in\{n,m\}}(-1)^{l_1+l_3+l_5+l_7} \\
            X_2 X_3 X_6 X_7 = & \prod_{l\in\{n,m\}}(-1)^{l_2+l_3+l_6+l_7}\\
            X_4 X_5 X_6 X_7 = & \prod_{l\in\{n,m\}}(-1)^{l_4+l_5+l_6+l_7} \ , 
    \end{split}
    \end{equation}
    with the measurement results $(m_j, n_j)$, post-select on shots where all three stabilisers return the trivial result, i.e. $X_1 X_3 X_5 X_7=X_2 X_3 X_6 X_7= X_4 X_5 X_6 X_7=1$. 
    \item If $\displaystyle\prod_{j=1}^{7} (-1)^{n_j+m_j} = 1$, there is an additional $Z_0$ Clifford offset to the $0^{\text{th}}$ subsystem's Pauli-frame, that needs to be tracked (notice the sign of $Y_0$ in equation \ref{eq:Y_state_consumption_G_mx}).
\end{enumerate}
Successful post-selection outputs a $\ket{Y}$ state with suppressed error: $p_{\text{out}}=\mathcal{O}(7p^3)$, given $\ket{Y}$ resource states with input error rate $p$ and a perfect distillation circuit. Assuming an error-free 7-to-1 distillation circuit, the full expression \cite{zhou2024algorithmicfaulttolerancefast} for $p_{\text{out}}$ is:
\begin{equation}
    p_{\text{out}} = \frac{7p^{3}(1-p)^4+p^7}{p^7+(1-p)^7+7(1-p)^3p^4+7(1-p)^4p^3} \ .
    \label{eq:Y_scaling_full}
\end{equation}
Due to the post-selection procedure, a certain number of shots/experiments will be discarded. The number of shots discarded should be proportional to:
\begin{equation}
      \text{discard}_{7\to1}(p)  = \frac{7}{8}\left[1 - (1-2p)^4\right]  = 7p  + \mathcal{O}(p^2) \ ,
\end{equation}
at low error rates \cite{zhou2024algorithmicfaulttolerancefast}. We will discuss simulation results concerning the 7-to-1 distillation circuit in section \ref{section:surface_code_sim} and confirming its error suppression and discard ratio scaling under circuit-level noise.

\subsection{15-to-1 distillation protocol}
The 15-to-1 distillation protocol is enabled by the $[[15,1,3]]$ Reed-Muller code \cite{eczoo_stab_15_1_3}, which allows for a transversal $T$ gate. This code has a geometric interpretation of qubits lying on the vertices of a tetrahedron complex, cellulated by four identical 6-sided polyhedron cells, in the configuration shown in figure \ref{fig:3D_colour_code}. This code has four weight-8 $X$ stabiliser checks specified by the cells of this tetrahedron. One example of a weight-8 $X$ stabiliser check is shown in figure \ref{fig:3D_colour_code_X_check}.
\begin{figure}[!h]
    \centering
\resizebox{0.65\linewidth}{!}{\tdplotsetmaincoords{30}{50}
\begin{tikzpicture}[line join = round, line cap = round,tdplot_main_coords, rotate = 30]
\pgfmathsetmacro{\factor}{1/sqrt(2)};
\coordinate [] (A) at (2,0,-2*\factor);
\coordinate [] (B) at (-2,0,-2*\factor);
\coordinate [] (C) at (0,2,2*\factor);
\coordinate [] (D) at (0,-2,2*\factor);

\draw[-, fill=none, opacity=.5] (A)--(D)--(B)--cycle;
\draw[-, fill=none, opacity=.5] (A) --(D)--(C)--cycle;
\draw[-, fill=none, opacity=.5] (B)--(D)--(C)--cycle;

\node[fill=white,shape=circle,draw=black] (NA) at (A) {};
\node[fill=white,shape=circle,draw=black] (NB) at (B) {};
\node[fill=white,shape=circle,draw=black] (NC) at (C) {};
\node[fill=white,shape=circle,draw=black] (ND) at (D) {};

\node[fill=white,shape=circle,draw=black] (NBnD) at ($(NB)!0.5!(ND)$) {};
\node[fill=white,shape=circle,draw=black] (NBnDnC) at ($(NBnD)!0.33!(NC)$) {};
\node[fill=white,shape=circle,draw=black] (NBnC) at ($(NB)!0.5!(NC)$) {};
\node[fill=white,shape=circle,draw=black] (NDnC) at ($(ND)!0.5!(NC)$) {};

\path [-,line width=0.01cm,black,opacity=1] (NBnDnC) edge node {} (NBnC);
\path [-,line width=0.01cm,black,opacity=1] (NBnDnC) edge node {} (NDnC);
\path [-,line width=0.01cm,black,opacity=1] (NBnDnC) edge node {} (NBnD);

\node[fill=white,shape=circle,draw=black] (NDnA) at ($(ND)!0.5!(NA)$) {};
\node[fill=white,shape=circle,draw=black] (NDnAnC) at ($(NDnA)!0.33!(NC)$) {};
\node[fill=white,shape=circle,draw=black] (NAnC) at ($(NA)!0.5!(NC)$) {};

\path [-,line width=0.01cm,black,opacity=1] (NDnA) edge node {} (NDnAnC);
\path [-,line width=0.01cm,black,opacity=1] (NAnC) edge node {} (NDnAnC);
\path [-,line width=0.01cm,black,opacity=1] (NDnC) edge node {} (NDnAnC);

\begin{pgfonlayer}{background}
\node[fill=white,shape=circle,draw=black] (NBnA) at ($(NB)!0.5!(NA)$) {};
\node[fill=white,shape=circle,draw=black] (NBnAnC) at ($(NBnA)!0.33!(NC)$) {};
\path [-,line width=0.01cm,black,opacity=1] (NBnA) edge node {} (NBnAnC);
\path [-,line width=0.01cm,black,opacity=1] (NAnC) edge node {} (NBnAnC);
\path [-,line width=0.01cm,black,opacity=1] (NBnC) edge node {} (NBnAnC);

\node[fill=white,shape=circle,draw=black] (NBnAnD) at ($(NBnA)!0.33!(ND)$) {};
\path [-,line width=0.01cm,black,opacity=1] (NBnD) edge node {} (NBnAnD);
\path [-,line width=0.01cm,black,opacity=1] (NDnA) edge node {} (NBnAnD);
\path [-,line width=0.01cm,black,opacity=1] (NBnA) edge node {} (NBnAnD);

\node[fill=white,shape=circle,draw=black] (NBnAnDnC) at ($(NBnAnD)!0.22!(NC)$) {};

\path [-,line width=0.01cm,black,opacity=1] (NBnAnDnC) edge node {} (NBnAnD);
\path [-,line width=0.01cm,black,opacity=1] (NBnAnDnC) edge node {} (NBnAnC);
\path [-,line width=0.01cm,black,opacity=1] (NBnAnDnC) edge node {} (NDnAnC);
\path [-,line width=0.01cm,black,opacity=1] (NBnAnDnC) edge node {} (NBnDnC);
\end{pgfonlayer}

\begin{pgfonlayer}{background}
\coordinate [] (BD) at (NBnD);
\coordinate [] (BDC) at (NBnDnC);

\coordinate [] (BC) at (NBnC);
\coordinate [] (DC) at (NDnC);
\coordinate [] (DA) at (NDnA);
\coordinate [] (DAC) at (NDnAnC);
\coordinate [] (AC) at (NAnC);

\coordinate [] (BA) at (NBnA);
\coordinate [] (BAC) at (NBnAnC);
\coordinate [] (BAD) at (NBnAnD);
\coordinate [] (BADC) at (NBnAnDnC);

\draw[fill=blue!30, opacity=.25,draw=none] (DA)--(DAC)--(AC)--(A)--cycle;
\draw[fill=blue!30, opacity=.25,draw=none] (DA)--(BAD)--(BADC)--(DAC)--cycle;
\draw[fill=blue!30, opacity=.25,draw=none] (BADC)--(BAD)--(BA)--(BAC)--cycle;
\draw[fill=blue!30, opacity=.25,draw=none] (BAC)--(AC)--(A)--(BA)--cycle;
\draw[fill=blue!30, opacity=.25,draw=none] (BAD)--(BA)--(A)--(DA)--cycle;
\draw[fill=blue!30, opacity=.25,draw=none] (BADC)--(BAC)--(AC)--(DAC)--cycle;

\draw[fill=yellow!30, opacity=.25,draw=none] (B)--(BD)--(BAD)--(BA)--cycle;
\draw[fill=yellow!30, opacity=.25,draw=none] (BDC)--(BADC)--(BAC)--(BC)--cycle;
\draw[fill=yellow!30, opacity=.25,draw=none] (BDC)--(BADC)--(BAD)--(BD)--cycle;
\draw[fill=yellow!30, opacity=.25,draw=none] (B)--(BA)--(BAC)--(BC)--cycle;
\draw[fill=yellow!30, opacity=.25,draw=none] (B)--(BC)--(BDC)--(BD)--cycle;
\draw[fill=yellow!30, opacity=.25,draw=none] (BAC)--(BADC)--(BAD)--(BA)--cycle;
\draw[fill=red!30, opacity=.25,draw=none] (BADC)--(DAC)--(DC)--(BDC)--cycle;
\draw[fill=red!30, opacity=.25,draw=none] (BADC)--(DAC)--(AC)--(BAC)--cycle;
\draw[fill=red!30, opacity=.25,draw=none] (BADC)--(BAC)--(BC)--(BDC)--cycle;
\draw[fill=red!30, opacity=.25,draw=none] (DAC)--(DC)--(C)--(AC)--cycle;
\draw[fill=red!30, opacity=.25,draw=none] (AC)--(BAC)--(BC)--(C)--cycle;
\draw[fill=red!30, opacity=.25,draw=none] (BDC)--(DC)--(C)--(BC)--cycle;
\draw[fill=green!30, opacity=.25,draw=none] (DA)--(D)--(DC)--(DAC)--cycle;
\draw[fill=green!30, opacity=.25,draw=none] (D)--(BD)--(BDC)--(DC)--cycle;
\draw[fill=green!30, opacity=.25,draw=none] (BD)--(BDC)--(BADC)--(BAD)--cycle;
\draw[fill=green!30, opacity=.25,draw=none] (BADC)--(DAC)--(DA)--(BAD)--cycle;
\draw[fill=green!30, opacity=.25,draw=none] (D)--(BD)--(BAD)--(DA)--cycle;
\draw[fill=green!30, opacity=.25,draw=none] (BADC)--(DAC)--(DC)--(BDC)--cycle;
\end{pgfonlayer}

\end{tikzpicture}}
    \caption{The $[[15,1,3]]$ Reed-Muller code's geometric representation with encoding qubits as circular nodes, rectangular stabiliser check nodes and its edges are excluded for representation purposes.}
    \label{fig:3D_colour_code}
\end{figure}

\begin{figure}[!h]
    \centering
\resizebox{0.65\linewidth}{!}{\tdplotsetmaincoords{30}{50}
\begin{tikzpicture}[line join = round, line cap = round,tdplot_main_coords, rotate = 30]
\pgfmathsetmacro{\factor}{1/sqrt(2)};
\coordinate [] (A) at (2,0,-2*\factor);
\coordinate [] (B) at (-2,0,-2*\factor);
\coordinate [] (C) at (0,2,2*\factor);
\coordinate [] (D) at (0,-2,2*\factor);

\draw[-, fill=none, opacity=.5] (A)--(D)--(B)--cycle;
\draw[-, fill=none, opacity=.5] (A) --(D)--(C)--cycle;
\draw[-, fill=none, opacity=.5] (B)--(D)--(C)--cycle;

\node[fill=white,shape=circle,draw=black] (NA) at (A) {};
\node[fill=none,shape=circle,draw=none] (NB) at (B) {};
\node[fill=none,shape=circle,draw=none] (NC) at (C) {};
\node[fill=none,shape=circle,draw=none] (ND) at (D) {};

\node[fill=none,shape=circle,draw=none] (NBnD) at ($(NB)!0.5!(ND)$) {};
\node[fill=none,shape=circle,draw=none] (NBnDnC) at ($(NBnD)!0.33!(NC)$) {};
\node[fill=none,shape=circle,draw=none] (NBnC) at ($(NB)!0.5!(NC)$) {};
\node[fill=none,shape=circle,draw=none] (NDnC) at ($(ND)!0.5!(NC)$) {};

\node[fill=white,shape=circle,draw=black] (NDnA) at ($(ND)!0.5!(NA)$) {};
\node[fill=white,shape=circle,draw=black] (NDnAnC) at ($(NDnA)!0.33!(NC)$) {};
\node[fill=white,shape=circle,draw=black] (NAnC) at ($(NA)!0.5!(NC)$) {};

\path [-,line width=0.01cm,black,opacity=1] (NDnA) edge node {} (NDnAnC);
\path [-,line width=0.01cm,black,opacity=1] (NAnC) edge node {} (NDnAnC);

\begin{pgfonlayer}{background}
\node[fill=white,shape=circle,draw=black] (NBnA) at ($(NB)!0.5!(NA)$) {};
\node[fill=white,shape=circle,draw=black] (NBnAnC) at ($(NBnA)!0.33!(NC)$) {};
\path [-,line width=0.01cm,black,opacity=1] (NBnA) edge node {} (NBnAnC);
\path [-,line width=0.01cm,black,opacity=1] (NAnC) edge node {} (NBnAnC);

\node[fill=white,shape=circle,draw=black] (NBnAnD) at ($(NBnA)!0.33!(ND)$) {};
\path [-,line width=0.01cm,black,opacity=1] (NDnA) edge node {} (NBnAnD);
\path [-,line width=0.01cm,black,opacity=1] (NBnA) edge node {} (NBnAnD);

\node[fill=white,shape=circle,draw=black] (NBnAnDnC) at ($(NBnAnD)!0.22!(NC)$) {};

\path [-,line width=0.01cm,black,opacity=1] (NBnAnDnC) edge node {} (NBnAnD);
\path [-,line width=0.01cm,black,opacity=1] (NBnAnDnC) edge node {} (NBnAnC);
\path [-,line width=0.01cm,black,opacity=1] (NBnAnDnC) edge node {} (NDnAnC);

\end{pgfonlayer}

\path [-,line width=0.01cm,black,opacity=1] (NDnA) edge node {} (NA);
\path [-,line width=0.01cm,black,opacity=1] (NA) edge node {} (NBnA);
\path [-,line width=0.01cm,black,opacity=1] (NA) edge node {} (NAnC);

\node[fill=white,shape=rectangle,draw=black] (ncheck) at ($(NBnAnDnC)!0.5!(NA)$) {};
\path [-,line width=0.04cm,black,opacity=1] (ncheck) edge node {} (NDnA);
\path [-,line width=0.04cm,black,opacity=1] (ncheck) edge node {} (NDnAnC);
\path [-,line width=0.04cm,black,opacity=1] (ncheck) edge node {} (NAnC);
\path [-,line width=0.04cm,black,opacity=1] (ncheck) edge node {} (NBnA);
\path [-,line width=0.04cm,black,opacity=1] (ncheck) edge node {} (NBnAnDnC);
\path [-,line width=0.04cm,black,opacity=1] (ncheck) edge node {} (NA);
\path [-,line width=0.04cm,black,opacity=1] (ncheck) edge node {} (NBnAnD);

\begin{pgfonlayer}{background}
\path [-,line width=0.04cm,black,opacity=1] (ncheck) edge node {} (NBnAnC);
\coordinate [] (BD) at (NBnD);
\coordinate [] (BDC) at (NBnDnC);

\coordinate [] (BC) at (NBnC);
\coordinate [] (DC) at (NDnC);
\coordinate [] (DA) at (NDnA);
\coordinate [] (DAC) at (NDnAnC);
\coordinate [] (AC) at (NAnC);

\coordinate [] (BA) at (NBnA);
\coordinate [] (BAC) at (NBnAnC);
\coordinate [] (BAD) at (NBnAnD);
\coordinate [] (BADC) at (NBnAnDnC);

\draw[fill=blue!30, opacity=.25,draw=none] (DA)--(DAC)--(AC)--(A)--cycle;
\draw[fill=blue!30, opacity=.25,draw=none] (DA)--(BAD)--(BADC)--(DAC)--cycle;
\draw[fill=blue!30, opacity=.25,draw=none] (BADC)--(BAD)--(BA)--(BAC)--cycle;
\draw[fill=blue!30, opacity=.25,draw=none] (BAC)--(AC)--(A)--(BA)--cycle;
\draw[fill=blue!30, opacity=.25,draw=none] (BAD)--(BA)--(A)--(DA)--cycle;
\draw[fill=blue!30, opacity=.25,draw=none] (BADC)--(BAC)--(AC)--(DAC)--cycle;

\end{pgfonlayer}

\end{tikzpicture}}
    \caption{One of the four weight-8 $X$ stabiliser check cells in the $[[15,1,3]]$ Reed-Muller code with the rectangular stabiliser check node explicitly shown.}
    \label{fig:3D_colour_code_X_check}
\end{figure}

The circuit in figure \ref{fig:T_state_distillation_circuit_T_consumption_optimised} is a construction of the 15-to-1 distillation protocol \cite{costofuni}. This circuit consists of 16 surface code patches initialised in the $\ket{+}$ or $\ket{0}$ state, before going through 4 parallel layers of CNOTs. This CNOT circuit (up until the blue dotted line labelled 4 in figure \ref{fig:T_state_distillation_circuit_T_consumption_optimised}) generates a $r=1$ Reed-Muller state, $\ket{\text{QRM}_{r}}$, from \cite{costofuni}. This 16 patch $\ket{\text{QRM}_{r}}$ state is a Bell state of the $0^{\text{th}}$ surface code patch entangled with the Reed-Muller code, stabilised by: 
\begin{equation}
    \Big\langle X_0\bar{X}_{\text{RM}}, \ Z_0\bar{Z}_{\text{RM}} \Big\rangle \cup \ \mathcal{G}_{\text{RM}} \ .
\end{equation}
The generators $\mathcal{G}_{\text{RM}}$ are the Reed-Muller code stabilisers and $\bar{X}_{\text{RM}}$ and $\bar{Z}_{\text{RM}}$ are the logical operators of the Reed-Muller code:
\begin{equation}
    \bar{X}_{\text{RM}} = \prod_{j\in \text{Face}} X_j = 
    \resizebox{0.25\linewidth}{!}{\tdplotsetmaincoords{30}{50}
\begin{tikzpicture}[baseline={(0, 0.5cm-\MathAxis pt)},line join = round, line cap = round,tdplot_main_coords, rotate = 30]
\pgfmathsetmacro{\factor}{1/sqrt(2)};
\coordinate [] (A) at (2,0,-2*\factor);
\coordinate [] (B) at (-2,0,-2*\factor);
\coordinate [] (C) at (0,2,2*\factor);
\coordinate [] (D) at (0,-2,2*\factor);

\draw[-, fill=none, opacity=.5] (A)--(D)--(B)--cycle;
\draw[-, fill=none, opacity=.5] (A) --(D)--(C)--cycle;
\draw[-, fill=none, opacity=.5] (B)--(D)--(C)--cycle;

\node[fill=white,shape=circle,draw=black] (NA) at (A) {};
\node[fill=white,shape=circle,draw=black] (NB) at (B) {};
\node[fill=white,shape=circle,draw=black] (NC) at (C) {};
\node[fill=white,shape=circle,draw=black] (ND) at (D) {};

\node[fill=white,shape=circle,draw=black] (NBnD) at ($(NB)!0.5!(ND)$) {};
\node[fill=white,shape=circle,draw=black] (NBnDnC) at ($(NBnD)!0.33!(NC)$) {};
\node[fill=white,shape=circle,draw=black] (NBnC) at ($(NB)!0.5!(NC)$) {};
\node[fill=white,shape=circle,draw=black] (NDnC) at ($(ND)!0.5!(NC)$) {};

\path [-,line width=0.01cm,black,opacity=1] (NBnDnC) edge node {} (NBnC);
\path [-,line width=0.01cm,black,opacity=1] (NBnDnC) edge node {} (NDnC);
\path [-,line width=0.01cm,black,opacity=1] (NBnDnC) edge node {} (NBnD);

\node[fill=white,shape=circle,draw=black] (NDnA) at ($(ND)!0.5!(NA)$) {};
\node[fill=white,shape=circle,draw=black] (NDnAnC) at ($(NDnA)!0.33!(NC)$) {};
\node[fill=white,shape=circle,draw=black] (NAnC) at ($(NA)!0.5!(NC)$) {};

\path [-,line width=0.01cm,black,opacity=1] (NDnA) edge node {} (NDnAnC);
\path [-,line width=0.01cm,black,opacity=1] (NAnC) edge node {} (NDnAnC);
\path [-,line width=0.01cm,black,opacity=1] (NDnC) edge node {} (NDnAnC);

\begin{pgfonlayer}{background}
\node[fill=white,shape=circle,draw=black] (NBnA) at ($(NB)!0.5!(NA)$) {};
\node[fill=white,shape=circle,draw=black] (NBnAnC) at ($(NBnA)!0.33!(NC)$) {};
\path [-,line width=0.01cm,black,opacity=1] (NBnA) edge node {} (NBnAnC);
\path [-,line width=0.01cm,black,opacity=1] (NAnC) edge node {} (NBnAnC);
\path [-,line width=0.01cm,black,opacity=1] (NBnC) edge node {} (NBnAnC);

\node[fill=white,shape=circle,draw=black] (NBnAnD) at ($(NBnA)!0.33!(ND)$) {};
\path [-,line width=0.01cm,black,opacity=1] (NBnD) edge node {} (NBnAnD);
\path [-,line width=0.01cm,black,opacity=1] (NDnA) edge node {} (NBnAnD);
\path [-,line width=0.01cm,black,opacity=1] (NBnA) edge node {} (NBnAnD);

\node[fill=white,shape=circle,draw=black] (NBnAnDnC) at ($(NBnAnD)!0.22!(NC)$) {};

\path [-,line width=0.01cm,black,opacity=1] (NBnAnDnC) edge node {} (NBnAnD);
\path [-,line width=0.01cm,black,opacity=1] (NBnAnDnC) edge node {} (NBnAnC);
\path [-,line width=0.01cm,black,opacity=1] (NBnAnDnC) edge node {} (NDnAnC);
\path [-,line width=0.01cm,black,opacity=1] (NBnAnDnC) edge node {} (NBnDnC);
\end{pgfonlayer}

\begin{pgfonlayer}{background}
\coordinate [] (BD) at (NBnD);
\coordinate [] (BDC) at (NBnDnC);

\coordinate [] (BC) at (NBnC);
\coordinate [] (DC) at (NDnC);
\coordinate [] (DA) at (NDnA);
\coordinate [] (DAC) at (NDnAnC);
\coordinate [] (AC) at (NAnC);

\coordinate [] (BA) at (NBnA);
\coordinate [] (BAC) at (NBnAnC);
\coordinate [] (BAD) at (NBnAnD);
\coordinate [] (BADC) at (NBnAnDnC);

\draw[fill=blue!30, opacity=.25,draw=none] (DA)--(DAC)--(AC)--(A)--cycle;
\draw[fill=blue!30, opacity=.25,draw=none] (DA)--(BAD)--(BADC)--(DAC)--cycle;
\draw[fill=blue!30, opacity=.25,draw=none] (BADC)--(BAD)--(BA)--(BAC)--cycle;
\draw[fill=blue!30, opacity=.25,draw=none] (BAC)--(AC)--(A)--(BA)--cycle;
\draw[fill=blue!30, opacity=.25,draw=none] (BAD)--(BA)--(A)--(DA)--cycle;
\draw[fill=blue!30, opacity=.25,draw=none] (BADC)--(BAC)--(AC)--(DAC)--cycle;

\draw[fill=yellow!30, opacity=.25,draw=none] (B)--(BD)--(BAD)--(BA)--cycle;
\draw[fill=yellow!30, opacity=.25,draw=none] (BDC)--(BADC)--(BAC)--(BC)--cycle;
\draw[fill=yellow!30, opacity=.25,draw=none] (BDC)--(BADC)--(BAD)--(BD)--cycle;
\draw[fill=yellow!30, opacity=.25,draw=none] (B)--(BA)--(BAC)--(BC)--cycle;
\draw[fill=yellow!30, opacity=.25,draw=none] (B)--(BC)--(BDC)--(BD)--cycle;
\draw[fill=yellow!30, opacity=.25,draw=none] (BAC)--(BADC)--(BAD)--(BA)--cycle;
\draw[fill=red!30, opacity=.25,draw=none] (BADC)--(DAC)--(DC)--(BDC)--cycle;
\draw[fill=red!30, opacity=.25,draw=none] (BADC)--(DAC)--(AC)--(BAC)--cycle;
\draw[fill=red!30, opacity=.25,draw=none] (BADC)--(BAC)--(BC)--(BDC)--cycle;
\draw[fill=red!30, opacity=.25,draw=none] (DAC)--(DC)--(C)--(AC)--cycle;
\draw[fill=red!30, opacity=.25,draw=none] (AC)--(BAC)--(BC)--(C)--cycle;
\draw[fill=red!30, opacity=.25,draw=none] (BDC)--(DC)--(C)--(BC)--cycle;
\draw[fill=green!30, opacity=.25,draw=none] (DA)--(D)--(DC)--(DAC)--cycle;
\draw[fill=green!30, opacity=.25,draw=none] (D)--(BD)--(BDC)--(DC)--cycle;
\draw[fill=green!30, opacity=.25,draw=none] (BD)--(BDC)--(BADC)--(BAD)--cycle;
\draw[fill=green!30, opacity=.25,draw=none] (BADC)--(DAC)--(DA)--(BAD)--cycle;
\draw[fill=green!30, opacity=.25,draw=none] (D)--(BD)--(BAD)--(DA)--cycle;
\draw[fill=green!30, opacity=.25,draw=none] (BADC)--(DAC)--(DC)--(BDC)--cycle;
\end{pgfonlayer}
\node[fill=red,shape=circle,draw=black] (C) at (C) {};
\node[fill=red,shape=circle,draw=black] (D) at (D) {};
\node[fill=red,shape=circle,draw=black] (A) at (A) {};
\node[fill=red,shape=circle,draw=black] (DC) at (DC) {};
\node[fill=red,shape=circle,draw=black] (DAC) at (DAC) {};
\node[fill=red,shape=circle,draw=black] (DA) at (DA) {};
\node[fill=red,shape=circle,draw=black] (AC) at (AC) {};
\end{tikzpicture}} \ ,
\end{equation}
\begin{equation}
    \bar{Z}_{\text{RM}} = \prod_{j\in \text{Edge}} Z_j = 
    \resizebox{0.25\linewidth}{!}{\tdplotsetmaincoords{30}{50}
\begin{tikzpicture}[baseline={(0, 0.5cm-\MathAxis pt)},line join = round, line cap = round,tdplot_main_coords, rotate = 30]
\pgfmathsetmacro{\factor}{1/sqrt(2)};
\coordinate [] (A) at (2,0,-2*\factor);
\coordinate [] (B) at (-2,0,-2*\factor);
\coordinate [] (C) at (0,2,2*\factor);
\coordinate [] (D) at (0,-2,2*\factor);

\draw[-, fill=none, opacity=.5] (A)--(D)--(B)--cycle;
\draw[-, fill=none, opacity=.5] (A) --(D)--(C)--cycle;
\draw[-, fill=none, opacity=.5] (B)--(D)--(C)--cycle;

\node[fill=white,shape=circle,draw=black] (NA) at (A) {};
\node[fill=white,shape=circle,draw=black] (NB) at (B) {};
\node[fill=white,shape=circle,draw=black] (NC) at (C) {};
\node[fill=white,shape=circle,draw=black] (ND) at (D) {};

\node[fill=white,shape=circle,draw=black] (NBnD) at ($(NB)!0.5!(ND)$) {};
\node[fill=white,shape=circle,draw=black] (NBnDnC) at ($(NBnD)!0.33!(NC)$) {};
\node[fill=white,shape=circle,draw=black] (NBnC) at ($(NB)!0.5!(NC)$) {};
\node[fill=white,shape=circle,draw=black] (NDnC) at ($(ND)!0.5!(NC)$) {};

\path [-,line width=0.01cm,black,opacity=1] (NBnDnC) edge node {} (NBnC);
\path [-,line width=0.01cm,black,opacity=1] (NBnDnC) edge node {} (NDnC);
\path [-,line width=0.01cm,black,opacity=1] (NBnDnC) edge node {} (NBnD);

\node[fill=white,shape=circle,draw=black] (NDnA) at ($(ND)!0.5!(NA)$) {};
\node[fill=white,shape=circle,draw=black] (NDnAnC) at ($(NDnA)!0.33!(NC)$) {};
\node[fill=white,shape=circle,draw=black] (NAnC) at ($(NA)!0.5!(NC)$) {};

\path [-,line width=0.01cm,black,opacity=1] (NDnA) edge node {} (NDnAnC);
\path [-,line width=0.01cm,black,opacity=1] (NAnC) edge node {} (NDnAnC);
\path [-,line width=0.01cm,black,opacity=1] (NDnC) edge node {} (NDnAnC);

\begin{pgfonlayer}{background}
\node[fill=white,shape=circle,draw=black] (NBnA) at ($(NB)!0.5!(NA)$) {};
\node[fill=white,shape=circle,draw=black] (NBnAnC) at ($(NBnA)!0.33!(NC)$) {};
\path [-,line width=0.01cm,black,opacity=1] (NBnA) edge node {} (NBnAnC);
\path [-,line width=0.01cm,black,opacity=1] (NAnC) edge node {} (NBnAnC);
\path [-,line width=0.01cm,black,opacity=1] (NBnC) edge node {} (NBnAnC);

\node[fill=white,shape=circle,draw=black] (NBnAnD) at ($(NBnA)!0.33!(ND)$) {};
\path [-,line width=0.01cm,black,opacity=1] (NBnD) edge node {} (NBnAnD);
\path [-,line width=0.01cm,black,opacity=1] (NDnA) edge node {} (NBnAnD);
\path [-,line width=0.01cm,black,opacity=1] (NBnA) edge node {} (NBnAnD);

\node[fill=white,shape=circle,draw=black] (NBnAnDnC) at ($(NBnAnD)!0.22!(NC)$) {};

\path [-,line width=0.01cm,black,opacity=1] (NBnAnDnC) edge node {} (NBnAnD);
\path [-,line width=0.01cm,black,opacity=1] (NBnAnDnC) edge node {} (NBnAnC);
\path [-,line width=0.01cm,black,opacity=1] (NBnAnDnC) edge node {} (NDnAnC);
\path [-,line width=0.01cm,black,opacity=1] (NBnAnDnC) edge node {} (NBnDnC);
\end{pgfonlayer}

\begin{pgfonlayer}{background}
\coordinate [] (BD) at (NBnD);
\coordinate [] (BDC) at (NBnDnC);

\coordinate [] (BC) at (NBnC);
\coordinate [] (DC) at (NDnC);
\coordinate [] (DA) at (NDnA);
\coordinate [] (DAC) at (NDnAnC);
\coordinate [] (AC) at (NAnC);

\coordinate [] (BA) at (NBnA);
\coordinate [] (BAC) at (NBnAnC);
\coordinate [] (BAD) at (NBnAnD);
\coordinate [] (BADC) at (NBnAnDnC);

\draw[fill=blue!30, opacity=.25,draw=none] (DA)--(DAC)--(AC)--(A)--cycle;
\draw[fill=blue!30, opacity=.25,draw=none] (DA)--(BAD)--(BADC)--(DAC)--cycle;
\draw[fill=blue!30, opacity=.25,draw=none] (BADC)--(BAD)--(BA)--(BAC)--cycle;
\draw[fill=blue!30, opacity=.25,draw=none] (BAC)--(AC)--(A)--(BA)--cycle;
\draw[fill=blue!30, opacity=.25,draw=none] (BAD)--(BA)--(A)--(DA)--cycle;
\draw[fill=blue!30, opacity=.25,draw=none] (BADC)--(BAC)--(AC)--(DAC)--cycle;

\draw[fill=yellow!30, opacity=.25,draw=none] (B)--(BD)--(BAD)--(BA)--cycle;
\draw[fill=yellow!30, opacity=.25,draw=none] (BDC)--(BADC)--(BAC)--(BC)--cycle;
\draw[fill=yellow!30, opacity=.25,draw=none] (BDC)--(BADC)--(BAD)--(BD)--cycle;
\draw[fill=yellow!30, opacity=.25,draw=none] (B)--(BA)--(BAC)--(BC)--cycle;
\draw[fill=yellow!30, opacity=.25,draw=none] (B)--(BC)--(BDC)--(BD)--cycle;
\draw[fill=yellow!30, opacity=.25,draw=none] (BAC)--(BADC)--(BAD)--(BA)--cycle;
\draw[fill=red!30, opacity=.25,draw=none] (BADC)--(DAC)--(DC)--(BDC)--cycle;
\draw[fill=red!30, opacity=.25,draw=none] (BADC)--(DAC)--(AC)--(BAC)--cycle;
\draw[fill=red!30, opacity=.25,draw=none] (BADC)--(BAC)--(BC)--(BDC)--cycle;
\draw[fill=red!30, opacity=.25,draw=none] (DAC)--(DC)--(C)--(AC)--cycle;
\draw[fill=red!30, opacity=.25,draw=none] (AC)--(BAC)--(BC)--(C)--cycle;
\draw[fill=red!30, opacity=.25,draw=none] (BDC)--(DC)--(C)--(BC)--cycle;
\draw[fill=green!30, opacity=.25,draw=none] (DA)--(D)--(DC)--(DAC)--cycle;
\draw[fill=green!30, opacity=.25,draw=none] (D)--(BD)--(BDC)--(DC)--cycle;
\draw[fill=green!30, opacity=.25,draw=none] (BD)--(BDC)--(BADC)--(BAD)--cycle;
\draw[fill=green!30, opacity=.25,draw=none] (BADC)--(DAC)--(DA)--(BAD)--cycle;
\draw[fill=green!30, opacity=.25,draw=none] (D)--(BD)--(BAD)--(DA)--cycle;
\draw[fill=green!30, opacity=.25,draw=none] (BADC)--(DAC)--(DC)--(BDC)--cycle;
\end{pgfonlayer}
\node[fill=blue,shape=circle,draw=black] (D) at (D) {};
\node[fill=blue,shape=circle,draw=black] (DC) at (DC) {};
\node[fill=blue,shape=circle,draw=black] (C) at (C) {};
\end{tikzpicture}} \ .
\end{equation}
The operators $\bar{X}_{\text{RM}}$ and $\bar{Z}_{\text{RM}}$ are any weight-7 $X$-faces and any weight-3 $Z$-edge strings of the entire tetrahedron \cite{eczoo_stab_15_1_3}. 

After the CNOT circuit, a logical $T$ gate is performed at the Reed-Muller code level via consuming 15 surface code $\ket{T}$ states and measurements in the $Z$ basis. Conditioned on the $Z$ basis measurements, $S$ gate(s) will be performed on the appropriate patch(es). This is followed by measurements in the $X$ basis, in order to construct all four weight-8 $X$ stabiliser checks of the Reed-Muller code:
\begin{equation}
\begin{split}
           \langle & X_{1} X_{3} X_{5} X_{7} X_{9} X_{11} X_{13} X_{15} ,\\
           & X_{2} X_{3} X_{6} X_{7} X_{10} X_{11} X_{14} X_{15} ,\\
           & X_{4} X_{5} X_{6} X_{7} X_{12} X_{13} X_{14} X_{15} ,\\
           & X_{8} X_{9} X_{10} X_{11} X_{12} X_{13} X_{14} X_{15}\rangle \ .
\end{split}
\end{equation}
The measurement results are then post-selected upon the trivial result on all four stabiliser checks. The value of $\bar{X}_{\text{RM}}$, given by $M({\bar{X}_{\text{RM}}})$ must also be constructed from these measurement results, as the final gate implemented on the $0^{\text{th}}$ output patch is $T^{3-2M({\bar{X}_{\text{RM}}})}$ \cite{costofuni}. 

\begin{figure}[!h]
    \centering
    \resizebox{0.99\linewidth}{!}{
    \begin{tikzpicture}
    \begin{yquant}
    qubit {0: $\ket{{+}}$} q[+1];
    qubit {} a[+1]; discard a[0];
    qubit {1: $\ket{{+}}$} q[+1];
    qubit {} a[+1]; discard a[1];
    qubit {2: $\ket{{+}}$} q[+1];
    qubit {} a[+1]; discard a[2];
    qubit {3: $\ket{{0}}$\hspace{0.1cm}} q[+1];
    qubit {} a[+1]; discard a[3];
    qubit {4: $\ket{{+}}$} q[+1];
    qubit {} a[+1]; discard a[4];
    qubit {5: $\ket{{0}}$\hspace{0.1cm}} q[+1];
    qubit {} a[+1]; discard a[5];
    qubit {6: $\ket{{0}}$\hspace{0.1cm}} q[+1];
    qubit {} a[+1]; discard a[6];
    qubit {7: $\ket{{0}}$\hspace{0.1cm}} q[+1];
    qubit {} a[+1]; discard a[7];
    qubit {8: $\ket{{+}}$} q[+1];
    qubit {} a[+1]; discard a[8];
    qubit {9: $\ket{{0}}$\hspace{0.1cm}} q[+1];
    qubit {} a[+1]; discard a[9];
    qubit {10: $\ket{{0}}$\hspace{0.1cm}} q[+1];
    qubit {} a[+1]; discard a[10];
    qubit {11: $\ket{{0}}$\hspace{0.1cm}} q[+1];
    qubit {} a[+1]; discard a[11];
    qubit {12: $\ket{{0}}$\hspace{0.1cm}} q[+1];
    qubit {} a[+1]; discard a[12];
    qubit {13: $\ket{{0}}$\hspace{0.1cm}} q[+1];
    qubit {} a[+1]; discard a[13];
    qubit {14: $\ket{{0}}$\hspace{0.1cm}} q[+1];
    qubit {} a[+1]; discard a[14];
    qubit {15: $\ket{{0}}$\hspace{0.1cm}} q[+1];
    
    [blue, thick, label=0]
    barrier (-);

    cnot q[9] | q[1];
    cnot q[10] | q[2];
    cnot q[12] | q[4];

    [blue, thick, label=1]
    barrier (-);

    align q;
    
    cnot q[4] | q[0];
    cnot q[5] | q[1];
    cnot q[6] | q[2];

    cnot q[12] | q[8];
    cnot q[13] | q[9];
    cnot q[14] | q[10];

    [blue, thick, label=2]
    barrier (-);

    align q;
    
    cnot q[2] | q[0];
    cnot q[6] | q[4];
    cnot q[10] | q[8];
    cnot q[14] | q[12];
    
    align q;
    
    cnot q[3] | q[1];
    cnot q[7] | q[5];
    cnot q[11] | q[9];
    cnot q[15] | q[13];
    
    [blue, thick, label=3]
    barrier (-);

    align q;
    
    cnot q[1] | q[0];
    cnot q[3] | q[2];
    cnot q[5] | q[4];
    cnot q[7] | q[6];

    cnot q[9] | q[8];
    cnot q[11] | q[10];
    cnot q[13] | q[12];
    cnot q[15] | q[14];
    
    [blue, thick, label=4]
    barrier (-);

    align a;

    init {$\ket{{T}}$} a[0];
    cnot a[0] | q[1];
    init {$\ket{{T}}$} a[1];
    cnot a[1] | q[2];
    init {$\ket{{T}}$} a[2];
    cnot a[2] | q[3];
    init {$\ket{{T}}$} a[3];
    cnot a[3] | q[4];
    init {$\ket{{T}}$} a[4];
    cnot a[4] | q[5];
    init {$\ket{{T}}$} a[5];
    cnot a[5] | q[6];
    init {$\ket{{T}}$} a[6];
    cnot a[6] | q[7];
    
    init {$\ket{{T}}$} a[7];
    cnot a[7] | q[8];
    init {$\ket{{T}}$} a[8];
    cnot a[8] | q[9];
    init {$\ket{{T}}$} a[9];
    cnot a[9] | q[10];
    init {$\ket{{T}}$} a[10];
    cnot a[10] | q[11];
    init {$\ket{{T}}$} a[11];
    cnot a[11] | q[12];
    init {$\ket{{T}}$} a[12];
    cnot a[12] | q[13];
    init {$\ket{{T}}$} a[13];
    cnot a[13] | q[14];
    init {$\ket{{T}}$} a[14];
    cnot a[14] | q[15];

    align a; align q;
    
    [blue, thick, label=5]
    barrier (-);

    align a,q;

measure a[0];
    box {$S$} (q[1]) | a[0];
    discard a[0];

 measure a[1];
    box {$S$} (q[2]) | a[1];
    discard a[1];

 measure a[2];
    box {$S$} (q[3]) | a[2];
    discard a[2];

 measure a[3];
    box {$S$} (q[4]) | a[3];
    discard a[3];

 measure a[4];
    box {$S$} (q[5]) | a[4];
    discard a[4];

 measure a[5];
    box {$S$} (q[6]) | a[5];
    discard a[5];

 measure a[6];
    box {$S$} (q[7]) | a[6];
    discard a[6];

 measure a[7];
    box {$S$} (q[8]) | a[7];
    discard a[7];

measure a[8];
    box {$S$} (q[9]) | a[8];
    discard a[8];
    
measure a[9];
    box {$S$} (q[10]) | a[9];
    discard a[9];

measure a[10];
    box {$S$} (q[11]) | a[10];
    discard a[10];

measure a[11];
    box {$S$} (q[12]) | a[11];
    discard a[11];

measure a[12];
    box {$S$} (q[13]) | a[12];
    discard a[12];

measure a[13];
    box {$S$} (q[14]) | a[13];
    discard a[13];

measure a[14];
    box {$S$} (q[15]) | a[14];
    discard a[14];

    hspace {10mm} - q[0]; text {$\sim\ket{{T}}$} q[0]; discard q[0];

    measure {\tiny{$M_{{X}}$}} q[1],q[2],q[3],q[4],q[5],q[6],q[7],q[8],q[9],q[10],q[11],q[12],q[13],q[14],q[15];

    \end{yquant}
    \end{tikzpicture}
    }
    \caption{The 15-to-1 ($\ket{T}$ state) distillation circuit from \cite{costofuni} with all the redundant CNOTs removed. 
    Three rounds of syndrome extraction are performed at every time step indicated by blue dotted lines in the surface code simulations.
    }
\label{fig:T_state_distillation_circuit_T_consumption_optimised}
\end{figure}

Given 15 faulty $\ket{T}$ states with input logical error probability $p$, the error-free 15-to-1 distillation circuit suppresses the error of the output state to $\mathcal{O}(35p^3)$, with the full expression below \cite{PhysRevA.71.022316}:
\begin{equation}
p_{\text{out}} = \frac{1-15(1-2p)^7+15(1-2p)^8-(1-2p)^{15}}{2(1+15(1-2p)^8)} \ .
    \label{eq:T_scaling_full}
\end{equation}
Similar to the 7-to-1 protocol, the post-selection procedure should discard 
\begin{equation}
      \text{discard}_{15\to1}(p)  = \frac{15}{16}\left[1 - (1-2p)^8\right] = 15p + \mathcal{O}(p^2)
\end{equation}
shots at low input error rates \cite{PhysRevA.71.022316,costofuni}. Both the error suppression and discard ratio scaling will be numerically confirmed in section \ref{section:surface_code_sim} later. 
Note that the 15-to-1 circuits with $\ket{-}/\ket{Y}$ state injected (to be distilled) does not require conditional feed-forward $S$ gates at the surface code level. This simplifies the numerical simulations, if a $\ket{-}/\ket{Y}$ state is consumed in place of a $\ket{T}$ state.

\section{Surface code simulations}
\label{section:surface_code_sim}
\textbf{This section and the next (Section \ref{section:surface_code_sim} and \ref{section:surface_code_non_clifford}) are the new contributions of this manuscript, and the results are entirely numerical.}

We present our various numerical findings on the performance of both distillation circuits. One of the aims of this paper is to investigate whether the distillation error suppression scaling ($p\rightarrow\mathcal{O}(p^3)$) holds using surface code patches entangled with transversal CNOTs. We employ the iterative decoding scheme from \cite{wan2024iterativetransversalcnotdecoder} to decode the entire CNOT circuits in figure \ref{fig:y_state_distillation_circuit_Y_consumption_optimised} and \ref{fig:T_state_distillation_circuit_T_consumption_optimised}. In order to characterise the performance of each protocol, we distil the logical $\ket{-}$ state as a stabiliser proxy for the performance of either distillation circuits. 
Then we distill logical $\ket{Y}$ states for both circuits using the inplace $\ket{Y}$ state preparation from~\cite{Gidney_2024}. The logical $\ket{Y}$ state acts as a stabiliser proxy to the 15-to-1 circuit, but it is the intended logical state the 7-to-1 circuit distills.  
In our simulations, we assume the following \cite{wan2024iterativetransversalcnotdecoder}:
\begin{enumerate}
    \item Error-free and instantaneous transversal CNOTs~\footnote{Note that transversal CNOT errors can be subsumed into a marginally higher circuit-level noise.}.
    \item Standard SD6 circuit-level depolarising noise model \cite{honeycomb2022gidney,ErrorCorrectionZoo} on all gates except the resource state initialisation errors.
    \item Artificial injection of $Z$ input errors \cite{zhou2024algorithmicfaulttolerancefast} (with \textbf{logical} error probability $p$) on all logical resource states to be consumed.
    \item {The $\ket{+}$ and $\ket{0}$ patches use three initial syndrome rounds. Inplace $\ket{Y}$ preparation and output read-out each use $\floor{d/2}+2$ additional rounds, including $\floor{d/2}$ boundary rounds; the transversally prepared $\ket{-}$ proxy needs no such padding. In principle, consumed $\ket{Y}$ and $\ket{T}$ resources could instead be supplied by a two-round, postselected injection protocol~\cite{Li_2015,LaoInjection2022,wan2025pauliwebyranglestate}, with subsequent distance growth counted separately.} A separate comparison turns off circuit-level noise throughout the inplace $\ket{Y}$ sub-routines while retaining the independently injected logical $Z$ errors. This is to emulate a logical $\ket{Y}$ state injected with logical error $p$.
\end{enumerate}

We demonstrate that the iterative decoding scheme from \cite{wan2024iterativetransversalcnotdecoder} can perform well in the context of state distillation. Furthermore, we propose a time complexity speed up given a re-configurable qubit platform such as trapped-ions \cite{Lekitsch_2017,Akhtar_2023} with the ability to perform long-range two qubit gates.

We use Stim \cite{gidney2021stim} to construct and sample the logical circuits from figures \ref{fig:y_state_distillation_circuit_Y_consumption_optimised} and \ref{fig:T_state_distillation_circuit_T_consumption_optimised}, we decode each surface code patch separately using PyMatching \cite{pymatchingv2} and propagate the Pauli-frames accordingly when a transversal logical CNOT is applied. No more than 3 global iterations of the algorithm provided in \cite{wan2024iterativetransversalcnotdecoder} were needed in all of the following simulations. Three rounds of syndrome extraction are inserted on all the logical qubits at every time step labelled with blue dotted lines in figures \ref{fig:y_state_distillation_circuit_Y_consumption_optimised} and \ref{fig:T_state_distillation_circuit_T_consumption_optimised}. We benchmark the circuits by computing the stabiliser proxies of distilling a logical $\ket{-}$ and $\ket{Y}$ state separately. {These proxy results provide an order-of-magnitude estimate for 15-to-1 $\ket{T}$ distillation; section~\ref{section:surface_code_non_clifford} presents the non-Clifford simulations.}

At each input error rate $p=p_{\rm inj}$, ordinary Monte Carlo sampling draws physical circuit faults and injected logical errors together, with a separate sample for each point. {Appendix~\ref{app:inplace_y_noise} compares noisy and noiseless inplace $Y$ access.} 

\begin{figure}
    \centering
    \includegraphics[width=0.95\linewidth]{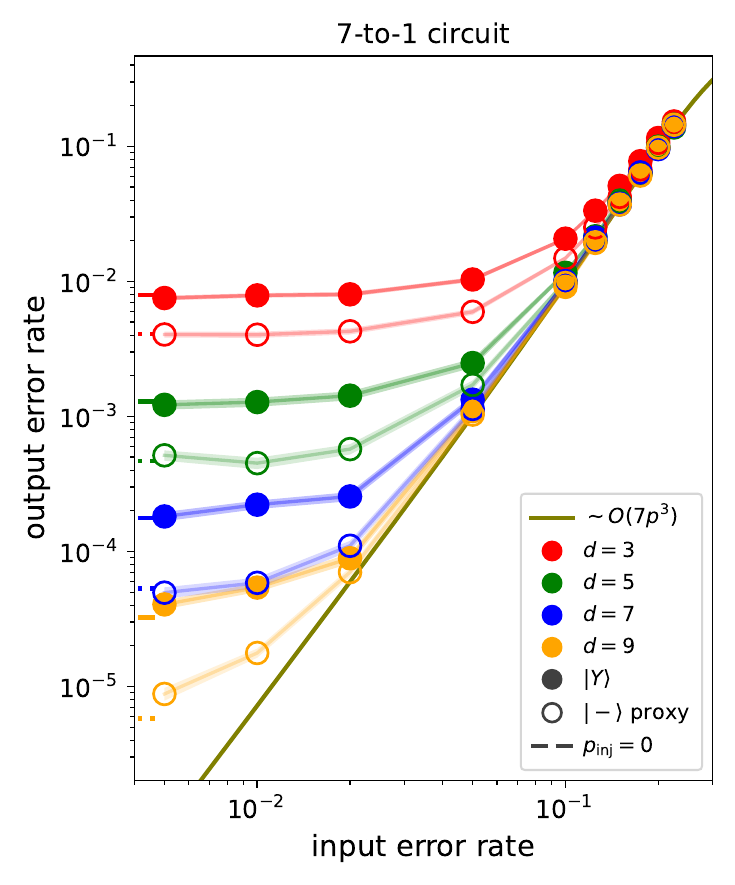}
    \caption{Surface code simulations of the 7-to-1 distillation output error rate. The output error scaling: $\mathcal{O}(7p^3)$ is confirmed across distances $d\in \{3,5,7,9\}$ at $0.1\%$ circuit-level noise.}
\label{fig:y_fac_stab_proxies}
\end{figure}

\begin{figure}
    \centering
    \includegraphics[width=0.95\linewidth]{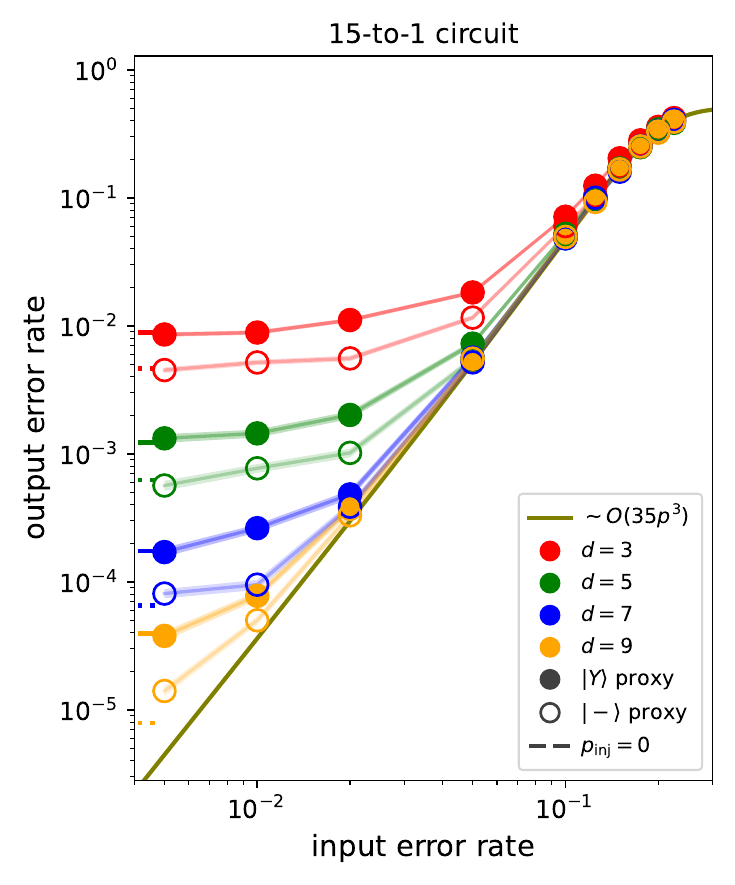}
    \caption{Surface code simulations of {the output error rates of the 15-to-1 stabiliser proxies}. The output error scaling: $\mathcal{O}(35p^3)$ is confirmed across distances $d\in \{3,5,7,9\}$ at $0.1\%$ circuit-level noise.}    \label{fig:t_fac_stab_proxies_H}
\end{figure}

\begin{figure}
    \centering
    \includegraphics[width=0.95\linewidth]{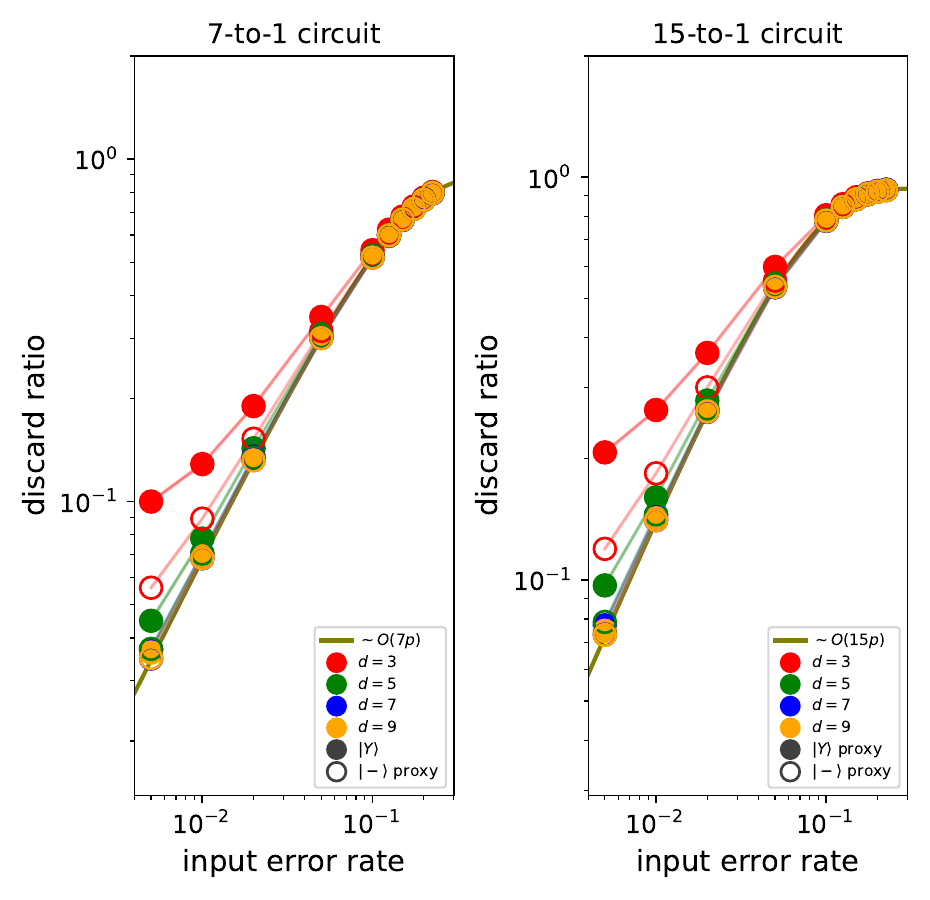}
    \caption{The discard ratio of the 7-to-1 circuit (left) and 15-to-1 circuit (right) stemming from the surface code simulations. We confirm the discard ratio of $\mathcal{O}(7p)$ and $\mathcal{O}(15p)$ at low input error rate $p$ for both distillation protocol.}
    \label{fig:y_t_discard}
\end{figure}

{The output error rates for the 7-to-1 circuit (figure \ref{fig:y_fac_stab_proxies}) and the 15-to-1 proxies (figure \ref{fig:t_fac_stab_proxies_H}) match the ideal distillation curves well, except at low $p_{\rm inj}$, where circuit-level noise produces a distance-dependent error floor.} We simulated the 7-to-1 circuit from \cite{zhou2024algorithmicfaulttolerancefast}, however we noticed that the first and fourth CNOTs applied in \cite{zhou2024algorithmicfaulttolerancefast} are redundant as they act on either $\ket{0,0}$ or $\ket{+,+}$, hence these gates are removed in our simulations. We confirmed the single level distillation factory results from \cite{zhou2024algorithmicfaulttolerancefast} at circuit-level noise $p_{\text{circuit}} = 0.1\%$. We noticed similar deviation from the zero circuit-level noise curve over a range of distances $d\in\{3,5,7,9\}$. Furthermore, we also notice similar deviations in our 15-to-1 circuit simulations. This is due to circuit-level noise imperfections similar to the findings from \cite{zhou2024algorithmicfaulttolerancefast}. The separately sampled $p=0$ output errors decrease with distance: for noisy $Y$ access they fall from $7.93\times10^{-3}$ to $3.26\times10^{-5}$ in the 7-to-1 circuit, and from $8.90\times10^{-3}$ to $3.94\times10^{-5}$ in the 15-to-1 proxy, between $d=3$ and $d=9$. A fixed nonzero circuit-noise floor prevents a pure cubic law as $p\to0$. Figure~\ref{fig:y_t_discard} reports the corresponding discard probabilities; Appendix~\ref{app:inplace_y_noise} compares noisy and noiseless inplace $Y$ access.

\begin{figure*}[!tbp]
\centering
  \includegraphics[width=0.65\linewidth]{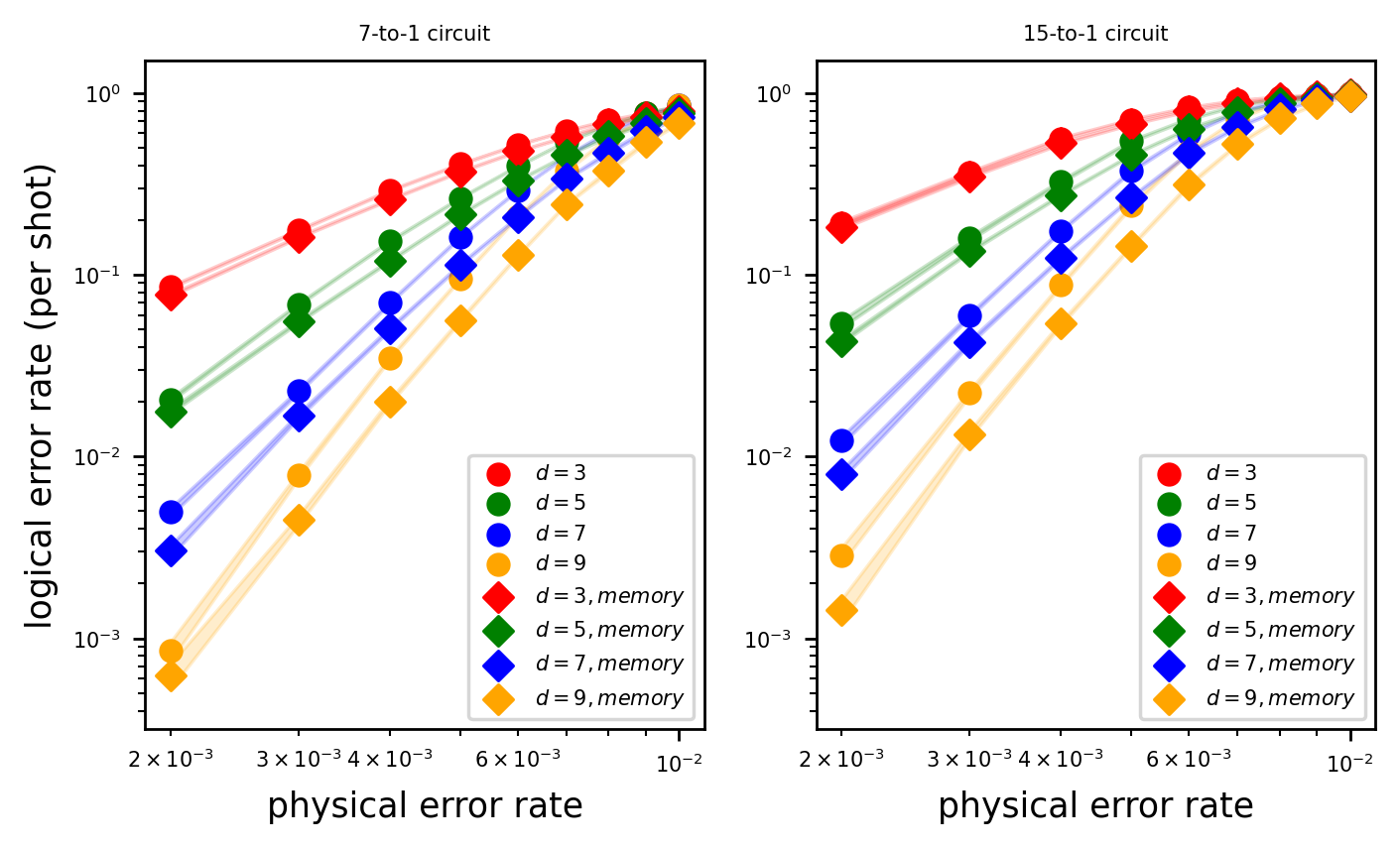}
  \caption{Logical error rate comparisons of the 7-to-1 (left) and 15-to-1 (right) pure CNOT sub-circuits against an equivalent memory experiment with the same number of syndrome extraction rounds. This separate benchmark uses one syndrome round per time step and excludes resource consumption and distillation post-selection.}
  \label{fig:7_and_15_to_1}
\end{figure*}

In a separate CNOT-only benchmark, we also computed the overall logical fidelity of the pure CNOT sub-circuit component of either distillation protocol. The logical error rate in figure \ref{fig:7_and_15_to_1} is defined to be the transversal CNOT sub-circuit logical failure rate benchmarked using stabiliser flow \cite{McEwen_2023}/Pauli web \cite{Bombin_2024,rodatz2024floquetifyingstabilisercodesdistancepreserving,rodatz2025faulttoleranceconstruction,rüsch2025completenessfaultequivalenceclifford}. See the appendices for discussions on using Pauli webs to benchmark these circuits. Close agreement of logical error rates compared to the equivalent idling memory experiment is observed in both distillation CNOT sub-circuits (figure \ref{fig:7_and_15_to_1}).

{The three-round schedule gives $15$ syndrome rounds for 7-to-1 and $18$ for the 15-to-1 proxies. Their data-qubit costs are $8d^2\times15+7d^2\times3=141d^2$ and $16d^2\times18+15d^2\times3=333d^2$ qubit-cycles respectively. 

{In the next section, the implementation based on faulty $T$ measurements~\cite{Litinski_2019_magic} uses $15$ syndrome rounds on $16$ patches, corresponding to $240d^2$ data-qubit cycles before shrinking and output verification.} These counts exclude resource preparation where required, physical measurement ancillas, decoder waiting and storage of the delivered output.}

\section{Surface code non-Clifford simulations}
\label{section:surface_code_non_clifford}

We next simulate the 15-to-1 circuit with genuine non-Clifford measurements.
We use the faulty $T$ measurement construction of
Ref.~\cite{Litinski_2019_magic} as shown in
Fig.~\ref{fig:nc_gadgets}. The purpose is to sidestep the conditional feedforward $S$ gate for the sake of simulations.

\begin{figure*}[!tbp]
\centering
\resizebox{.86\textwidth}{!}{%
\begin{tabular}{@{}c@{\qquad}c@{\qquad}c@{}}
{\small (a) State injection} & & {\small (b) Faulty $T$ measurement}\\[1mm]
\begin{tikzpicture}[font=\small,baseline={($(isc-0.center)!0.5!(iz-0.center)$)}]
\begin{yquant}[operator/separation=3mm,register/separation=5mm]
qubit {$\ket{\psi}$} q;
qubit {$\ket{+}$} a;
[nc gate] box {$T$} a;
[nc error] box {$E$} a;
cnot a | q;
[name=iz] measure {$Z$} a;
[name=isc,nc correction] box {$S$} (q) | a;
text {$m$} a;
discard a;
output {} q;
\end{yquant}
\end{tikzpicture}
& $\overset{E=I}{=}$ &
\begin{tikzpicture}[font=\small,baseline={($(fzc-0.center)!0.5!(fx-0.center)$)}]
\begin{yquant}[operator/separation=3mm,register/separation=5mm]
qubit {$\ket{\psi}$} q;
qubit {$\ket{0}$} a;
cnot a | q;
[name=fe,nc error] box {$E$} a;
[name=ft,nc gate] box {$T$} a;
[name=fx] measure {$X$} a;
[name=fzc,nc correction] box {$Z$} (q) | a;
text {$m$} a;
discard a;
output {} q;
\end{yquant}
\node[draw=black!50,dashed,rounded corners=1mm,fit=(fe-0)(ft-0)(fx-0),inner sep=1.6mm] (fbox) {};
\node[ft note,anchor=north] at ($(fbox.south)+(0,-1mm)$) {faulty $T$ measurement};
\end{tikzpicture}
\end{tabular}}
\caption{Two implementations of $T$ when $E=I$. The order of the fault $E$ and
the entangling operation is different in the two constructions.
In (b), $X,Y,Z$ faults before $T$ give residual $S^\dagger,S,Z$
errors after the corrected gate.}
\label{fig:nc_gadgets}
\end{figure*}

Figure~\ref{fig:nc_factory} uses the same four encoder CNOT layers as
the 15-to-1 circuit, followed by fifteen faulty $T^\dagger$ measurements, giving a $\ket{T}$ output in our convention.
Sixteen surface code patches undergo five blocks of three syndrome rounds,
including the initial block, resulting in $15$ rounds in total before shrinking.
There is no separate resource-consumption CNOT layer in this construction.
We use $p_{\rm circuit}=10^{-3}$ and noiseless transversal CNOTs.

\begin{figure}[htbp]
\centering
\resizebox{\linewidth}{!}{\begin{tikzpicture}[font=\small]
\begin{yquant}[operator/separation=1.4mm,register/separation=1.7mm]
qubit {$q_0:\ket{+}$} q[+1];
qubit {$q_1:\ket{+}$} q[+1];
qubit {$q_2:\ket{+}$} q[+1];
qubit {$q_3:\ket{0}$} q[+1];
qubit {$q_4:\ket{+}$} q[+1];
qubit {$q_5:\ket{0}$} q[+1];
qubit {$q_6:\ket{0}$} q[+1];
qubit {$q_7:\ket{0}$} q[+1];
qubit {$q_8:\ket{+}$} q[+1];
qubit {$q_9:\ket{0}$} q[+1];
qubit {$q_{10}:\ket{0}$} q[+1];
qubit {$q_{11}:\ket{0}$} q[+1];
qubit {$q_{12}:\ket{0}$} q[+1];
qubit {$q_{13}:\ket{0}$} q[+1];
qubit {$q_{14}:\ket{0}$} q[+1];
qubit {$q_{15}:\ket{0}$} q[+1];
[draw=ncblue,draw opacity=.7,label=0] barrier (-);
cnot q[9] | q[1];
cnot q[10] | q[2];
cnot q[12] | q[4];
[draw=ncblue,draw opacity=.7,label=1] barrier (-);
align -;
cnot q[4] | q[0];
cnot q[5] | q[1];
cnot q[6] | q[2];
cnot q[12] | q[8];
cnot q[13] | q[9];
cnot q[14] | q[10];
[draw=ncblue,draw opacity=.7,label=2] barrier (-);
align -;
cnot q[2] | q[0];
cnot q[6] | q[4];
cnot q[10] | q[8];
cnot q[14] | q[12];
align -;
cnot q[3] | q[1];
cnot q[7] | q[5];
cnot q[11] | q[9];
cnot q[15] | q[13];
[draw=ncblue,draw opacity=.7,label=3] barrier (-);
align -;
cnot q[1] | q[0];
cnot q[3] | q[2];
cnot q[5] | q[4];
cnot q[7] | q[6];
cnot q[9] | q[8];
cnot q[11] | q[10];
cnot q[13] | q[12];
cnot q[15] | q[14];
[draw=ncblue,draw opacity=.7,label=4] barrier (-);
align -;
[nc error] box {$Z^{e_j}$} q[1-15];
[nc gate] box {$T^\dagger$} q[1-15];
[name=fmx] measure {$X$} q[1-15];
align q;
text {$m_1$} q[1];
text {$m_2$} q[2];
text {$m_3$} q[3];
text {$m_4$} q[4];
text {$m_5$} q[5];
text {$m_6$} q[6];
text {$m_7$} q[7];
text {$m_8$} q[8];
text {$m_9$} q[9];
text {$m_{10}$} q[10];
text {$m_{11}$} q[11];
text {$m_{12}$} q[12];
text {$m_{13}$} q[13];
text {$m_{14}$} q[14];
text {$m_{15}$} q[15];
discard q[1-15];
[name=fout,nc correction] box {$Z^\mu$} q[0];
output {$\ket{\psi_{\rm out}}$} q[0];
\end{yquant}
\node[ft note,anchor=south] (parity) at ($(fout-0.north)+(0,3mm)$) {$\mu=\bigoplus_j m_j$};
\draw[ft classical] (parity.south) -- (fout-0.north);
\end{tikzpicture}}
\caption{The contracted 15-to-1 circuit. Each blue separator denotes
three syndrome rounds.}
\label{fig:nc_factory}
\end{figure}

The fifteen measured patches are shrunk to central physical qubits by
simultaneous $X$ and $Z$ measurements of the surrounding data qubits.
The iterative decoder matches each patch separately, propagates its Pauli
frame across the transversal CNOTs, and uses three global iterations.
The input corrections use only records available before the non-Clifford
measurements. After accounting for the known shrinking byproducts, apply
$X^{\hat x_j}$ before $T^\dagger$ and report
$m_j=m_j^{\rm raw}\oplus\hat z_j$.
Figure~\ref{fig:nc_boundary} shows this decoder boundary.

\begin{figure}[htbp]
\centering
\resizebox{\linewidth}{!}{\begin{tikzpicture}[font=\small]
\begin{yquant}[operator/separation=4mm]
qubit {$\ket{\psi_L}$ patch} q;
[name=shrink,nc access] box {shrink} q;
[name=xframe,nc correction] box {$X^{\hat x}$} q;
[nc error] box {$E$} q;
[nc gate] box {$T^\dagger$} q;
[name=terminal] measure {$X$} q;
output {$m$} q;
\end{yquant}
\node[nc records,anchor=north] (decoder) at ($(shrink-0.south)+(0,-9mm)$) {iterative per-patch decoder\\available prefix records};
\draw[ft classical] (shrink-0.south) -- (decoder.north);
\draw[ft classical]
  (xframe-0.south |- decoder.north)
  -- node[midway,right,font=\scriptsize] {$\hat x$}
  (xframe-0.south);
\node[nc report] (correctbit) at (terminal-0.center |- decoder.center) {report $m\oplus\hat z$};
\draw[ft classical] (decoder.east) -- node[below,font=\scriptsize] {$\hat z$} (correctbit.west);
\draw[ft classical] (terminal-0.south) -- node[right,font=\scriptsize] {$m$} (correctbit.north);
\end{tikzpicture}}
\caption{Decoder corrections for a faulty $T^\dagger$ measurement.
The decoder uses syndrome and shrinking records available before
$T^\dagger$ to determine $\hat x$ and $\hat z$.
{The displayed $X^{\hat x}$ correction can be
implemented virtually: choose $T^\dagger$ for $\hat x=0$ or $T$
for $\hat x=1$, then measure $X$ and report $m\oplus\hat z$.}
$E$ represents residual access faults and the sampled injection error.}
\label{fig:nc_boundary}
\end{figure}

We use ordinary Monte Carlo sampling throughout. Stim~\cite{gidney2021stim}
samples the noisy Clifford circuit and shrinking records. The resulting
conditional stabiliser state, with the inferred corrections applied, is
passed to Clifft~\cite{chase2026clifftfastexactsimulation} for actual
$T^\dagger$ measurements. Each trajectory receives a fresh quantum
measurement sample, and different $p_{\rm inj}$ points use independent
samples.

\begin{figure*}[t]
\centering
\includegraphics[width=.94\textwidth]{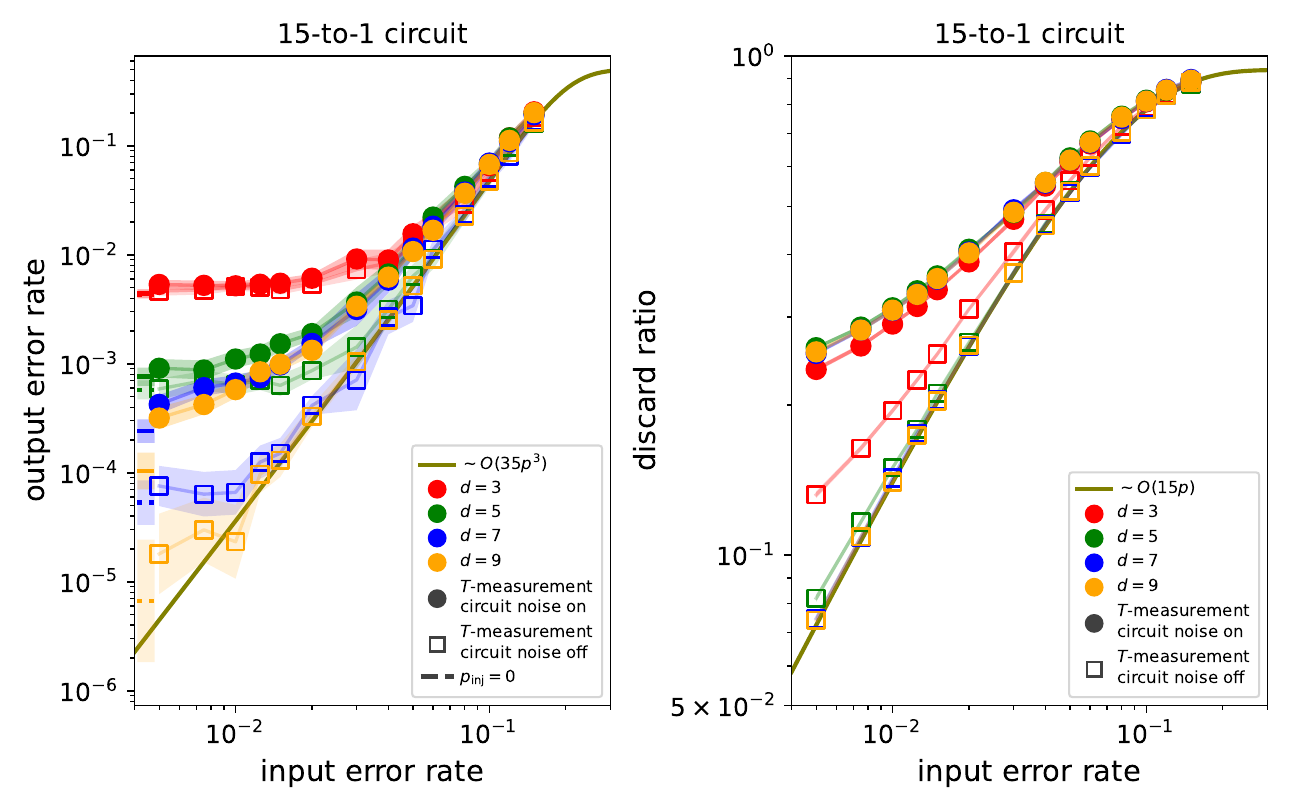}
\caption{Non-Clifford 15-to-1 surface code simulations at $d=3,5,7,9$,
with $p_{\rm circuit}=10^{-3}$ and three syndrome rounds per time step.
All sampled probabilities are shown together. {Filled circles include circuit-level noise in the
$T$-measurement routines; open squares turn this noise off.}}
\label{fig:nc_results}
\end{figure*}

{Figure~\ref{fig:nc_results} shows the accepted output error
and discard probabilities from $17.9$ million Monte Carlo shots at
$d=3,5,7,9$. {With $T$-measurement circuit noise off, we idealise the
final syndrome round on the fifteen measured patches, their shrinking
measurements and their terminal gates, while retaining earlier circuit
noise and all $p_{\rm inj}$ channels.}}

\section{Discussions and outlook}
The numerical simulation results in this manuscript address the feasibility of using long-range transversal CNOT operations to perform magic state distillation with the surface code.

The crux of this study is to confirm the error suppression scaling: $p \xrightarrow[]{\text{15-to-1}}35p^3$ and $p \xrightarrow[]{\text{7-to-1}}7p^3$, under circuit-level noise with additional injected $Z$ input logical error of strength $p$. Our results show that a single-level distillation factory can indeed suppress error, and distill resource states when paired with the decoder from \cite{wan2024iterativetransversalcnotdecoder}. We remark that each surface code patch is decoded separately, with appropriate propagation of Pauli-frames. {The iterative decoder is capped at three global iterations. We also simulate the non-Clifford 15-to-1 protocol using faulty $T$ measurements and causal decoder corrections. The output verifier is ideal, and decoder waiting noise is not included.}

This work provides a step towards realistic implementation of time-optimal magic state distillation on re-configurable hardware architectures. 
Future works on benchmarking the 15-to-1 circuit with actual logical non-Clifford $\ket{T}$ states injected~\cite{Li_2015,LaoInjection2022,wan2025pauliwebyranglestate} {along with conditional feedforward $S$ gates (potentially using distilled $\ket{Y}$ states)}, using low-rank non-Clifford simulators such as Clifft~\cite{chase2026clifftfastexactsimulation}, SOFT~\cite{li2025softhighperformancesimulatoruniversal}, \texttt{tsim}~\cite{haenel2026tsimfastuniversalsimulator}, Pauli-propagation~\cite{rudolph2026paulipropagationcomputationalframework,wan2026exactlogicalerrorrates} or tensor network contractions~\cite{Fishman_2022,wan2026exactlogicalerrorrates}, {will give a more complete benchmark of this injection and feedforward implementation}.
In addition, analysis of hardware realistic connectivity error models is under preparation \cite{updating_impact}. 

Recent works on magic state cultivation \cite{gidney2024magicstatecultivationgrowing,chen2025efficientmagicstatecultivation,vaknin2025magicstatecultivationsurface,lee2025lowoverheadmagicstatedistillation,jacoby2025magicstateinjectionerasure} promises high quality injected (`cultivated') magic state with logical error rate as low as $2\times 10^{-9}$ and expected spacetime volume of $5 \times 10^{3}$ qubit-cycles at $0.1\%$ circuit-level noise \cite{chen2025efficientmagicstatecultivation}. This may potentially make multi-level magic state distillation obsolete. However, many fault-tolerant quantum algorithms require T-count greater than $\frac{1}{2\times10^{-9}}=5\times10^{8}$ \cite{dalzell2023quantumalgorithmssurveyapplications}. Single-level magic state distillation might still be vital in suppressing error rates from $2\times 10^{-9}\xrightarrow[]{\text{15-to-1}}2.8\times 10^{-25}$. Incorporating an end-to-end analysis of magic state cultivation followed by transversal CNOT enabled magic state distillation is a fruitful topic to study. 

\begin{acknowledgments}
We acknowledge discussions with Zhenghao Zhong on Pauli webs, Austin Fowler on magic state distillations and Mark Webber on early manuscript feedback. We want to thank our colleagues at Universal Quantum Ltd for their support. {The numerical simulations were performed on a 2023 Dell XPS17 laptop with an Intel Core i7-12700H processor (14 cores) and 64 GB of physical memory, and an Apple MacBook M4 (14 cores) with 24 GB of physical memory. LLMs were used for minor refinement of text and methods between arXiv versions $2$ and $3$.} The tikzpicture code used to construct the tetrahedron can be found in \cite{tetrahedron_stackexchange}. Most of the ZX-diagrams in the appendices were generated using $\mathtt{pyzx}$ \cite{kissinger2020Pyzx}.
\end{acknowledgments}

\bibliographystyle{apsrev4-2}
\bibliography{main}

\appendix
\onecolumngrid
\clearpage
\section{Stabiliser-proxy circuits}
\label{app:proxy_circuit_overview}

\begin{figure}[h!]
\centering
{\small
\tikz[baseline=(yprep.base)]
  \node[rectangle,rounded corners=1mm,draw=orange!75!black,fill=orange!13,inner xsep=2mm,inner ysep=.4mm]
  (yprep) {$\ket{Y}$};
\quad $\lfloor d/2\rfloor+2$ rounds of inplace preparation.\par}
\medskip

\begin{minipage}[t]{.485\linewidth}
\centering
\textbf{(a) 15-to-1: $\ket{Y}$ proxy}\\[2mm]
\resizebox{.95\linewidth}{!}{\begin{tikzpicture}
    \begin{yquant}
    qubit {0: $\ket{{+}}$} q[+1];
    qubit {} a[+1]; discard a[0];
    qubit {1: $\ket{{+}}$} q[+1];
    qubit {} a[+1]; discard a[1];
    qubit {2: $\ket{{+}}$} q[+1];
    qubit {} a[+1]; discard a[2];
    qubit {3: $\ket{{0}}$\hspace{0.1cm}} q[+1];
    qubit {} a[+1]; discard a[3];
    qubit {4: $\ket{{+}}$} q[+1];
    qubit {} a[+1]; discard a[4];
    qubit {5: $\ket{{0}}$\hspace{0.1cm}} q[+1];
    qubit {} a[+1]; discard a[5];
    qubit {6: $\ket{{0}}$\hspace{0.1cm}} q[+1];
    qubit {} a[+1]; discard a[6];
    qubit {7: $\ket{{0}}$\hspace{0.1cm}} q[+1];
    qubit {} a[+1]; discard a[7];
    qubit {8: $\ket{{+}}$} q[+1];
    qubit {} a[+1]; discard a[8];
    qubit {9: $\ket{{0}}$\hspace{0.1cm}} q[+1];
    qubit {} a[+1]; discard a[9];
    qubit {10: $\ket{{0}}$\hspace{0.1cm}} q[+1];
    qubit {} a[+1]; discard a[10];
    qubit {11: $\ket{{0}}$\hspace{0.1cm}} q[+1];
    qubit {} a[+1]; discard a[11];
    qubit {12: $\ket{{0}}$\hspace{0.1cm}} q[+1];
    qubit {} a[+1]; discard a[12];
    qubit {13: $\ket{{0}}$\hspace{0.1cm}} q[+1];
    qubit {} a[+1]; discard a[13];
    qubit {14: $\ket{{0}}$\hspace{0.1cm}} q[+1];
    qubit {} a[+1]; discard a[14];
    qubit {15: $\ket{{0}}$\hspace{0.1cm}} q[+1];
    
    [blue, thick, label=0]
    barrier (-);

    cnot q[9] | q[1];
    cnot q[10] | q[2];
    cnot q[12] | q[4];

    [blue, thick, label=1]
    barrier (-);

    align q;
    
    cnot q[4] | q[0];
    cnot q[5] | q[1];
    cnot q[6] | q[2];

    cnot q[12] | q[8];
    cnot q[13] | q[9];
    cnot q[14] | q[10];

    [blue, thick, label=2]
    barrier (-);

    align q;
    
    cnot q[2] | q[0];
    cnot q[6] | q[4];
    cnot q[10] | q[8];
    cnot q[14] | q[12];
    
    align q;
    
    cnot q[3] | q[1];
    cnot q[7] | q[5];
    cnot q[11] | q[9];
    cnot q[15] | q[13];
    
    [blue, thick, label=3]
    barrier (-);

    align q;
    
    cnot q[1] | q[0];
    cnot q[3] | q[2];
    cnot q[5] | q[4];
    cnot q[7] | q[6];

    cnot q[9] | q[8];
    cnot q[11] | q[10];
    cnot q[13] | q[12];
    cnot q[15] | q[14];
    
    [blue, thick, label=4]
    barrier (-);

    align a;

    [name=inj0] init {$\ket{{Y}}$} a[0];
    cnot a[0] | q[1];
    [name=inj1] init {$\ket{{Y}}$} a[1];
    cnot a[1] | q[2];
    [name=inj2] init {$\ket{{Y}}$} a[2];
    cnot a[2] | q[3];
    [name=inj3] init {$\ket{{Y}}$} a[3];
    cnot a[3] | q[4];
    [name=inj4] init {$\ket{{Y}}$} a[4];
    cnot a[4] | q[5];
    [name=inj5] init {$\ket{{Y}}$} a[5];
    cnot a[5] | q[6];
    [name=inj6] init {$\ket{{Y}}$} a[6];
    cnot a[6] | q[7];
    
    [name=inj7] init {$\ket{{Y}}$} a[7];
    cnot a[7] | q[8];
    [name=inj8] init {$\ket{{Y}}$} a[8];
    cnot a[8] | q[9];
    [name=inj9] init {$\ket{{Y}}$} a[9];
    cnot a[9] | q[10];
    [name=inj10] init {$\ket{{Y}}$} a[10];
    cnot a[10] | q[11];
    [name=inj11] init {$\ket{{Y}}$} a[11];
    cnot a[11] | q[12];
    [name=inj12] init {$\ket{{Y}}$} a[12];
    cnot a[12] | q[13];
    [name=inj13] init {$\ket{{Y}}$} a[13];
    cnot a[13] | q[14];
    [name=inj14] init {$\ket{{Y}}$} a[14];
    cnot a[14] | q[15];

    align a; align q;
    
    [blue, thick, label=5]
    barrier (-);

    align a,q;

measure a;
    text {$n_{1}$} a[0];
    text {$n_{2}$} a[1];
    text {$n_{3}$} a[2];
    text {$n_{4}$} a[3];
    text {$n_{5}$} a[4];
    text {$n_{6}$} a[5];
    text {$n_{7}$} a[6];
    text {$n_{8}$} a[7];
    text {$n_{9}$} a[8];
    text {$n_{10}$} a[9];
    text {$n_{11}$} a[10];
    text {$n_{12}$} a[11];
    text {$n_{13}$} a[12];
    text {$n_{14}$} a[13];
    text {$n_{15}$} a[14];
    discard a;
    align q;
    measure {\tiny{$M_{{X}}$}} q[1-15];
    text {$m_{1}$} q[1];
    text {$m_{2}$} q[2];
    text {$m_{3}$} q[3];
    text {$m_{4}$} q[4];
    text {$m_{5}$} q[5];
    text {$m_{6}$} q[6];
    text {$m_{7}$} q[7];
    text {$m_{8}$} q[8];
    text {$m_{9}$} q[9];
    text {$m_{10}$} q[10];
    text {$m_{11}$} q[11];
    text {$m_{12}$} q[12];
    text {$m_{13}$} q[13];
    text {$m_{14}$} q[14];
    text {$m_{15}$} q[15];
    text {$\sim\ket{{Y}}$} q[0];
    discard q;
    \end{yquant}
    \begin{pgfonlayer}{background}
\node[rectangle,rounded corners=1mm,draw=orange!75!black,fill=orange!13,line width=.4pt,inner xsep=1mm,inner ysep=.35mm,fit=(inj0-0)] {};
\node[rectangle,rounded corners=1mm,draw=orange!75!black,fill=orange!13,line width=.4pt,inner xsep=1mm,inner ysep=.35mm,fit=(inj1-0)] {};
\node[rectangle,rounded corners=1mm,draw=orange!75!black,fill=orange!13,line width=.4pt,inner xsep=1mm,inner ysep=.35mm,fit=(inj2-0)] {};
\node[rectangle,rounded corners=1mm,draw=orange!75!black,fill=orange!13,line width=.4pt,inner xsep=1mm,inner ysep=.35mm,fit=(inj3-0)] {};
\node[rectangle,rounded corners=1mm,draw=orange!75!black,fill=orange!13,line width=.4pt,inner xsep=1mm,inner ysep=.35mm,fit=(inj4-0)] {};
\node[rectangle,rounded corners=1mm,draw=orange!75!black,fill=orange!13,line width=.4pt,inner xsep=1mm,inner ysep=.35mm,fit=(inj5-0)] {};
\node[rectangle,rounded corners=1mm,draw=orange!75!black,fill=orange!13,line width=.4pt,inner xsep=1mm,inner ysep=.35mm,fit=(inj6-0)] {};
\node[rectangle,rounded corners=1mm,draw=orange!75!black,fill=orange!13,line width=.4pt,inner xsep=1mm,inner ysep=.35mm,fit=(inj7-0)] {};
\node[rectangle,rounded corners=1mm,draw=orange!75!black,fill=orange!13,line width=.4pt,inner xsep=1mm,inner ysep=.35mm,fit=(inj8-0)] {};
\node[rectangle,rounded corners=1mm,draw=orange!75!black,fill=orange!13,line width=.4pt,inner xsep=1mm,inner ysep=.35mm,fit=(inj9-0)] {};
\node[rectangle,rounded corners=1mm,draw=orange!75!black,fill=orange!13,line width=.4pt,inner xsep=1mm,inner ysep=.35mm,fit=(inj10-0)] {};
\node[rectangle,rounded corners=1mm,draw=orange!75!black,fill=orange!13,line width=.4pt,inner xsep=1mm,inner ysep=.35mm,fit=(inj11-0)] {};
\node[rectangle,rounded corners=1mm,draw=orange!75!black,fill=orange!13,line width=.4pt,inner xsep=1mm,inner ysep=.35mm,fit=(inj12-0)] {};
\node[rectangle,rounded corners=1mm,draw=orange!75!black,fill=orange!13,line width=.4pt,inner xsep=1mm,inner ysep=.35mm,fit=(inj13-0)] {};
\node[rectangle,rounded corners=1mm,draw=orange!75!black,fill=orange!13,line width=.4pt,inner xsep=1mm,inner ysep=.35mm,fit=(inj14-0)] {};
\end{pgfonlayer}
\end{tikzpicture}}
\end{minipage}\hfill
\begin{minipage}[t]{.485\linewidth}
\centering
\textbf{(b) 15-to-1: $\ket{-}$ proxy}\\[2mm]
\resizebox{.95\linewidth}{!}{\begin{tikzpicture}
    \begin{yquant}
    qubit {0: $\ket{{+}}$} q[+1];
    qubit {} a[+1]; discard a[0];
    qubit {1: $\ket{{+}}$} q[+1];
    qubit {} a[+1]; discard a[1];
    qubit {2: $\ket{{+}}$} q[+1];
    qubit {} a[+1]; discard a[2];
    qubit {3: $\ket{{0}}$\hspace{0.1cm}} q[+1];
    qubit {} a[+1]; discard a[3];
    qubit {4: $\ket{{+}}$} q[+1];
    qubit {} a[+1]; discard a[4];
    qubit {5: $\ket{{0}}$\hspace{0.1cm}} q[+1];
    qubit {} a[+1]; discard a[5];
    qubit {6: $\ket{{0}}$\hspace{0.1cm}} q[+1];
    qubit {} a[+1]; discard a[6];
    qubit {7: $\ket{{0}}$\hspace{0.1cm}} q[+1];
    qubit {} a[+1]; discard a[7];
    qubit {8: $\ket{{+}}$} q[+1];
    qubit {} a[+1]; discard a[8];
    qubit {9: $\ket{{0}}$\hspace{0.1cm}} q[+1];
    qubit {} a[+1]; discard a[9];
    qubit {10: $\ket{{0}}$\hspace{0.1cm}} q[+1];
    qubit {} a[+1]; discard a[10];
    qubit {11: $\ket{{0}}$\hspace{0.1cm}} q[+1];
    qubit {} a[+1]; discard a[11];
    qubit {12: $\ket{{0}}$\hspace{0.1cm}} q[+1];
    qubit {} a[+1]; discard a[12];
    qubit {13: $\ket{{0}}$\hspace{0.1cm}} q[+1];
    qubit {} a[+1]; discard a[13];
    qubit {14: $\ket{{0}}$\hspace{0.1cm}} q[+1];
    qubit {} a[+1]; discard a[14];
    qubit {15: $\ket{{0}}$\hspace{0.1cm}} q[+1];
    
    [blue, thick, label=0]
    barrier (-);

    cnot q[9] | q[1];
    cnot q[10] | q[2];
    cnot q[12] | q[4];

    [blue, thick, label=1]
    barrier (-);

    align q;
    
    cnot q[4] | q[0];
    cnot q[5] | q[1];
    cnot q[6] | q[2];

    cnot q[12] | q[8];
    cnot q[13] | q[9];
    cnot q[14] | q[10];

    [blue, thick, label=2]
    barrier (-);

    align q;
    
    cnot q[2] | q[0];
    cnot q[6] | q[4];
    cnot q[10] | q[8];
    cnot q[14] | q[12];
    
    align q;
    
    cnot q[3] | q[1];
    cnot q[7] | q[5];
    cnot q[11] | q[9];
    cnot q[15] | q[13];
    
    [blue, thick, label=3]
    barrier (-);

    align q;
    
    cnot q[1] | q[0];
    cnot q[3] | q[2];
    cnot q[5] | q[4];
    cnot q[7] | q[6];

    cnot q[9] | q[8];
    cnot q[11] | q[10];
    cnot q[13] | q[12];
    cnot q[15] | q[14];
    
    [blue, thick, label=4]
    barrier (-);

    align a;

    init {$\ket{{-}}$} a[0];
    cnot a[0] | q[1];
    init {$\ket{{-}}$} a[1];
    cnot a[1] | q[2];
    init {$\ket{{-}}$} a[2];
    cnot a[2] | q[3];
    init {$\ket{{-}}$} a[3];
    cnot a[3] | q[4];
    init {$\ket{{-}}$} a[4];
    cnot a[4] | q[5];
    init {$\ket{{-}}$} a[5];
    cnot a[5] | q[6];
    init {$\ket{{-}}$} a[6];
    cnot a[6] | q[7];
    
    init {$\ket{{-}}$} a[7];
    cnot a[7] | q[8];
    init {$\ket{{-}}$} a[8];
    cnot a[8] | q[9];
    init {$\ket{{-}}$} a[9];
    cnot a[9] | q[10];
    init {$\ket{{-}}$} a[10];
    cnot a[10] | q[11];
    init {$\ket{{-}}$} a[11];
    cnot a[11] | q[12];
    init {$\ket{{-}}$} a[12];
    cnot a[12] | q[13];
    init {$\ket{{-}}$} a[13];
    cnot a[13] | q[14];
    init {$\ket{{-}}$} a[14];
    cnot a[14] | q[15];

    align a; align q;
    
    [blue, thick, label=5]
    barrier (-);

    align a,q;

measure a;
    text {$n_{1}$} a[0];
    text {$n_{2}$} a[1];
    text {$n_{3}$} a[2];
    text {$n_{4}$} a[3];
    text {$n_{5}$} a[4];
    text {$n_{6}$} a[5];
    text {$n_{7}$} a[6];
    text {$n_{8}$} a[7];
    text {$n_{9}$} a[8];
    text {$n_{10}$} a[9];
    text {$n_{11}$} a[10];
    text {$n_{12}$} a[11];
    text {$n_{13}$} a[12];
    text {$n_{14}$} a[13];
    text {$n_{15}$} a[14];
    discard a;
    align q;
    measure {\tiny{$M_{{X}}$}} q[1-15];
    text {$m_{1}$} q[1];
    text {$m_{2}$} q[2];
    text {$m_{3}$} q[3];
    text {$m_{4}$} q[4];
    text {$m_{5}$} q[5];
    text {$m_{6}$} q[6];
    text {$m_{7}$} q[7];
    text {$m_{8}$} q[8];
    text {$m_{9}$} q[9];
    text {$m_{10}$} q[10];
    text {$m_{11}$} q[11];
    text {$m_{12}$} q[12];
    text {$m_{13}$} q[13];
    text {$m_{14}$} q[14];
    text {$m_{15}$} q[15];
    text {$\sim\ket{{-}}$} q[0];
    discard q;
    \end{yquant}
    \end{tikzpicture}}
\end{minipage}
\par\medskip
\begin{minipage}[t]{.485\linewidth}
\centering
\textbf{(c) 7-to-1: $\ket{Y}$}\\[2mm]
\resizebox{.75\linewidth}{!}{\begin{tikzpicture}
    \begin{yquant}
    qubit {0: $\ket{{+}}$} q[+1];
    qubit {} a[+1]; discard a[0];
    qubit {1: $\ket{{+}}$} q[+1];
    qubit {} a[+1]; discard a[1];
    qubit {2: $\ket{{+}}$} q[+1];
    qubit {} a[+1]; discard a[2];
    qubit {3: $\ket{{0}}$\hspace{0.1cm}} q[+1];
    qubit {} a[+1]; discard a[3];
    qubit {4: $\ket{{+}}$} q[+1];
    qubit {} a[+1]; discard a[4];
    qubit {5: $\ket{{0}}$\hspace{0.1cm}} q[+1];
    qubit {} a[+1]; discard a[5];
    qubit {6: $\ket{{0}}$\hspace{0.1cm}} q[+1];
    qubit {} a[+1]; discard a[6];
    qubit {7: $\ket{{0}}$\hspace{0.1cm}} q[+1];

    [blue, thick, label=0]
    barrier (-);
    
    cnot q[5] | q[1];
    cnot q[6] | q[2];

    [blue, thick, label=1]
    barrier (-);

    align q;
    
    cnot q[2] | q[0];
    cnot q[6] | q[4];
    
    align q;
    
    cnot q[3] | q[1];
    cnot q[7] | q[5];

    [blue, thick, label=2]
    barrier (-);

    align q;
    
    cnot q[1] | q[0];
    cnot q[3] | q[2];
    cnot q[5] | q[4];
    cnot q[7] | q[6];
    
    [blue, thick, label=3]
    barrier (-);

    align a;

    [name=inj0] init {$\ket{{Y}}$} a[0];
    cnot a[0] | q[1];
    [name=inj1] init {$\ket{{Y}}$} a[1];
    cnot a[1] | q[2];
    [name=inj2] init {$\ket{{Y}}$} a[2];
    cnot a[2] | q[3];
    [name=inj3] init {$\ket{{Y}}$} a[3];
    cnot a[3] | q[4];
    [name=inj4] init {$\ket{{Y}}$} a[4];
    cnot a[4] | q[5];
    [name=inj5] init {$\ket{{Y}}$} a[5];
    cnot a[5] | q[6];
    [name=inj6] init {$\ket{{Y}}$} a[6];
    cnot a[6] | q[7];
    [blue, thick, label=4]
    barrier (-);
    
    hspace {-2mm} a;
    
    measure a;
    
    text {$n_1$} a[0];
    text {$n_2$} a[1];
    text {$n_3$} a[2];
    text {$n_4$} a[3];
    text {$n_5$} a[4];
    text {$n_6$} a[5];
    text {$n_7$} a[6];
    align a, q;
    
    discard a;
    hspace {-7mm} q;

    measure {\tiny{$M_{{X}}$}} q[1],q[2],q[3],q[4],q[5],q[6],q[7];

    text {$m_1$} q[1];
    text {$m_2$} q[2];
    text {$m_3$} q[3];
    text {$m_4$} q[4];
    text {$m_5$} q[5];
    text {$m_6$} q[6];
    text {$m_7$} q[7];
    
    text {$\sim\ket{{Y}}$} q[0];

    discard q;
    \end{yquant}
    \begin{pgfonlayer}{background}
\node[rectangle,rounded corners=1mm,draw=orange!75!black,fill=orange!13,line width=.4pt,inner xsep=1mm,inner ysep=.35mm,fit=(inj0-0)] {};
\node[rectangle,rounded corners=1mm,draw=orange!75!black,fill=orange!13,line width=.4pt,inner xsep=1mm,inner ysep=.35mm,fit=(inj1-0)] {};
\node[rectangle,rounded corners=1mm,draw=orange!75!black,fill=orange!13,line width=.4pt,inner xsep=1mm,inner ysep=.35mm,fit=(inj2-0)] {};
\node[rectangle,rounded corners=1mm,draw=orange!75!black,fill=orange!13,line width=.4pt,inner xsep=1mm,inner ysep=.35mm,fit=(inj3-0)] {};
\node[rectangle,rounded corners=1mm,draw=orange!75!black,fill=orange!13,line width=.4pt,inner xsep=1mm,inner ysep=.35mm,fit=(inj4-0)] {};
\node[rectangle,rounded corners=1mm,draw=orange!75!black,fill=orange!13,line width=.4pt,inner xsep=1mm,inner ysep=.35mm,fit=(inj5-0)] {};
\node[rectangle,rounded corners=1mm,draw=orange!75!black,fill=orange!13,line width=.4pt,inner xsep=1mm,inner ysep=.35mm,fit=(inj6-0)] {};
\end{pgfonlayer}
\end{tikzpicture}}
\end{minipage}\hfill
\begin{minipage}[t]{.485\linewidth}
\centering
\textbf{(d) 7-to-1: $\ket{-}$ proxy}\\[2mm]
\resizebox{.75\linewidth}{!}{\begin{tikzpicture}
    \begin{yquant}
    qubit {0: $\ket{{+}}$} q[+1];
    qubit {} a[+1]; discard a[0];
    qubit {1: $\ket{{+}}$} q[+1];
    qubit {} a[+1]; discard a[1];
    qubit {2: $\ket{{+}}$} q[+1];
    qubit {} a[+1]; discard a[2];
    qubit {3: $\ket{{0}}$\hspace{0.1cm}} q[+1];
    qubit {} a[+1]; discard a[3];
    qubit {4: $\ket{{+}}$} q[+1];
    qubit {} a[+1]; discard a[4];
    qubit {5: $\ket{{0}}$\hspace{0.1cm}} q[+1];
    qubit {} a[+1]; discard a[5];
    qubit {6: $\ket{{0}}$\hspace{0.1cm}} q[+1];
    qubit {} a[+1]; discard a[6];
    qubit {7: $\ket{{0}}$\hspace{0.1cm}} q[+1];

    [blue, thick, label=0]
    barrier (-);
    
    cnot q[5] | q[1];
    cnot q[6] | q[2];

    [blue, thick, label=1]
    barrier (-);

    align q;
    
    cnot q[2] | q[0];
    cnot q[6] | q[4];
    
    align q;
    
    cnot q[3] | q[1];
    cnot q[7] | q[5];

    [blue, thick, label=2]
    barrier (-);

    align q;
    
    cnot q[1] | q[0];
    cnot q[3] | q[2];
    cnot q[5] | q[4];
    cnot q[7] | q[6];
    
    [blue, thick, label=3]
    barrier (-);

    align a;

    init {$\ket{{-}}$} a[0];
    cnot a[0] | q[1];
    init {$\ket{{-}}$} a[1];
    cnot a[1] | q[2];
    init {$\ket{{-}}$} a[2];
    cnot a[2] | q[3];
    init {$\ket{{-}}$} a[3];
    cnot a[3] | q[4];
    init {$\ket{{-}}$} a[4];
    cnot a[4] | q[5];
    init {$\ket{{-}}$} a[5];
    cnot a[5] | q[6];
    init {$\ket{{-}}$} a[6];
    cnot a[6] | q[7];
    [blue, thick, label=4]
    barrier (-);
    
    hspace {-2mm} a;
    
    measure a;
    
    text {$n_1$} a[0];
    text {$n_2$} a[1];
    text {$n_3$} a[2];
    text {$n_4$} a[3];
    text {$n_5$} a[4];
    text {$n_6$} a[5];
    text {$n_7$} a[6];
    align a, q;
    
    discard a;
    hspace {-7mm} q;

    measure {\tiny{$M_{{X}}$}} q[1],q[2],q[3],q[4],q[5],q[6],q[7];

    text {$m_1$} q[1];
    text {$m_2$} q[2];
    text {$m_3$} q[3];
    text {$m_4$} q[4];
    text {$m_5$} q[5];
    text {$m_6$} q[6];
    text {$m_7$} q[7];
    
    text {$\sim\ket{{-}}$} q[0];

    discard q;
    \end{yquant}
    \end{tikzpicture}}
\end{minipage}
\caption{Stabiliser-proxy circuits.
(a,b) As Fig.~\ref{fig:T_state_distillation_circuit_T_consumption_optimised},
with $\ket{Y}$ or $\ket{-}$ resources and no conditional $S$ gates.
(c,d) As Fig.~\ref{fig:y_state_distillation_circuit_Y_consumption_optimised}.
Output Pauli frames are tracked.
In all four circuits, $n_j$ labels the resource $Z$ measurement and
$m_j$ the data $X$ measurement.}
\label{fig:proxy_circuits}
\end{figure}
\clearpage

\section{Figures of merit and benchmarking logical failure}
It may be more convenient to use ZX-calculus \cite{KissingerWetering2024Book} and Pauli webs \cite{rodatz2024floquetifyingstabilisercodesdistancepreserving,rodatz2025faulttoleranceconstruction,rüsch2025completenessfaultequivalenceclifford} to represent the stabiliser structures of the logical distillation circuits studied in the main text; and to benchmark their corresponding stabiliser proxies ofr logical error rate analysis. Below we shall briefly describe how to map the 7-to-1 protocol's quantum circuit to its ZX-diagram, before illustrating how Pauli webs can be used to benchmark the logical error rate of this protocol.
\subsection{Translations from quantum circuit to ZX-diagram}
The following equations (equation \ref{eq:cnot} to \ref{eq:x_meas}) represent the mapping of quantum circuit diagrams to ZX-gadgets. Namely, CNOT, various Clifford state initialisation and measurements with specific outcomes.
\begin{center}
\begin{equation}
\text{CNOT}=\scalebox{1.5}{\begin{tikzpicture}
    \begin{pgfonlayer}{nodelayer}
        \node [style=none] (0) at (1.00, 0.5) {};
        \node [style=none] (1) at (1.00, -0.5) {};
        \node [style=Z dot] (4) at (2.00, 0.5) {};
        \node [style=X dot] (5) at (2.00, -0.5) {};
        \node [style=none] (8) at (3.00, 0.5) {};
        \node [style=none] (9) at (3.00, -0.5) {};
    \end{pgfonlayer}
    \begin{pgfonlayer}{edgelayer}
        \draw (0) to (4);
        \draw (1) to (5);
        \draw (4) to (5);
        \draw (4) to (8);
        \draw (5) to (9);
    \end{pgfonlayer}
\end{tikzpicture}}
\label{eq:cnot}
\end{equation}

\vspace{2mm}

\begin{minipage}{0.48\linewidth}
\begin{equation}
\ket{0}\propto \scalebox{1.5}{
\begin{tikzpicture}
    \begin{pgfonlayer}{nodelayer}
        \node [style=X dot] (0) at (0.00,0) {};
        \node [style=none] (1) at (1.00, 0) {};
    \end{pgfonlayer}
    \begin{pgfonlayer}{edgelayer}
        \draw (0) to (1);
    \end{pgfonlayer}
\end{tikzpicture}
}
\label{eq:zero_state}
\end{equation}
\end{minipage}\hfill
\begin{minipage}{0.48\linewidth}
\begin{equation}
\ket{+}\propto \scalebox{1.5}{
\begin{tikzpicture}
    \begin{pgfonlayer}{nodelayer}
        \node [style=Z dot] (0) at (0.00,0) {};
        \node [style=none] (1) at (1.00, 0) {};
    \end{pgfonlayer}
    \begin{pgfonlayer}{edgelayer}
        \draw (0) to (1);
    \end{pgfonlayer}
\end{tikzpicture}
}
\label{eq:plus_state}
\end{equation}
\end{minipage}

\vspace{2mm}

\begin{minipage}{0.48\linewidth}
\begin{equation}
\ket{1}\propto \scalebox{1.5}{
\begin{tikzpicture}
    \begin{pgfonlayer}{nodelayer}
        \node [style=X phase dot] (0) at (0.00,0) {$\pi$};
        \node [style=none] (1) at (1.00, 0) {};
    \end{pgfonlayer}
    \begin{pgfonlayer}{edgelayer}
        \draw (0) to (1);
    \end{pgfonlayer}
\end{tikzpicture}
}
\label{eq:one_state}
\end{equation}
\end{minipage}\hfill
\begin{minipage}{0.48\linewidth}
\begin{equation}
\ket{-}\propto \scalebox{1.5}{
\begin{tikzpicture}
    \begin{pgfonlayer}{nodelayer}
        \node [style=Z phase dot] (0) at (0.00,0) {$\pi$};
        \node [style=none] (1) at (1.00, 0) {};
    \end{pgfonlayer}
    \begin{pgfonlayer}{edgelayer}
        \draw (0) to (1);
    \end{pgfonlayer}
\end{tikzpicture}
}
\label{eq:minus_state}
\end{equation}
\end{minipage}

\vspace{2mm}

\begin{minipage}{0.48\linewidth}
\begin{equation}
\ket{Y}\propto \scalebox{1.5}{
\begin{tikzpicture}
    \begin{pgfonlayer}{nodelayer}
        \node [style=Z phase dot] (0) at (0.00,0) {$\frac{\pi}{2}$};
        \node [style=none] (1) at (1.00, 0) {};
    \end{pgfonlayer}
    \begin{pgfonlayer}{edgelayer}
        \draw (0) to (1);
    \end{pgfonlayer}
\end{tikzpicture}
}
\label{eq:y_state}
\end{equation}
\end{minipage}\hfill
\begin{minipage}{0.48\linewidth}
\begin{equation}
S = \scalebox{1.5}{
\begin{tikzpicture}
    \begin{pgfonlayer}{nodelayer}
        \node [style=none] (0) at (1.00,0) {};
        \node [style=Z phase dot] (1) at (0.00, 0) {$\frac{\pi}{2}$};
        \node [style=none] (2) at (-1.00, 0) {};
    \end{pgfonlayer}
    \begin{pgfonlayer}{edgelayer}
        \draw (0) to (1);
        \draw (1) to (2);
    \end{pgfonlayer}
\end{tikzpicture}
}
\label{eq:s_gate}
\end{equation}
\end{minipage}

\vspace{2mm}

\begin{minipage}{0.48\linewidth}
\begin{equation}
\raisebox{-0.35\height}{%
\begin{tikzpicture}
\begin{yquantgroup}
\registers{qubit {} q[+1];}
\circuit{
hspace {5mm} -;
measure q[0];
output {$n_j\in\{0,1\}$} q[0];
}
\end{yquantgroup}
\end{tikzpicture}}
\rightarrow\scalebox{1.5}{
\begin{tikzpicture}
    \begin{pgfonlayer}{nodelayer}
        \node [style=X phase dot] (0) at (1.00,0) {$n_j\pi$};
        \node [style=none] (1) at (-0.50, 0) {};
    \end{pgfonlayer}
    \begin{pgfonlayer}{edgelayer}
        \draw (0) to (1);
    \end{pgfonlayer}
\end{tikzpicture}
}
\label{eq:z_meas}
\end{equation}
\end{minipage}\hfill
\begin{minipage}{0.48\linewidth}
\begin{equation}
\raisebox{-0.35\height}{%
\begin{tikzpicture}
\begin{yquantgroup}
\registers{qubit {} q[+1];}
\circuit{
hspace {5mm} -;
measure {\tiny{$M_{{X}}$}} q[0];
output {$m_j\in\{0,1\}$} q[0];
}
\end{yquantgroup}
\end{tikzpicture}}
\rightarrow\scalebox{1.5}{
\begin{tikzpicture}
    \begin{pgfonlayer}{nodelayer}
        \node [style=Z phase dot] (0) at (1.00,0) {$m_j\pi$};
        \node [style=none] (1) at (-0.50, 0) {};
    \end{pgfonlayer}
    \begin{pgfonlayer}{edgelayer}
        \draw (0) to (1);
    \end{pgfonlayer}
\end{tikzpicture}
}
\label{eq:x_meas}
\end{equation}
\end{minipage}

\end{center}
Along with the non-Clifford $T$ gate and $\ket{T}$ state.
\begin{center}
\begin{minipage}{0.48\linewidth}
\begin{equation}
\ket{T}\propto \scalebox{1.5}{
\begin{tikzpicture}
    \begin{pgfonlayer}{nodelayer}
        \node [style=Z phase dot] (0) at (0.00,0) {$\frac{\pi}{4}$};
        \node [style=none] (1) at (1.00, 0) {};
    \end{pgfonlayer}
    \begin{pgfonlayer}{edgelayer}
        \draw (0) to (1);
    \end{pgfonlayer}
\end{tikzpicture}
}
\label{eq:t_state}
\end{equation}
\end{minipage}\hfill
\begin{minipage}{0.48\linewidth}
\begin{equation}
T = \scalebox{1.5}{
\begin{tikzpicture}
    \begin{pgfonlayer}{nodelayer}
        \node [style=none] (0) at (1.00,0) {};
        \node [style=Z phase dot] (1) at (0.00, 0) {$\frac{\pi}{4}$};
        \node [style=none] (2) at (-1.00, 0) {};
    \end{pgfonlayer}
    \begin{pgfonlayer}{edgelayer}
        \draw (0) to (1);
        \draw (1) to (2);
    \end{pgfonlayer}
\end{tikzpicture}
}
\label{eq:t_gate}
\end{equation}
\end{minipage}
\end{center}

\subsection{ZX-diagram interpretation for the 7-to-1 $\ket{Y}$ protocol}
\label{app:zx_yfac}
ZX-calculus provides a convenient bookkeeping language for stabiliser
structure (including CNOT circuits, Pauli state
preparations, and Pauli measurements) for quantum circuits represented as ZX-diagrams. In our case, the
7-to-1 protocol can be viewed as a Clifford entangling circuit that prepares a Bell pair between the output patch $0$ and an encoded Steane block (patches $1,\dots,7$), followed by a transversal $S$ on the Steane block implemented by $\ket{Y}$-state consumption, and finally $X$-basis readout of the Steane code. The resulting quantum circuit to ZX-diagram mapping in equation \ref{eq:ycir_to_yfac} shows:
\begin{enumerate}
    \item \textbf{CNOT layers $\rightarrow$ connected graph.}
    Each CNOT corresponds to a fixed ZX gadget ($\scalebox{0.75}{\begin{tikzpicture}
    \begin{pgfonlayer}{nodelayer}
        \node [style=none] (0) at (1.00, 0.5) {};
        \node [style=none] (1) at (1.00, -0.5) {};
        \node [style=Z dot] (4) at (2.00, 0.5) {};
        \node [style=X dot] (5) at (2.00, -0.5) {};
        \node [style=none] (8) at (3.00, 0.5) {};
        \node [style=none] (9) at (3.00, -0.5) {};
    \end{pgfonlayer}
    \begin{pgfonlayer}{edgelayer}
        \draw (0) to (4);
        \draw (1) to (5);
        \draw (4) to (5);
        \draw (4) to (8);
        \draw (5) to (9);
    \end{pgfonlayer}
\end{tikzpicture}}$, see equation \ref{eq:cnot}), and the
    three parallel CNOT layers of figure \ref{fig:y_state_distillation_circuit_Y_consumption_optimised}
    translate to a purely graphical wiring pattern among vertices.

    \item \textbf{State preparations and measurements $\rightarrow$ boundary conditions of the ZX-diagram.}
    The initialisations $\ket{+}$ ($\scalebox{0.75}{
\begin{tikzpicture}
    \begin{pgfonlayer}{nodelayer}
        \node [style=Z dot] (0) at (0.00,0) {};
        \node [style=none] (1) at (1.00, 0) {};
    \end{pgfonlayer}
    \begin{pgfonlayer}{edgelayer}
        \draw (0) to (1);
    \end{pgfonlayer}
\end{tikzpicture}
}$) and $\ket{0}$ ($\scalebox{0.75}{
\begin{tikzpicture}
    \begin{pgfonlayer}{nodelayer}
        \node [style=X dot] (0) at (0.00,0) {};
        \node [style=none] (1) at (1.00, 0) {};
    \end{pgfonlayer}
    \begin{pgfonlayer}{edgelayer}
        \draw (0) to (1);
    \end{pgfonlayer}
\end{tikzpicture}
}$) correspond to boundary vertices (equation \ref{eq:plus_state} and \ref{eq:zero_state}) which fix the stabilisers at the input of the circuit. Similarly, the final $Z$- ($\scalebox{0.75}{
\begin{tikzpicture}
    \begin{pgfonlayer}{nodelayer}
        \node [style=X phase dot] (0) at (1.00,0) {$n_j\pi$};
        \node [style=none] (1) at (-0.50, 0) {};
    \end{pgfonlayer}
    \begin{pgfonlayer}{edgelayer}
        \draw (0) to (1);
    \end{pgfonlayer}
\end{tikzpicture}
}$) and $X$-basis ($\scalebox{0.75}{
\begin{tikzpicture}
    \begin{pgfonlayer}{nodelayer}
        \node [style=Z phase dot] (0) at (1.00,0) {$m_j\pi$};
        \node [style=none] (1) at (-0.50, 0) {};
    \end{pgfonlayer}
    \begin{pgfonlayer}{edgelayer}
        \draw (0) to (1);
    \end{pgfonlayer}
\end{tikzpicture}
}$) measurements (equation \ref{eq:z_meas} and \ref{eq:x_meas}) terminate wires and generate measurement outcomes $n_j$ and $m_j$.
    \item \textbf{$\ket{Y}$ ($\scalebox{0.75}{
\begin{tikzpicture}
    \begin{pgfonlayer}{nodelayer}
        \node [style=Z phase dot] (0) at (0.00,0) {$\frac{\pi}{2}$};
        \node [style=none] (1) at (1.00, 0) {};
    \end{pgfonlayer}
    \begin{pgfonlayer}{edgelayer}
        \draw (0) to (1);
    \end{pgfonlayer}
\end{tikzpicture}
}$) consumption realises a transversal $S$ ($\scalebox{0.75}{
\begin{tikzpicture}
    \begin{pgfonlayer}{nodelayer}
        \node [style=none] (0) at (1.00,0) {};
        \node [style=Z phase dot] (1) at (0.00, 0) {$\frac{\pi}{2}$};
        \node [style=none] (2) at (-1.00, 0) {};
    \end{pgfonlayer}
    \begin{pgfonlayer}{edgelayer}
        \draw (0) to (1);
        \draw (1) to (2);
    \end{pgfonlayer}
\end{tikzpicture}
}$) on patches $1$ to $7$.}
    Consuming $\ket{Y}$ via the teleportation gadget (equation \ref{eq:y_state} with a CNOT and
    $Z$ measurement) implements $S$ up to a Pauli byproduct conditioned on the
    measurement outcome.
\end{enumerate}

\begin{equation}
    \label{eq:ycir_to_yfac}
        \raisebox{-0.475\height}{\resizebox{0.441\linewidth}{!}{
    \begin{tikzpicture}
    \begin{yquant}
    qubit {0: $\ket{{+}}$} q[+1];
    qubit {} a[+1]; discard a[0];
    qubit {1: $\ket{{+}}$} q[+1];
    qubit {} a[+1]; discard a[1];
    qubit {2: $\ket{{+}}$} q[+1];
    qubit {} a[+1]; discard a[2];
    qubit {3: $\ket{{0}}$\hspace{0.1cm}} q[+1];
    qubit {} a[+1]; discard a[3];
    qubit {4: $\ket{{+}}$} q[+1];
    qubit {} a[+1]; discard a[4];
    qubit {5: $\ket{{0}}$\hspace{0.1cm}} q[+1];
    qubit {} a[+1]; discard a[5];
    qubit {6: $\ket{{0}}$\hspace{0.1cm}} q[+1];
    qubit {} a[+1]; discard a[6];
    qubit {7: $\ket{{0}}$\hspace{0.1cm}} q[+1];

    [blue, thick, label=0]
    barrier (-);
    
    cnot q[5] | q[1];
    cnot q[6] | q[2];

    [blue, thick, label=1]
    barrier (-);

    align q;
    
    cnot q[2] | q[0];
    cnot q[6] | q[4];
    
    align q;
    
    cnot q[3] | q[1];
    cnot q[7] | q[5];

    [blue, thick, label=2]
    barrier (-);

    align q;
    
    cnot q[1] | q[0];
    cnot q[3] | q[2];
    cnot q[5] | q[4];
    cnot q[7] | q[6];
    
    [blue, thick, label=3]
    barrier (-);

    align a;

    init {$\ket{{Y}}$} a[0];
    cnot a[0] | q[1];
    init {$\ket{{Y}}$} a[1];
    cnot a[1] | q[2];
    init {$\ket{{Y}}$} a[2];
    cnot a[2] | q[3];
    init {$\ket{{Y}}$} a[3];
    cnot a[3] | q[4];
    init {$\ket{{Y}}$} a[4];
    cnot a[4] | q[5];
    init {$\ket{{Y}}$} a[5];
    cnot a[5] | q[6];
    init {$\ket{{Y}}$} a[6];
    cnot a[6] | q[7];
    [blue, thick, label=4]
    barrier (-);
    
    hspace {-2mm} a;
    
    measure a;
    
    text {$n_1$} a[0];
    text {$n_2$} a[1];
    text {$n_3$} a[2];
    text {$n_4$} a[3];
    text {$n_5$} a[4];
    text {$n_6$} a[5];
    text {$n_7$} a[6];
    align a, q;
    
    discard a;
    hspace {-7mm} q;

    measure {\tiny{$M_{{X}}$}} q[1],q[2],q[3],q[4],q[5],q[6],q[7];

    text {$m_1$} q[1];
    text {$m_2$} q[2];
    text {$m_3$} q[3];
    text {$m_4$} q[4];
    text {$m_5$} q[5];
    text {$m_6$} q[6];
    text {$m_7$} q[7];
    
    text {$\sim\ket{{Y}}$} q[0];

    discard q;
    \end{yquant}
    \end{tikzpicture}
    }}
 \hspace{.5cm}\xrightarrow[\hspace{.5cm}]{} \hspace{.5cm}
    \raisebox{18\height}{\scalebox{.9}{\begin{tikzpicture}
    \begin{pgfonlayer}{nodelayer}
        \node [style=Z dot] (0) at (0.00, 0.00) {};
        \node [style=Z dot] (1) at (0.00, -2.00) {};
        \node [style=Z dot] (2) at (0.00, -4.00) {};
        \node [style=X dot] (3) at (0.00, -6.00) {};
        \node [style=Z dot] (4) at (0.00, -8.00) {};
        \node [style=X dot] (5) at (0.00, -10.00) {};
        \node [style=X dot] (6) at (0.00, -12.00) {};
        \node [style=X dot] (7) at (0.00, -14.00) {};
        \node [style=Z dot] (8) at (1.00, -2.00) {};
        \node [style=X dot] (9) at (1.00, -10.00) {};
        \node [style=Z dot] (10) at (2.00, -4.00) {};
        \node [style=X dot] (11) at (2.00, -12.00) {};
        \node [style=Z dot] (12) at (3.00, 0.00) {};
        \node [style=X dot] (13) at (3.00, -4.00) {};
        \node [style=Z dot] (14) at (3.00, -8.00) {};
        \node [style=X dot] (15) at (3.00, -12.00) {};
        \node [style=Z dot] (16) at (4.00, -2.00) {};
        \node [style=X dot] (17) at (4.00, -6.00) {};
        \node [style=Z dot] (18) at (4.00, -10.00) {};
        \node [style=X dot] (19) at (4.00, -14.00) {};
        \node [style=Z dot] (20) at (5.00, 0.00) {};
        \node [style=X dot] (21) at (5.00, -2.00) {};
        \node [style=Z dot] (22) at (5.00, -4.00) {};
        \node [style=X dot] (23) at (5.00, -6.00) {};
        \node [style=Z dot] (24) at (5.00, -8.00) {};
        \node [style=X dot] (25) at (5.00, -10.00) {};
        \node [style=Z dot] (26) at (5.00, -12.00) {};
        \node [style=X dot] (27) at (5.00, -14.00) {};
        \node [style=none] (28) at (11.00, 0.00) {};
        \node [style=Z phase dot] (29) at (10.00, -2.00) {$m_1\pi$};
        \node [style=Z phase dot] (30) at (10.00, -4.00) {$m_2\pi$};
        \node [style=Z phase dot] (31) at (10.00, -6.00) {$m_3\pi$};
        \node [style=Z phase dot] (32) at (10.00, -8.00) {$m_4\pi$};
        \node [style=Z phase dot] (33) at (10.00, -10.00) {$m_5\pi$};
        \node [style=Z phase dot] (34) at (10.00, -12.00) {$m_6\pi$};
        \node [style=Z phase dot] (35) at (10.00, -14.00) {$m_7\pi$};
        \node [style=Z phase dot] (36) at (6.00, -1.00) {$\frac{\pi}{2}$};
        \node [style=Z phase dot] (37) at (6.00, -3.00) {$\frac{\pi}{2}$};
        \node [style=Z phase dot] (38) at (6.00, -5.00) {$\frac{\pi}{2}$};
        \node [style=Z phase dot] (39) at (6.00, -7.00) {$\frac{\pi}{2}$};
        \node [style=Z phase dot] (40) at (6.00, -9.00) {$\frac{\pi}{2}$};
        \node [style=Z phase dot] (41) at (6.00, -11.00) {$\frac{\pi}{2}$};
        \node [style=Z phase dot] (42) at (6.00, -13.00) {$\frac{\pi}{2}$};
        \node [style=X dot] (43) at (7.00, -1.00) {};
        \node [style=Z dot] (44) at (7.00, -2.00) {};
        \node [style=X dot] (45) at (7.00, -3.00) {};
        \node [style=Z dot] (46) at (7.00, -4.00) {};
        \node [style=X dot] (47) at (7.00, -5.00) {};
        \node [style=Z dot] (48) at (7.00, -6.00) {};
        \node [style=X dot] (49) at (7.00, -7.00) {};
        \node [style=Z dot] (50) at (7.00, -8.00) {};
        \node [style=X dot] (51) at (7.00, -9.00) {};
        \node [style=Z dot] (52) at (7.00, -10.00) {};
        \node [style=X dot] (53) at (7.00, -11.00) {};
        \node [style=Z dot] (54) at (7.00, -12.00) {};
        \node [style=X dot] (55) at (7.00, -13.00) {};
        \node [style=Z dot] (56) at (7.00, -14.00) {};
        \node [style=X phase dot] (57) at (8.00, -1.00) {$n_1\pi$};
        \node [style=X phase dot] (58) at (8.00, -3.00) {$n_2\pi$};
        \node [style=X phase dot] (59) at (8.00, -5.00) {$n_3\pi$};
        \node [style=X phase dot] (60) at (8.00, -7.00) {$n_4\pi$};
        \node [style=X phase dot] (61) at (8.00, -9.00) {$n_5\pi$};
        \node [style=X phase dot] (62) at (8.00, -11.00) {$n_6\pi$};
        \node [style=X phase dot] (63) at (8.00, -13.00) {$n_7\pi$};
    \end{pgfonlayer}
    \begin{pgfonlayer}{edgelayer}
        \draw (0) to (12);
        \draw (1) to (8);
        \draw (2) to (10);
        \draw (3) to (17);
        \draw (4) to (14);
        \draw (5) to (9);
        \draw (6) to (11);
        \draw (7) to (19);
        \draw (8) to (9);
        \draw (8) to (16);
        \draw (9) to (18);
        \draw (10) to (11);
        \draw (10) to (13);
        \draw (11) to (15);
        \draw (12) to (13);
        \draw (12) to (20);
        \draw (13) to (22);
        \draw (14) to (15);
        \draw (14) to (24);
        \draw (15) to (26);
        \draw (16) to (17);
        \draw (16) to (21);
        \draw (17) to (23);
        \draw (18) to (19);
        \draw (18) to (25);
        \draw (19) to (27);
        \draw (20) to (21);
        \draw (20) to (28);
        \draw (21) to (44);
        \draw (22) to (23);
        \draw (22) to (46);
        \draw (23) to (48);
        \draw (24) to (25);
        \draw (24) to (50);
        \draw (25) to (52);
        \draw (26) to (27);
        \draw (26) to (54);
        \draw (27) to (56);
        \draw (29) to (44);
        \draw (30) to (46);
        \draw (31) to (48);
        \draw (32) to (50);
        \draw (33) to (52);
        \draw (34) to (54);
        \draw (35) to (56);
        \draw (36) to (43);
        \draw (37) to (45);
        \draw (38) to (47);
        \draw (39) to (49);
        \draw (40) to (51);
        \draw (41) to (53);
        \draw (42) to (55);
        \draw (43) to (57);
        \draw (43) to (44);
        \draw (45) to (58);
        \draw (45) to (46);
        \draw (47) to (59);
        \draw (47) to (48);
        \draw (49) to (60);
        \draw (49) to (50);
        \draw (51) to (61);
        \draw (51) to (52);
        \draw (53) to (62);
        \draw (53) to (54);
        \draw (55) to (63);
        \draw (55) to (56);
    \end{pgfonlayer}
\end{tikzpicture}}} 
\end{equation}

\subsection{Pauli webs for 7-to-1 distillation protocol}
\label{app:pauli_webs_yfac}
Pauli webs \cite{Bombin_2024} (also called stabiliser flows for quantum circuits \cite{McEwen_2023}) are decorations on Clifford ZX-diagrams, providing a compact description of how a Pauli observable ``propagates'' through a Clifford circuit. Or in other words, it reveals the underlying stabiliser structure of the corresponding ZX-diagram/resultant quantum state. As an example, the 3-GHZ state with a single $S$ gate applied ($SII(\ket{000}+\ket{111})$) can be represented as the following ZX-diagram with the following stabiliser generators: 
\begin{equation}
\label{eq:3ghz_s}
\begin{split}
        \scalebox{1}{\begin{tikzpicture}[scale=2]
	\begin{pgfonlayer}{nodelayer}
		\node [style=none] (0) at (0.8, 0.65) {};
		\node [style=none] (3) at (0.8, 0) {};
		\node [style=none] (6) at (0.8, -0.65) {};
		\node [style=Z phase dot] (9) at (0, 0) {$\frac{\pi}{2}$};
	\end{pgfonlayer}
	\begin{pgfonlayer}{edgelayer}
		\draw [in=-175, out=90] (9) to (0);
		\draw (9) to (3);
		\draw [in=175, out=-90] (9) to (6);
	\end{pgfonlayer}
\end{tikzpicture}} \propto SII(\ket{000}+\ket{111}) & \ , \ \hspace{1cm} \langle ZZI, IZZ, YXX \rangle  \ , \\
\end{split}
\end{equation}
where the following open Pauli webs (edge highlights) describe its stabilisers of this state at the boundary of the ZX-diagram (Pauli-${\color{red}X}{\color{blue}Y}{\color{green}Z}$ coloured accordingly).
\begin{equation}
\label{eq:3ghz_s_webs}
\setlength{\arraycolsep}{4pt}
\begin{array}{ccc}
\Bigg{\langle}\scalebox{1}{\begin{tikzpicture}[scale=2]
	\begin{pgfonlayer}{nodelayer}
		\node [style=none] (0) at (0.8, 0.65) {{\color{green}$Z$}};
		\node [style=none] (3) at (0.8, 0) {{\color{green}$Z$}};
		\node [style=none] (6) at (0.8, -0.65) {};
		\node [style=Z phase dot] (9) at (0, 0) {$\frac{\pi}{2}$};
	\end{pgfonlayer}
	\begin{pgfonlayer}{edgelayer}
		\draw [in=-175, out=90] (9) to (0);
		\draw (9) to (3);
		\draw [in=175, out=-90] (9) to (6);
	\end{pgfonlayer}
    \begin{pgfonlayer}{edgelayer}
    \draw[green, line width=2.5pt, opacity=0.4,in=-175, out=90] (9) to (0);
    \draw[green, line width=2.5pt, opacity=0.4] (9) to (3);
    \end{pgfonlayer}

\end{tikzpicture}} \ , &
\scalebox{1}{\begin{tikzpicture}[scale=2]
	\begin{pgfonlayer}{nodelayer}
		\node [style=none] (0) at (0.8, 0.65) {};
		\node [style=none] (3) at (0.8, 0) {{\color{green}$Z$}};
		\node [style=none] (6) at (0.8, -0.65) {{\color{green}$Z$}};
		\node [style=Z phase dot] (9) at (0, 0) {$\frac{\pi}{2}$};
	\end{pgfonlayer}
	\begin{pgfonlayer}{edgelayer}
		\draw [in=-175, out=90] (9) to (0);
		\draw (9) to (3);
		\draw [in=175, out=-90] (9) to (6);
	\end{pgfonlayer}
    \begin{pgfonlayer}{edgelayer}
    \draw[green, line width=2.5pt, opacity=0.4] (9) to (3);
    \draw[green, line width=2.5pt, opacity=0.4,in=175, out=-90] (9) to (6);
    \end{pgfonlayer}

\end{tikzpicture}} \ , &
\scalebox{1}{\begin{tikzpicture}[scale=2]
	\begin{pgfonlayer}{nodelayer}
		\node [style=none] (0) at (0.8, 0.65) {{\color{blue}$Y$}};
		\node [style=none] (3) at (0.8, 0) {{\color{red}$X$}};
		\node [style=none] (6) at (0.8, -0.65) {{\color{red}$X$}};
		\node [style=Z phase dot] (9) at (0, 0) {$\frac{\pi}{2}$};
	\end{pgfonlayer}
	\begin{pgfonlayer}{edgelayer}
		\draw [in=-175, out=90] (9) to (0);
		\draw (9) to (3);
		\draw [in=175, out=-90] (9) to (6);
	\end{pgfonlayer}
    \begin{pgfonlayer}{edgelayer}
    \draw[blue, line width=2.5pt, opacity=0.4,in=-175, out=90] (9) to (0);
    \draw[red, line width=2.5pt, opacity=0.4] (9) to (3);
    \draw[red, line width=2.5pt, opacity=0.4,in=175, out=-90] (9) to (6);
    \end{pgfonlayer}

\end{tikzpicture}}\Bigg{\rangle} \\
&
&
\\
\scriptsize \langle {\color{green}Z}{\color{green}Z}I, &
\scriptsize I{\color{green}Z}{\color{green}Z}, &
\scriptsize {\color{blue}Y}{\color{red}X}{\color{red}X} \rangle
\end{array}
\end{equation}
For the 7-to-1 protocol's ZX-diagram, there are four Pauli webs, shown in equation \ref{eq:yfac_all}. These are the logical Pauli webs, where each subsystem corresponds to a surface code patch in our case.
\begin{equation}
\setlength{\arraycolsep}{4pt}
\begin{array}{cccc}
\scalebox{.65}{\input{yfac_logical.tikz}} &
\scalebox{.65}{\input{yfac_check_1357.tikz}} &
\scalebox{.65}{\input{yfac_check_2367.tikz}} &
\scalebox{.65}{\input{yfac_check_4567.tikz}} \\
&
&
&
\\
\text{logical }  Y \text{ membrane}&
 X_1X_3X_5X_7 &
 X_2X_3X_6X_7 &
 X_4X_5X_6X_7
\end{array}
\label{eq:yfac_all}
\end{equation}

\begin{itemize}
    \item \textbf{Logical $Y$.}
    The web labelled ``Logical $Y$ membrane'' is the Pauli web/stabiliser flow corresponding to
    the desired output stabiliser on patch $0$ (for $\ket{Y}$ distillation, the
    target is $Y_0$ up to a known sign). The sign of the output stabiliser depends on a global parity of the classical outcomes, as shown in the boundary of these Pauli webs:
    \begin{equation}
        Y_0 \;\; \mapsto \;\; (-1)^{1+\sum_{k=1}^{7}(m_k+n_k)}\,Y_0 ,
        \label{eq:yfac_global_parity}
    \end{equation}
    matching equation \ref{eq:Y_state_consumption_G_mx}. In practice, we track this sign as a Pauli-frame update (or equivalently, a potential $Z_0$ correction), rather than applying a physical correction.

    \item \textbf{Steane code $X$-stabiliser check Pauli webs.}
    The remaining three webs in equation \ref{eq:yfac_all} correspond to the three independent weight-4 Steane $X$ stabilisers. In
    terms of the measurement outcomes, the three parities are
    \begin{align}
        s_{1357} &= (m_1\oplus n_1)\oplus(m_3\oplus n_3)\oplus(m_5\oplus n_5)\oplus(m_7\oplus n_7),\\
        s_{2367} &= (m_2\oplus n_2)\oplus(m_3\oplus n_3)\oplus(m_6\oplus n_6)\oplus(m_7\oplus n_7),\\
        s_{4567} &= (m_4\oplus n_4)\oplus(m_5\oplus n_5)\oplus(m_6\oplus n_6)\oplus(m_7\oplus n_7).
        \label{eq:yfac_steane_parities}
    \end{align}
    The post-selection rule is simply
    \begin{equation}
        s_{1357}=s_{2367}=s_{4567}=0 \ ,
        \label{eq:yfac_postselect_rule}
    \end{equation}
    equivalent to checking for all the Steane code $X$-stabiliser check Pauli webs to be unviolated.
\end{itemize}

\subsection{Benchmarking the 7-to-1 protocol and logical failure}
\label{app:benchmark_yfac}
A Pauli web corresponds to a Pauli observable propagated through the circuit. Two webs that differ by multiplication with a check web (namely a stabiliser flow that is fixed to $+1$ upon post-selection) represent the same logical observable. This is the analogue of multiplying a logical operator by a stabiliser in a stabiliser code. In the 7-to-1 protocol, post-selection enforces the three Steane $X$-checks in equation \ref{eq:yfac_steane_parities} to be trivial. Consequently, the logical-$Y$ membrane web in equation \ref{eq:yfac_all} can be freely multiplied by any of the three check webs without changing its action on the codespace. Multiplying the logical $Y$ membrane by the $X_4X_5X_6X_7$ web yields an equivalent, smaller weight Pauli web representation:
\begin{equation}
\raisebox{20.5\height}{\scalebox{.65}{\begin{tikzpicture}
    \begin{pgfonlayer}{nodelayer}
        \node [style=Z dot] (0) at (0.00, 0.00) {};
        \node [style=Z dot] (1) at (0.00, -2.00) {};
        \node [style=Z dot] (2) at (0.00, -4.00) {};
        \node [style=X dot] (3) at (0.00, -6.00) {};
        \node [style=Z dot] (4) at (0.00, -8.00) {};
        \node [style=X dot] (5) at (0.00, -10.00) {};
        \node [style=X dot] (6) at (0.00, -12.00) {};
        \node [style=X dot] (7) at (0.00, -14.00) {};
        \node [style=Z dot] (8) at (1.00, -2.00) {};
        \node [style=X dot] (9) at (1.00, -10.00) {};
        \node [style=Z dot] (10) at (2.00, -4.00) {};
        \node [style=X dot] (11) at (2.00, -12.00) {};
        \node [style=Z dot] (12) at (3.00, 0.00) {};
        \node [style=X dot] (13) at (3.00, -4.00) {};
        \node [style=Z dot] (14) at (3.00, -8.00) {};
        \node [style=X dot] (15) at (3.00, -12.00) {};
        \node [style=Z dot] (16) at (4.00, -2.00) {};
        \node [style=X dot] (17) at (4.00, -6.00) {};
        \node [style=Z dot] (18) at (4.00, -10.00) {};
        \node [style=X dot] (19) at (4.00, -14.00) {};
        \node [style=Z dot] (20) at (5.00, 0.00) {};
        \node [style=X dot] (21) at (5.00, -2.00) {};
        \node [style=Z dot] (22) at (5.00, -4.00) {};
        \node [style=X dot] (23) at (5.00, -6.00) {};
        \node [style=Z dot] (24) at (5.00, -8.00) {};
        \node [style=X dot] (25) at (5.00, -10.00) {};
        \node [style=Z dot] (26) at (5.00, -12.00) {};
        \node [style=X dot] (27) at (5.00, -14.00) {};
        \node [style=none] (28) at (11.00, 0.00) {};
        \node [style=Z phase dot] (29) at (10.00, -2.00) {$m_1\pi$};
        \node [style=Z phase dot] (30) at (10.00, -4.00) {$m_2\pi$};
        \node [style=Z phase dot] (31) at (10.00, -6.00) {$m_3\pi$};
        \node [style=Z phase dot] (32) at (10.00, -8.00) {$m_4\pi$};
        \node [style=Z phase dot] (33) at (10.00, -10.00) {$m_5\pi$};
        \node [style=Z phase dot] (34) at (10.00, -12.00) {$m_6\pi$};
        \node [style=Z phase dot] (35) at (10.00, -14.00) {$m_7\pi$};
        \node [style=Z phase dot] (36) at (6.00, -1.00) {$\frac{\pi}{2}$};
        \node [style=Z phase dot] (37) at (6.00, -3.00) {$\frac{\pi}{2}$};
        \node [style=Z phase dot] (38) at (6.00, -5.00) {$\frac{\pi}{2}$};
        \node [style=Z phase dot] (39) at (6.00, -7.00) {$\frac{\pi}{2}$};
        \node [style=Z phase dot] (40) at (6.00, -9.00) {$\frac{\pi}{2}$};
        \node [style=Z phase dot] (41) at (6.00, -11.00) {$\frac{\pi}{2}$};
        \node [style=Z phase dot] (42) at (6.00, -13.00) {$\frac{\pi}{2}$};
        \node [style=X dot] (43) at (7.00, -1.00) {};
        \node [style=Z dot] (44) at (7.00, -2.00) {};
        \node [style=X dot] (45) at (7.00, -3.00) {};
        \node [style=Z dot] (46) at (7.00, -4.00) {};
        \node [style=X dot] (47) at (7.00, -5.00) {};
        \node [style=Z dot] (48) at (7.00, -6.00) {};
        \node [style=X dot] (49) at (7.00, -7.00) {};
        \node [style=Z dot] (50) at (7.00, -8.00) {};
        \node [style=X dot] (51) at (7.00, -9.00) {};
        \node [style=Z dot] (52) at (7.00, -10.00) {};
        \node [style=X dot] (53) at (7.00, -11.00) {};
        \node [style=Z dot] (54) at (7.00, -12.00) {};
        \node [style=X dot] (55) at (7.00, -13.00) {};
        \node [style=Z dot] (56) at (7.00, -14.00) {};
        \node [style=X phase dot] (57) at (8.00, -1.00) {$n_1\pi$};
        \node [style=X phase dot] (58) at (8.00, -3.00) {$n_2\pi$};
        \node [style=X phase dot] (59) at (8.00, -5.00) {$n_3\pi$};
        \node [style=X phase dot] (60) at (8.00, -7.00) {$n_4\pi$};
        \node [style=X phase dot] (61) at (8.00, -9.00) {$n_5\pi$};
        \node [style=X phase dot] (62) at (8.00, -11.00) {$n_6\pi$};
        \node [style=X phase dot] (63) at (8.00, -13.00) {$n_7\pi$};
    \end{pgfonlayer}
    \begin{pgfonlayer}{edgelayer}
        \draw (0) to (12);
        \draw (1) to (8);
        \draw (2) to (10);
        \draw (3) to (17);
        \draw (4) to (14);
        \draw (5) to (9);
        \draw (6) to (11);
        \draw (7) to (19);
        \draw (8) to (9);
        \draw (8) to (16);
        \draw (9) to (18);
        \draw (10) to (11);
        \draw (10) to (13);
        \draw (11) to (15);
        \draw (12) to (13);
        \draw (12) to (20);
        \draw (13) to (22);
        \draw (14) to (15);
        \draw (14) to (24);
        \draw (15) to (26);
        \draw (16) to (17);
        \draw (16) to (21);
        \draw (17) to (23);
        \draw (18) to (19);
        \draw (18) to (25);
        \draw (19) to (27);
        \draw (20) to (21);
        \draw (20) to (28);
        \draw (21) to (44);
        \draw (22) to (23);
        \draw (22) to (46);
        \draw (23) to (48);
        \draw (24) to (25);
        \draw (24) to (50);
        \draw (25) to (52);
        \draw (26) to (27);
        \draw (26) to (54);
        \draw (27) to (56);
        \draw (29) to (44);
        \draw (30) to (46);
        \draw (31) to (48);
        \draw (32) to (50);
        \draw (33) to (52);
        \draw (34) to (54);
        \draw (35) to (56);
        \draw (36) to (43);
        \draw (37) to (45);
        \draw (38) to (47);
        \draw (39) to (49);
        \draw (40) to (51);
        \draw (41) to (53);
        \draw (42) to (55);
        \draw (43) to (57);
        \draw (43) to (44);
        \draw (45) to (58);
        \draw (45) to (46);
        \draw (47) to (59);
        \draw (47) to (48);
        \draw (49) to (60);
        \draw (49) to (50);
        \draw (51) to (61);
        \draw (51) to (52);
        \draw (53) to (62);
        \draw (53) to (54);
        \draw (55) to (63);
        \draw (55) to (56);
    \end{pgfonlayer}

\begin{pgfonlayer}{edgelayer}
\draw[blue, line width=2.5pt, opacity=0.4] (20) -- ($(20)!0.5!(21)$);
\draw[blue, line width=2.5pt, opacity=0.4] (21) -- ($(21)!0.5!(20)$);
\draw[blue, line width=2.5pt, opacity=0.4] (20) -- ($(20)!0.5!(28)$);
\draw[blue, line width=2.5pt, opacity=0.4] (28) -- ($(28)!0.5!(20)$);
\draw[red, line width=2.5pt, opacity=0.4] (12) -- ($(12)!0.5!(13)$);
\draw[red, line width=2.5pt, opacity=0.4] (13) -- ($(13)!0.5!(12)$);
\draw[green, line width=2.5pt, opacity=0.4] (17) -- ($(17)!0.5!(16)$);
\draw[green, line width=2.5pt, opacity=0.4] (16) -- ($(16)!0.5!(17)$);
\draw[green, line width=2.5pt, opacity=0.4] (21) -- ($(21)!0.5!(16)$);
\draw[green, line width=2.5pt, opacity=0.4] (16) -- ($(16)!0.5!(21)$);
\draw[blue, line width=2.5pt, opacity=0.4] (21) -- ($(21)!0.5!(44)$);
\draw[blue, line width=2.5pt, opacity=0.4] (44) -- ($(44)!0.5!(21)$);
\draw[red, line width=2.5pt, opacity=0.4] (22) -- ($(22)!0.5!(13)$);
\draw[red, line width=2.5pt, opacity=0.4] (13) -- ($(13)!0.5!(22)$);
\draw[blue, line width=2.5pt, opacity=0.4] (22) -- ($(22)!0.5!(23)$);
\draw[blue, line width=2.5pt, opacity=0.4] (23) -- ($(23)!0.5!(22)$);
\draw[blue, line width=2.5pt, opacity=0.4] (23) -- ($(23)!0.5!(48)$);
\draw[blue, line width=2.5pt, opacity=0.4] (48) -- ($(48)!0.5!(23)$);
\draw[blue, line width=2.5pt, opacity=0.4] (43) -- ($(43)!0.5!(36)$);
\draw[blue, line width=2.5pt, opacity=0.4] (36) -- ($(36)!0.5!(43)$);
\draw[blue, line width=2.5pt, opacity=0.4] (43) -- ($(43)!0.5!(44)$);
\draw[blue, line width=2.5pt, opacity=0.4] (44) -- ($(44)!0.5!(43)$);
\draw[blue, line width=2.5pt, opacity=0.4] (45) -- ($(45)!0.5!(37)$);
\draw[blue, line width=2.5pt, opacity=0.4] (37) -- ($(37)!0.5!(45)$);
\draw[blue, line width=2.5pt, opacity=0.4] (45) -- ($(45)!0.5!(46)$);
\draw[blue, line width=2.5pt, opacity=0.4] (46) -- ($(46)!0.5!(45)$);
\draw[blue, line width=2.5pt, opacity=0.4] (47) -- ($(47)!0.5!(38)$);
\draw[blue, line width=2.5pt, opacity=0.4] (38) -- ($(38)!0.5!(47)$);
\draw[blue, line width=2.5pt, opacity=0.4] (47) -- ($(47)!0.5!(48)$);
\draw[blue, line width=2.5pt, opacity=0.4] (48) -- ($(48)!0.5!(47)$);
\draw[red, line width=2.5pt, opacity=0.4] (0) -- ($(0)!0.5!(12)$);
\draw[red, line width=2.5pt, opacity=0.4] (12) -- ($(12)!0.5!(0)$);
\draw[green, line width=2.5pt, opacity=0.4] (3) -- ($(3)!0.5!(17)$);
\draw[green, line width=2.5pt, opacity=0.4] (17) -- ($(17)!0.5!(3)$);
\draw[red, line width=2.5pt, opacity=0.4] (12) -- ($(12)!0.5!(20)$);
\draw[red, line width=2.5pt, opacity=0.4] (20) -- ($(20)!0.5!(12)$);
\draw[green, line width=2.5pt, opacity=0.4] (17) -- ($(17)!0.5!(23)$);
\draw[green, line width=2.5pt, opacity=0.4] (23) -- ($(23)!0.5!(17)$);
\draw[blue, line width=2.5pt, opacity=0.4] (22) -- ($(22)!0.5!(46)$);
\draw[blue, line width=2.5pt, opacity=0.4] (46) -- ($(46)!0.5!(22)$);
\draw[red, line width=2.5pt, opacity=0.4] (29) -- ($(29)!0.5!(44)$);
\draw[red, line width=2.5pt, opacity=0.4] (44) -- ($(44)!0.5!(29)$);
\draw[red, line width=2.5pt, opacity=0.4] (30) -- ($(30)!0.5!(46)$);
\draw[red, line width=2.5pt, opacity=0.4] (46) -- ($(46)!0.5!(30)$);
\draw[red, line width=2.5pt, opacity=0.4] (31) -- ($(31)!0.5!(48)$);
\draw[red, line width=2.5pt, opacity=0.4] (48) -- ($(48)!0.5!(31)$);
\draw[green, line width=2.5pt, opacity=0.4] (43) -- ($(43)!0.5!(57)$);
\draw[green, line width=2.5pt, opacity=0.4] (57) -- ($(57)!0.5!(43)$);
\draw[green, line width=2.5pt, opacity=0.4] (45) -- ($(45)!0.5!(58)$);
\draw[green, line width=2.5pt, opacity=0.4] (58) -- ($(58)!0.5!(45)$);
\draw[green, line width=2.5pt, opacity=0.4] (47) -- ($(47)!0.5!(59)$);
\draw[green, line width=2.5pt, opacity=0.4] (59) -- ($(59)!0.5!(47)$);
\end{pgfonlayer}
\end{tikzpicture}}}
\;=\;
\raisebox{20.5\height}{\scalebox{.65}{\input{yfac_logical.tikz}}}
\;\times\;
\raisebox{20.5\height}{\scalebox{.65}{\input{yfac_check_4567.tikz}}}\, .
\label{eq:y_web_reduction}
\end{equation}

We may benchmark this stabiliser circuit by estimating the post-selected logical failure rate on the output patch $0$. For each sampled shot, we:
\begin{enumerate}
    \item decode all surface-code patches (iteratively, when transversal CNOTs are present) to obtain logical measurement outcomes $(m_j)$ and consumption outcomes $(n_j)$, together with a consistent Pauli frame, then
    \item apply the post-selection rule: $s_{1357}=s_{2367}=s_{4567}=0$
    from equation \ref{eq:yfac_postselect_rule},
    \item for accepted shots, then compute the predicted sign of the output $Y_0$
    stabiliser using the global parity in equation \ref{eq:yfac_global_parity}
    (equivalently, update the output Pauli frame by the corresponding Clifford
    correction), and finally
    \item declare a logical failure if the decoded output is inconsistent with the target $\ket{Y}$ eigenvalue after applying the tracked frame update.
\end{enumerate}
Or alternatively, if the Pauli web on the LHS of equation \ref{eq:y_web_reduction} is violated, then it implies a logical error.

\section{Benchmarking the stabiliser proxy to the 15-to-1 protocol}
\label{app:benchmark_15to1_proxy}

The 15-to-1 protocol is a non-Clifford distillation routine when implemented with
$\ket{T}$ state consumptions, hence direct large distance end to end simulation is challenging. Instead, one can benchmark
a stabiliser proxy obtained by replacing all the non-Clifford resource state by a Clifford state (for example, $\ket{Y}$ states are consumed, shown in equation \ref{eq:tcir_to_tfac}), implementing a transversal $S$ up to Pauli byproducts. This retains a similar CNOT encoding sub-circuit and, crucially, preserves a similar post-selection structure based on Reed-Muller $X$-stabiliser checks.
\begin{equation}
    \label{eq:tcir_to_tfac}
        \raisebox{-0.475\height}{\resizebox{0.39\linewidth}{!}{
        \begin{tikzpicture}
    \begin{yquant}
    qubit {0: $\ket{{+}}$} q[+1];
    qubit {} a[+1]; discard a[0];
    qubit {1: $\ket{{+}}$} q[+1];
    qubit {} a[+1]; discard a[1];
    qubit {2: $\ket{{+}}$} q[+1];
    qubit {} a[+1]; discard a[2];
    qubit {3: $\ket{{0}}$\hspace{0.1cm}} q[+1];
    qubit {} a[+1]; discard a[3];
    qubit {4: $\ket{{+}}$} q[+1];
    qubit {} a[+1]; discard a[4];
    qubit {5: $\ket{{0}}$\hspace{0.1cm}} q[+1];
    qubit {} a[+1]; discard a[5];
    qubit {6: $\ket{{0}}$\hspace{0.1cm}} q[+1];
    qubit {} a[+1]; discard a[6];
    qubit {7: $\ket{{0}}$\hspace{0.1cm}} q[+1];
    qubit {} a[+1]; discard a[7];
    qubit {8: $\ket{{+}}$} q[+1];
    qubit {} a[+1]; discard a[8];
    qubit {9: $\ket{{0}}$\hspace{0.1cm}} q[+1];
    qubit {} a[+1]; discard a[9];
    qubit {10: $\ket{{0}}$\hspace{0.1cm}} q[+1];
    qubit {} a[+1]; discard a[10];
    qubit {11: $\ket{{0}}$\hspace{0.1cm}} q[+1];
    qubit {} a[+1]; discard a[11];
    qubit {12: $\ket{{0}}$\hspace{0.1cm}} q[+1];
    qubit {} a[+1]; discard a[12];
    qubit {13: $\ket{{0}}$\hspace{0.1cm}} q[+1];
    qubit {} a[+1]; discard a[13];
    qubit {14: $\ket{{0}}$\hspace{0.1cm}} q[+1];
    qubit {} a[+1]; discard a[14];
    qubit {15: $\ket{{0}}$\hspace{0.1cm}} q[+1];
    
    [blue, thick, label=0]
    barrier (-);

    cnot q[9] | q[1];
    cnot q[10] | q[2];
    cnot q[12] | q[4];

    [blue, thick, label=1]
    barrier (-);

    align q;
    
    cnot q[4] | q[0];
    cnot q[5] | q[1];
    cnot q[6] | q[2];

    cnot q[12] | q[8];
    cnot q[13] | q[9];
    cnot q[14] | q[10];

    [blue, thick, label=2]
    barrier (-);

    align q;
    
    cnot q[2] | q[0];
    cnot q[6] | q[4];
    cnot q[10] | q[8];
    cnot q[14] | q[12];
    
    align q;
    
    cnot q[3] | q[1];
    cnot q[7] | q[5];
    cnot q[11] | q[9];
    cnot q[15] | q[13];
    
    [blue, thick, label=3]
    barrier (-);

    align q;
    
    cnot q[1] | q[0];
    cnot q[3] | q[2];
    cnot q[5] | q[4];
    cnot q[7] | q[6];

    cnot q[9] | q[8];
    cnot q[11] | q[10];
    cnot q[13] | q[12];
    cnot q[15] | q[14];
    
    [blue, thick, label=4]
    barrier (-);

    align a;

    init {$\ket{{Y}}$} a[0];
    cnot a[0] | q[1];
    init {$\ket{{Y}}$} a[1];
    cnot a[1] | q[2];
    init {$\ket{{Y}}$} a[2];
    cnot a[2] | q[3];
    init {$\ket{{Y}}$} a[3];
    cnot a[3] | q[4];
    init {$\ket{{Y}}$} a[4];
    cnot a[4] | q[5];
    init {$\ket{{Y}}$} a[5];
    cnot a[5] | q[6];
    init {$\ket{{Y}}$} a[6];
    cnot a[6] | q[7];
    
    init {$\ket{{Y}}$} a[7];
    cnot a[7] | q[8];
    init {$\ket{{Y}}$} a[8];
    cnot a[8] | q[9];
    init {$\ket{{Y}}$} a[9];
    cnot a[9] | q[10];
    init {$\ket{{Y}}$} a[10];
    cnot a[10] | q[11];
    init {$\ket{{Y}}$} a[11];
    cnot a[11] | q[12];
    init {$\ket{{Y}}$} a[12];
    cnot a[12] | q[13];
    init {$\ket{{Y}}$} a[13];
    cnot a[13] | q[14];
    init {$\ket{{Y}}$} a[14];
    cnot a[14] | q[15];

    align a; align q;
    
    [blue, thick, label=5]
    barrier (-);

    align a,q;

  hspace {10mm} - q[0]; text {$\sim\ket{{Y}}$} q[0]; discard q[0];
  
measure {} a[0];
text {$n_1$} a[0];

 measure {} a[1];
 text {$n_2$} a[1];

 measure {} a[2];
 text {$n_3$} a[2];

 measure {} a[3];
  text {$n_4$} a[3];

 measure {} a[4];
  text {$n_5$} a[4];

 measure {} a[5];
  text {$n_6$} a[5];

 measure {} a[6];
  text {$n_7$} a[6];

 measure {} a[7];
  text {$n_8$} a[7];

measure {} a[8];
 text {$n_9$} a[8];
    
measure {} a[9];
 text {$n_{10}$} a[9];

measure {} a[10];
 text {$n_{11}$} a[10];

measure {} a[11];
 text {$n_{12}$} a[11];

measure {} a[12]; 
text {$n_{13}$} a[12];

measure {} a[13];
 text {$n_{14}$} a[13];

measure {} a[14];
 text {$n_{15}$} a[14];

 measure {\tiny{$M_{{X}}$}} q[1];
 text {$m_1$} q[1];

 measure {\tiny{$M_{{X}}$}} q[2];
 text {$m_2$} q[2];

 measure {\tiny{$M_{{X}}$}} q[3];
  text {$m_3$} q[3];

 measure {\tiny{$M_{{X}}$}} q[4];
  text {$m_4$} q[4];

 measure {\tiny{$M_{{X}}$}} q[5];
  text {$m_5$} q[5];

 measure {\tiny{$M_{{X}}$}} q[6];
  text {$m_6$} q[6];

 measure {\tiny{$M_{{X}}$}} q[7];
  text {$m_7$} q[7];

measure {\tiny{$M_{{X}}$}} q[8];
 text {$m_8$} q[8];
    
measure {\tiny{$M_{{X}}$}} q[9];
 text {$m_{9}$} q[9];

measure {\tiny{$M_{{X}}$}} q[10];
 text {$m_{10}$} q[10];

measure {\tiny{$M_{{X}}$}} q[11];
 text {$m_{11}$} q[11];

measure {\tiny{$M_{{X}}$}} q[12]; 
text {$m_{12}$} q[12];

measure {\tiny{$M_{{X}}$}} q[13];
 text {$m_{13}$} q[13];

measure {\tiny{$M_{{X}}$}} q[14];
 text {$m_{14}$} q[14];

measure {\tiny{$M_{{X}}$}} q[15];
text {$m_{15}$} q[15];

    discard a,q;

    \end{yquant}
    \end{tikzpicture}

    }}
 \hspace{.5cm}\xrightarrow[\hspace{.5cm}]{} \hspace{.5cm}
    \raisebox{40\height}{\scalebox{.62}{\input{tfac.tikz}}} 
\end{equation}
In the Pauli web language, the proxy circuit admits four independent weight-8 Reed-Muller code $X$-type stabilisers as Pauli webs (see equation \ref{eq:tfac_all}) and one logical membrane web supported on the output of patch $0$ (shown on the RHS of equation \ref{eq:tcir_to_tfac}). We post-select on the trivial values of all four $X$-check webs, and then track the Pauli-frame on patch $0$. Conditioned on successful post-selection, a logical failure occurs when the logical membrane web is violated (equivalently, the decoded output is inconsistent with the target stabiliser sign after Pauli frame update).
\begin{equation}
\setlength{\arraycolsep}{4pt}
\begin{array}{cccc}
\scalebox{.45}{\input{tc0.tikz}} &
\scalebox{.45}{\input{tc1.tikz}} &
\scalebox{.45}{\input{tc2.tikz}} &
\scalebox{.45}{\input{tc3.tikz}} \\
&&&\\[-2pt]
\resizebox{.15\linewidth}{!}{$X_{1} X_{3} X_{5} X_{7} X_{9} X_{11} X_{13} X_{15}$} &
\resizebox{.15\linewidth}{!}{$X_{2} X_{3} X_{6} X_{7} X_{10} X_{11} X_{14} X_{15}$} &
\resizebox{.15\linewidth}{!}{$X_{4} X_{5} X_{6} X_{7} X_{12} X_{13} X_{14} X_{15}$} &
\resizebox{.15\linewidth}{!}{$X_{8} X_{9} X_{10} X_{11} X_{12} X_{13} X_{14} X_{15}$}
\end{array}\, 
\label{eq:tfac_all}
\end{equation}

\section{Pauli webs for the $\ket{-}$ state proxies}
\label{app:main_text_proxy_note}

The main text benchmarks both the 7-to-1 and 15-to-1 circuits using $\ket{-}$ and $\ket{Y}$ resources. Here we describe the $\ket{-}$ stabiliser proxies, with injected $Z$ faults of \textbf{logical} error probability $p$ on the consumed logical resource states. These proxies isolate the post-selection and decoding behaviour of the factory and allow us to confirm the
expected cubic suppression scaling under circuit-level noise. See below (equation \ref{eq:sims_proxy}) for the logical membranes of the $\ket{-}$ state proxies to the 7-to-1 and 15-to-1 protocols.

\begin{equation}
\setlength{\arraycolsep}{4pt}
\begin{array}{cc}
\raisebox{-41\height}{%
\scalebox{.6}{\begin{tikzpicture}
    \begin{pgfonlayer}{nodelayer}
        \node [style=Z dot] (0) at (0.00, 0.00) {};
        \node [style=Z dot] (1) at (0.00, -2.00) {};
        \node [style=Z dot] (2) at (0.00, -4.00) {};
        \node [style=X dot] (3) at (0.00, -6.00) {};
        \node [style=Z dot] (4) at (0.00, -8.00) {};
        \node [style=X dot] (5) at (0.00, -10.00) {};
        \node [style=X dot] (6) at (0.00, -12.00) {};
        \node [style=X dot] (7) at (0.00, -14.00) {};
        \node [style=Z dot] (8) at (1.00, -2.00) {};
        \node [style=X dot] (9) at (1.00, -10.00) {};
        \node [style=Z dot] (10) at (2.00, -4.00) {};
        \node [style=X dot] (11) at (2.00, -12.00) {};
        \node [style=Z dot] (12) at (3.00, 0.00) {};
        \node [style=X dot] (13) at (3.00, -4.00) {};
        \node [style=Z dot] (14) at (3.00, -8.00) {};
        \node [style=X dot] (15) at (3.00, -12.00) {};
        \node [style=Z dot] (16) at (4.00, -2.00) {};
        \node [style=X dot] (17) at (4.00, -6.00) {};
        \node [style=Z dot] (18) at (4.00, -10.00) {};
        \node [style=X dot] (19) at (4.00, -14.00) {};
        \node [style=Z dot] (20) at (5.00, 0.00) {};
        \node [style=X dot] (21) at (5.00, -2.00) {};
        \node [style=Z dot] (22) at (5.00, -4.00) {};
        \node [style=X dot] (23) at (5.00, -6.00) {};
        \node [style=Z dot] (24) at (5.00, -8.00) {};
        \node [style=X dot] (25) at (5.00, -10.00) {};
        \node [style=Z dot] (26) at (5.00, -12.00) {};
        \node [style=X dot] (27) at (5.00, -14.00) {};
        \node [style=none] (28) at (11.00, 0.00) {};
        \node [style=Z phase dot] (29) at (10.00, -2.00) {$m_1\pi$};
        \node [style=Z phase dot] (30) at (10.00, -4.00) {$m_2\pi$};
        \node [style=Z phase dot] (31) at (10.00, -6.00) {$m_3\pi$};
        \node [style=Z phase dot] (32) at (10.00, -8.00) {$m_4\pi$};
        \node [style=Z phase dot] (33) at (10.00, -10.00) {$m_5\pi$};
        \node [style=Z phase dot] (34) at (10.00, -12.00) {$m_6\pi$};
        \node [style=Z phase dot] (35) at (10.00, -14.00) {$m_7\pi$};
        \node [style=Z phase dot] (36) at (6.00, -1.00) {$\pi$};
        \node [style=Z phase dot] (37) at (6.00, -3.00) {$\pi$};
        \node [style=Z phase dot] (38) at (6.00, -5.00) {$\pi$};
        \node [style=Z phase dot] (39) at (6.00, -7.00) {$\pi$};
        \node [style=Z phase dot] (40) at (6.00, -9.00) {$\pi$};
        \node [style=Z phase dot] (41) at (6.00, -11.00) {$\pi$};
        \node [style=Z phase dot] (42) at (6.00, -13.00) {$\pi$};
        \node [style=X dot] (43) at (7.00, -1.00) {};
        \node [style=Z dot] (44) at (7.00, -2.00) {};
        \node [style=X dot] (45) at (7.00, -3.00) {};
        \node [style=Z dot] (46) at (7.00, -4.00) {};
        \node [style=X dot] (47) at (7.00, -5.00) {};
        \node [style=Z dot] (48) at (7.00, -6.00) {};
        \node [style=X dot] (49) at (7.00, -7.00) {};
        \node [style=Z dot] (50) at (7.00, -8.00) {};
        \node [style=X dot] (51) at (7.00, -9.00) {};
        \node [style=Z dot] (52) at (7.00, -10.00) {};
        \node [style=X dot] (53) at (7.00, -11.00) {};
        \node [style=Z dot] (54) at (7.00, -12.00) {};
        \node [style=X dot] (55) at (7.00, -13.00) {};
        \node [style=Z dot] (56) at (7.00, -14.00) {};
        \node [style=X phase dot] (57) at (8.00, -1.00) {$n_1\pi$};
        \node [style=X phase dot] (58) at (8.00, -3.00) {$n_2\pi$};
        \node [style=X phase dot] (59) at (8.00, -5.00) {$n_3\pi$};
        \node [style=X phase dot] (60) at (8.00, -7.00) {$n_4\pi$};
        \node [style=X phase dot] (61) at (8.00, -9.00) {$n_5\pi$};
        \node [style=X phase dot] (62) at (8.00, -11.00) {$n_6\pi$};
        \node [style=X phase dot] (63) at (8.00, -13.00) {$n_7\pi$};
    \end{pgfonlayer}
    \begin{pgfonlayer}{edgelayer}
        \draw (0) to (12);
        \draw (1) to (8);
        \draw (2) to (10);
        \draw (3) to (17);
        \draw (4) to (14);
        \draw (5) to (9);
        \draw (6) to (11);
        \draw (7) to (19);
        \draw (8) to (9);
        \draw (8) to (16);
        \draw (9) to (18);
        \draw (10) to (11);
        \draw (10) to (13);
        \draw (11) to (15);
        \draw (12) to (13);
        \draw (12) to (20);
        \draw (13) to (22);
        \draw (14) to (15);
        \draw (14) to (24);
        \draw (15) to (26);
        \draw (16) to (17);
        \draw (16) to (21);
        \draw (17) to (23);
        \draw (18) to (19);
        \draw (18) to (25);
        \draw (19) to (27);
        \draw (20) to (21);
        \draw (20) to (28);
        \draw (21) to (44);
        \draw (22) to (23);
        \draw (22) to (46);
        \draw (23) to (48);
        \draw (24) to (25);
        \draw (24) to (50);
        \draw (25) to (52);
        \draw (26) to (27);
        \draw (26) to (54);
        \draw (27) to (56);
        \draw (29) to (44);
        \draw (30) to (46);
        \draw (31) to (48);
        \draw (32) to (50);
        \draw (33) to (52);
        \draw (34) to (54);
        \draw (35) to (56);
        \draw (36) to (43);
        \draw (37) to (45);
        \draw (38) to (47);
        \draw (39) to (49);
        \draw (40) to (51);
        \draw (41) to (53);
        \draw (42) to (55);
        \draw (43) to (57);
        \draw (43) to (44);
        \draw (45) to (58);
        \draw (45) to (46);
        \draw (47) to (59);
        \draw (47) to (48);
        \draw (49) to (60);
        \draw (49) to (50);
        \draw (51) to (61);
        \draw (51) to (52);
        \draw (53) to (62);
        \draw (53) to (54);
        \draw (55) to (63);
        \draw (55) to (56);
    \end{pgfonlayer}

\begin{pgfonlayer}{edgelayer}
\draw[red, line width=2.5pt, opacity=0.4] (20) -- ($(20)!0.5!(21)$);
\draw[red, line width=2.5pt, opacity=0.4] (21) -- ($(21)!0.5!(20)$);
\draw[red, line width=2.5pt, opacity=0.4] (20) -- ($(20)!0.5!(28)$);
\draw[red, line width=2.5pt, opacity=0.4] (28) -- ($(28)!0.5!(20)$);
\draw[red, line width=2.5pt, opacity=0.4] (12) -- ($(12)!0.5!(13)$);
\draw[red, line width=2.5pt, opacity=0.4] (13) -- ($(13)!0.5!(12)$);
\draw[red, line width=2.5pt, opacity=0.4] (22) -- ($(22)!0.5!(13)$);
\draw[red, line width=2.5pt, opacity=0.4] (13) -- ($(13)!0.5!(22)$);
\draw[red, line width=2.5pt, opacity=0.4] (22) -- ($(22)!0.5!(23)$);
\draw[red, line width=2.5pt, opacity=0.4] (23) -- ($(23)!0.5!(22)$);
\draw[red, line width=2.5pt, opacity=0.4] (36) -- ($(36)!0.5!(43)$);
\draw[red, line width=2.5pt, opacity=0.4] (43) -- ($(43)!0.5!(36)$);
\draw[red, line width=2.5pt, opacity=0.4] (37) -- ($(37)!0.5!(45)$);
\draw[red, line width=2.5pt, opacity=0.4] (45) -- ($(45)!0.5!(37)$);
\draw[red, line width=2.5pt, opacity=0.4] (38) -- ($(38)!0.5!(47)$);
\draw[red, line width=2.5pt, opacity=0.4] (47) -- ($(47)!0.5!(38)$);
\draw[red, line width=2.5pt, opacity=0.4] (44) -- ($(44)!0.5!(21)$);
\draw[red, line width=2.5pt, opacity=0.4] (21) -- ($(21)!0.5!(44)$);
\draw[red, line width=2.5pt, opacity=0.4] (44) -- ($(44)!0.5!(43)$);
\draw[red, line width=2.5pt, opacity=0.4] (43) -- ($(43)!0.5!(44)$);
\draw[red, line width=2.5pt, opacity=0.4] (46) -- ($(46)!0.5!(45)$);
\draw[red, line width=2.5pt, opacity=0.4] (45) -- ($(45)!0.5!(46)$);
\draw[red, line width=2.5pt, opacity=0.4] (48) -- ($(48)!0.5!(23)$);
\draw[red, line width=2.5pt, opacity=0.4] (23) -- ($(23)!0.5!(48)$);
\draw[red, line width=2.5pt, opacity=0.4] (48) -- ($(48)!0.5!(47)$);
\draw[red, line width=2.5pt, opacity=0.4] (47) -- ($(47)!0.5!(48)$);
\draw[red, line width=2.5pt, opacity=0.4] (0) -- ($(0)!0.5!(12)$);
\draw[red, line width=2.5pt, opacity=0.4] (12) -- ($(12)!0.5!(0)$);
\draw[red, line width=2.5pt, opacity=0.4] (12) -- ($(12)!0.5!(20)$);
\draw[red, line width=2.5pt, opacity=0.4] (20) -- ($(20)!0.5!(12)$);
\draw[red, line width=2.5pt, opacity=0.4] (22) -- ($(22)!0.5!(46)$);
\draw[red, line width=2.5pt, opacity=0.4] (46) -- ($(46)!0.5!(22)$);
\draw[red, line width=2.5pt, opacity=0.4] (29) -- ($(29)!0.5!(44)$);
\draw[red, line width=2.5pt, opacity=0.4] (44) -- ($(44)!0.5!(29)$);
\draw[red, line width=2.5pt, opacity=0.4] (30) -- ($(30)!0.5!(46)$);
\draw[red, line width=2.5pt, opacity=0.4] (46) -- ($(46)!0.5!(30)$);
\draw[red, line width=2.5pt, opacity=0.4] (31) -- ($(31)!0.5!(48)$);
\draw[red, line width=2.5pt, opacity=0.4] (48) -- ($(48)!0.5!(31)$);
\end{pgfonlayer}
\end{tikzpicture}}
} & \scalebox{.60}{\input{neg_state_proxy15.tikz}} \\
& \\
\text{7-to-1} & \text{15-to-1}
\end{array} 
\label{eq:sims_proxy}
\end{equation}
Post-selecting on the following logical check webs for the 7-to-1 protocol,
\begin{equation}
\setlength{\arraycolsep}{4pt}
\begin{array}{cccc}
\scalebox{.65}{\begin{tikzpicture}
    \begin{pgfonlayer}{nodelayer}
        \node [style=Z dot] (0) at (0.00, 0.00) {};
        \node [style=Z dot] (1) at (0.00, -2.00) {};
        \node [style=Z dot] (2) at (0.00, -4.00) {};
        \node [style=X dot] (3) at (0.00, -6.00) {};
        \node [style=Z dot] (4) at (0.00, -8.00) {};
        \node [style=X dot] (5) at (0.00, -10.00) {};
        \node [style=X dot] (6) at (0.00, -12.00) {};
        \node [style=X dot] (7) at (0.00, -14.00) {};
        \node [style=Z dot] (8) at (1.00, -2.00) {};
        \node [style=X dot] (9) at (1.00, -10.00) {};
        \node [style=Z dot] (10) at (2.00, -4.00) {};
        \node [style=X dot] (11) at (2.00, -12.00) {};
        \node [style=Z dot] (12) at (3.00, 0.00) {};
        \node [style=X dot] (13) at (3.00, -4.00) {};
        \node [style=Z dot] (14) at (3.00, -8.00) {};
        \node [style=X dot] (15) at (3.00, -12.00) {};
        \node [style=Z dot] (16) at (4.00, -2.00) {};
        \node [style=X dot] (17) at (4.00, -6.00) {};
        \node [style=Z dot] (18) at (4.00, -10.00) {};
        \node [style=X dot] (19) at (4.00, -14.00) {};
        \node [style=Z dot] (20) at (5.00, 0.00) {};
        \node [style=X dot] (21) at (5.00, -2.00) {};
        \node [style=Z dot] (22) at (5.00, -4.00) {};
        \node [style=X dot] (23) at (5.00, -6.00) {};
        \node [style=Z dot] (24) at (5.00, -8.00) {};
        \node [style=X dot] (25) at (5.00, -10.00) {};
        \node [style=Z dot] (26) at (5.00, -12.00) {};
        \node [style=X dot] (27) at (5.00, -14.00) {};
        \node [style=none] (28) at (11.00, 0.00) {};
        \node [style=Z phase dot] (29) at (10.00, -2.00) {$m_1\pi$};
        \node [style=Z phase dot] (30) at (10.00, -4.00) {$m_2\pi$};
        \node [style=Z phase dot] (31) at (10.00, -6.00) {$m_3\pi$};
        \node [style=Z phase dot] (32) at (10.00, -8.00) {$m_4\pi$};
        \node [style=Z phase dot] (33) at (10.00, -10.00) {$m_5\pi$};
        \node [style=Z phase dot] (34) at (10.00, -12.00) {$m_6\pi$};
        \node [style=Z phase dot] (35) at (10.00, -14.00) {$m_7\pi$};
        \node [style=Z phase dot] (36) at (6.00, -1.00) {$\pi$};
        \node [style=Z phase dot] (37) at (6.00, -3.00) {$\pi$};
        \node [style=Z phase dot] (38) at (6.00, -5.00) {$\pi$};
        \node [style=Z phase dot] (39) at (6.00, -7.00) {$\pi$};
        \node [style=Z phase dot] (40) at (6.00, -9.00) {$\pi$};
        \node [style=Z phase dot] (41) at (6.00, -11.00) {$\pi$};
        \node [style=Z phase dot] (42) at (6.00, -13.00) {$\pi$};
        \node [style=X dot] (43) at (7.00, -1.00) {};
        \node [style=Z dot] (44) at (7.00, -2.00) {};
        \node [style=X dot] (45) at (7.00, -3.00) {};
        \node [style=Z dot] (46) at (7.00, -4.00) {};
        \node [style=X dot] (47) at (7.00, -5.00) {};
        \node [style=Z dot] (48) at (7.00, -6.00) {};
        \node [style=X dot] (49) at (7.00, -7.00) {};
        \node [style=Z dot] (50) at (7.00, -8.00) {};
        \node [style=X dot] (51) at (7.00, -9.00) {};
        \node [style=Z dot] (52) at (7.00, -10.00) {};
        \node [style=X dot] (53) at (7.00, -11.00) {};
        \node [style=Z dot] (54) at (7.00, -12.00) {};
        \node [style=X dot] (55) at (7.00, -13.00) {};
        \node [style=Z dot] (56) at (7.00, -14.00) {};
        \node [style=X phase dot] (57) at (8.00, -1.00) {$n_1\pi$};
        \node [style=X phase dot] (58) at (8.00, -3.00) {$n_2\pi$};
        \node [style=X phase dot] (59) at (8.00, -5.00) {$n_3\pi$};
        \node [style=X phase dot] (60) at (8.00, -7.00) {$n_4\pi$};
        \node [style=X phase dot] (61) at (8.00, -9.00) {$n_5\pi$};
        \node [style=X phase dot] (62) at (8.00, -11.00) {$n_6\pi$};
        \node [style=X phase dot] (63) at (8.00, -13.00) {$n_7\pi$};
    \end{pgfonlayer}
    \begin{pgfonlayer}{edgelayer}
        \draw (0) to (12);
        \draw (1) to (8);
        \draw (2) to (10);
        \draw (3) to (17);
        \draw (4) to (14);
        \draw (5) to (9);
        \draw (6) to (11);
        \draw (7) to (19);
        \draw (8) to (9);
        \draw (8) to (16);
        \draw (9) to (18);
        \draw (10) to (11);
        \draw (10) to (13);
        \draw (11) to (15);
        \draw (12) to (13);
        \draw (12) to (20);
        \draw (13) to (22);
        \draw (14) to (15);
        \draw (14) to (24);
        \draw (15) to (26);
        \draw (16) to (17);
        \draw (16) to (21);
        \draw (17) to (23);
        \draw (18) to (19);
        \draw (18) to (25);
        \draw (19) to (27);
        \draw (20) to (21);
        \draw (20) to (28);
        \draw (21) to (44);
        \draw (22) to (23);
        \draw (22) to (46);
        \draw (23) to (48);
        \draw (24) to (25);
        \draw (24) to (50);
        \draw (25) to (52);
        \draw (26) to (27);
        \draw (26) to (54);
        \draw (27) to (56);
        \draw (29) to (44);
        \draw (30) to (46);
        \draw (31) to (48);
        \draw (32) to (50);
        \draw (33) to (52);
        \draw (34) to (54);
        \draw (35) to (56);
        \draw (36) to (43);
        \draw (37) to (45);
        \draw (38) to (47);
        \draw (39) to (49);
        \draw (40) to (51);
        \draw (41) to (53);
        \draw (42) to (55);
        \draw (43) to (57);
        \draw (43) to (44);
        \draw (45) to (58);
        \draw (45) to (46);
        \draw (47) to (59);
        \draw (47) to (48);
        \draw (49) to (60);
        \draw (49) to (50);
        \draw (51) to (61);
        \draw (51) to (52);
        \draw (53) to (62);
        \draw (53) to (54);
        \draw (55) to (63);
        \draw (55) to (56);
    \end{pgfonlayer}

\begin{pgfonlayer}{edgelayer}
\draw[red, line width=2.5pt, opacity=0.4] (8) -- ($(8)!0.5!(9)$);
\draw[red, line width=2.5pt, opacity=0.4] (9) -- ($(9)!0.5!(8)$);
\draw[red, line width=2.5pt, opacity=0.4] (16) -- ($(16)!0.5!(17)$);
\draw[red, line width=2.5pt, opacity=0.4] (17) -- ($(17)!0.5!(16)$);
\draw[red, line width=2.5pt, opacity=0.4] (16) -- ($(16)!0.5!(21)$);
\draw[red, line width=2.5pt, opacity=0.4] (21) -- ($(21)!0.5!(16)$);
\draw[red, line width=2.5pt, opacity=0.4] (18) -- ($(18)!0.5!(9)$);
\draw[red, line width=2.5pt, opacity=0.4] (9) -- ($(9)!0.5!(18)$);
\draw[red, line width=2.5pt, opacity=0.4] (18) -- ($(18)!0.5!(19)$);
\draw[red, line width=2.5pt, opacity=0.4] (19) -- ($(19)!0.5!(18)$);
\draw[red, line width=2.5pt, opacity=0.4] (18) -- ($(18)!0.5!(25)$);
\draw[red, line width=2.5pt, opacity=0.4] (25) -- ($(25)!0.5!(18)$);
\draw[red, line width=2.5pt, opacity=0.4] (36) -- ($(36)!0.5!(43)$);
\draw[red, line width=2.5pt, opacity=0.4] (43) -- ($(43)!0.5!(36)$);
\draw[red, line width=2.5pt, opacity=0.4] (38) -- ($(38)!0.5!(47)$);
\draw[red, line width=2.5pt, opacity=0.4] (47) -- ($(47)!0.5!(38)$);
\draw[red, line width=2.5pt, opacity=0.4] (40) -- ($(40)!0.5!(51)$);
\draw[red, line width=2.5pt, opacity=0.4] (51) -- ($(51)!0.5!(40)$);
\draw[red, line width=2.5pt, opacity=0.4] (42) -- ($(42)!0.5!(55)$);
\draw[red, line width=2.5pt, opacity=0.4] (55) -- ($(55)!0.5!(42)$);
\draw[red, line width=2.5pt, opacity=0.4] (44) -- ($(44)!0.5!(21)$);
\draw[red, line width=2.5pt, opacity=0.4] (21) -- ($(21)!0.5!(44)$);
\draw[red, line width=2.5pt, opacity=0.4] (44) -- ($(44)!0.5!(43)$);
\draw[red, line width=2.5pt, opacity=0.4] (43) -- ($(43)!0.5!(44)$);
\draw[red, line width=2.5pt, opacity=0.4] (48) -- ($(48)!0.5!(23)$);
\draw[red, line width=2.5pt, opacity=0.4] (23) -- ($(23)!0.5!(48)$);
\draw[red, line width=2.5pt, opacity=0.4] (48) -- ($(48)!0.5!(47)$);
\draw[red, line width=2.5pt, opacity=0.4] (47) -- ($(47)!0.5!(48)$);
\draw[red, line width=2.5pt, opacity=0.4] (52) -- ($(52)!0.5!(25)$);
\draw[red, line width=2.5pt, opacity=0.4] (25) -- ($(25)!0.5!(52)$);
\draw[red, line width=2.5pt, opacity=0.4] (52) -- ($(52)!0.5!(51)$);
\draw[red, line width=2.5pt, opacity=0.4] (51) -- ($(51)!0.5!(52)$);
\draw[red, line width=2.5pt, opacity=0.4] (56) -- ($(56)!0.5!(27)$);
\draw[red, line width=2.5pt, opacity=0.4] (27) -- ($(27)!0.5!(56)$);
\draw[red, line width=2.5pt, opacity=0.4] (56) -- ($(56)!0.5!(55)$);
\draw[red, line width=2.5pt, opacity=0.4] (55) -- ($(55)!0.5!(56)$);
\draw[red, line width=2.5pt, opacity=0.4] (1) -- ($(1)!0.5!(8)$);
\draw[red, line width=2.5pt, opacity=0.4] (8) -- ($(8)!0.5!(1)$);
\draw[red, line width=2.5pt, opacity=0.4] (8) -- ($(8)!0.5!(16)$);
\draw[red, line width=2.5pt, opacity=0.4] (16) -- ($(16)!0.5!(8)$);
\draw[red, line width=2.5pt, opacity=0.4] (17) -- ($(17)!0.5!(23)$);
\draw[red, line width=2.5pt, opacity=0.4] (23) -- ($(23)!0.5!(17)$);
\draw[red, line width=2.5pt, opacity=0.4] (19) -- ($(19)!0.5!(27)$);
\draw[red, line width=2.5pt, opacity=0.4] (27) -- ($(27)!0.5!(19)$);
\draw[red, line width=2.5pt, opacity=0.4] (29) -- ($(29)!0.5!(44)$);
\draw[red, line width=2.5pt, opacity=0.4] (44) -- ($(44)!0.5!(29)$);
\draw[red, line width=2.5pt, opacity=0.4] (31) -- ($(31)!0.5!(48)$);
\draw[red, line width=2.5pt, opacity=0.4] (48) -- ($(48)!0.5!(31)$);
\draw[red, line width=2.5pt, opacity=0.4] (33) -- ($(33)!0.5!(52)$);
\draw[red, line width=2.5pt, opacity=0.4] (52) -- ($(52)!0.5!(33)$);
\draw[red, line width=2.5pt, opacity=0.4] (35) -- ($(35)!0.5!(56)$);
\draw[red, line width=2.5pt, opacity=0.4] (56) -- ($(56)!0.5!(35)$);
\end{pgfonlayer}
\end{tikzpicture}} &
\scalebox{.65}{\begin{tikzpicture}
    \begin{pgfonlayer}{nodelayer}
        \node [style=Z dot] (0) at (0.00, 0.00) {};
        \node [style=Z dot] (1) at (0.00, -2.00) {};
        \node [style=Z dot] (2) at (0.00, -4.00) {};
        \node [style=X dot] (3) at (0.00, -6.00) {};
        \node [style=Z dot] (4) at (0.00, -8.00) {};
        \node [style=X dot] (5) at (0.00, -10.00) {};
        \node [style=X dot] (6) at (0.00, -12.00) {};
        \node [style=X dot] (7) at (0.00, -14.00) {};
        \node [style=Z dot] (8) at (1.00, -2.00) {};
        \node [style=X dot] (9) at (1.00, -10.00) {};
        \node [style=Z dot] (10) at (2.00, -4.00) {};
        \node [style=X dot] (11) at (2.00, -12.00) {};
        \node [style=Z dot] (12) at (3.00, 0.00) {};
        \node [style=X dot] (13) at (3.00, -4.00) {};
        \node [style=Z dot] (14) at (3.00, -8.00) {};
        \node [style=X dot] (15) at (3.00, -12.00) {};
        \node [style=Z dot] (16) at (4.00, -2.00) {};
        \node [style=X dot] (17) at (4.00, -6.00) {};
        \node [style=Z dot] (18) at (4.00, -10.00) {};
        \node [style=X dot] (19) at (4.00, -14.00) {};
        \node [style=Z dot] (20) at (5.00, 0.00) {};
        \node [style=X dot] (21) at (5.00, -2.00) {};
        \node [style=Z dot] (22) at (5.00, -4.00) {};
        \node [style=X dot] (23) at (5.00, -6.00) {};
        \node [style=Z dot] (24) at (5.00, -8.00) {};
        \node [style=X dot] (25) at (5.00, -10.00) {};
        \node [style=Z dot] (26) at (5.00, -12.00) {};
        \node [style=X dot] (27) at (5.00, -14.00) {};
        \node [style=none] (28) at (11.00, 0.00) {};
        \node [style=Z phase dot] (29) at (10.00, -2.00) {$m_1\pi$};
        \node [style=Z phase dot] (30) at (10.00, -4.00) {$m_2\pi$};
        \node [style=Z phase dot] (31) at (10.00, -6.00) {$m_3\pi$};
        \node [style=Z phase dot] (32) at (10.00, -8.00) {$m_4\pi$};
        \node [style=Z phase dot] (33) at (10.00, -10.00) {$m_5\pi$};
        \node [style=Z phase dot] (34) at (10.00, -12.00) {$m_6\pi$};
        \node [style=Z phase dot] (35) at (10.00, -14.00) {$m_7\pi$};
        \node [style=Z phase dot] (36) at (6.00, -1.00) {$\pi$};
        \node [style=Z phase dot] (37) at (6.00, -3.00) {$\pi$};
        \node [style=Z phase dot] (38) at (6.00, -5.00) {$\pi$};
        \node [style=Z phase dot] (39) at (6.00, -7.00) {$\pi$};
        \node [style=Z phase dot] (40) at (6.00, -9.00) {$\pi$};
        \node [style=Z phase dot] (41) at (6.00, -11.00) {$\pi$};
        \node [style=Z phase dot] (42) at (6.00, -13.00) {$\pi$};
        \node [style=X dot] (43) at (7.00, -1.00) {};
        \node [style=Z dot] (44) at (7.00, -2.00) {};
        \node [style=X dot] (45) at (7.00, -3.00) {};
        \node [style=Z dot] (46) at (7.00, -4.00) {};
        \node [style=X dot] (47) at (7.00, -5.00) {};
        \node [style=Z dot] (48) at (7.00, -6.00) {};
        \node [style=X dot] (49) at (7.00, -7.00) {};
        \node [style=Z dot] (50) at (7.00, -8.00) {};
        \node [style=X dot] (51) at (7.00, -9.00) {};
        \node [style=Z dot] (52) at (7.00, -10.00) {};
        \node [style=X dot] (53) at (7.00, -11.00) {};
        \node [style=Z dot] (54) at (7.00, -12.00) {};
        \node [style=X dot] (55) at (7.00, -13.00) {};
        \node [style=Z dot] (56) at (7.00, -14.00) {};
        \node [style=X phase dot] (57) at (8.00, -1.00) {$n_1\pi$};
        \node [style=X phase dot] (58) at (8.00, -3.00) {$n_2\pi$};
        \node [style=X phase dot] (59) at (8.00, -5.00) {$n_3\pi$};
        \node [style=X phase dot] (60) at (8.00, -7.00) {$n_4\pi$};
        \node [style=X phase dot] (61) at (8.00, -9.00) {$n_5\pi$};
        \node [style=X phase dot] (62) at (8.00, -11.00) {$n_6\pi$};
        \node [style=X phase dot] (63) at (8.00, -13.00) {$n_7\pi$};
    \end{pgfonlayer}
    \begin{pgfonlayer}{edgelayer}
        \draw (0) to (12);
        \draw (1) to (8);
        \draw (2) to (10);
        \draw (3) to (17);
        \draw (4) to (14);
        \draw (5) to (9);
        \draw (6) to (11);
        \draw (7) to (19);
        \draw (8) to (9);
        \draw (8) to (16);
        \draw (9) to (18);
        \draw (10) to (11);
        \draw (10) to (13);
        \draw (11) to (15);
        \draw (12) to (13);
        \draw (12) to (20);
        \draw (13) to (22);
        \draw (14) to (15);
        \draw (14) to (24);
        \draw (15) to (26);
        \draw (16) to (17);
        \draw (16) to (21);
        \draw (17) to (23);
        \draw (18) to (19);
        \draw (18) to (25);
        \draw (19) to (27);
        \draw (20) to (21);
        \draw (20) to (28);
        \draw (21) to (44);
        \draw (22) to (23);
        \draw (22) to (46);
        \draw (23) to (48);
        \draw (24) to (25);
        \draw (24) to (50);
        \draw (25) to (52);
        \draw (26) to (27);
        \draw (26) to (54);
        \draw (27) to (56);
        \draw (29) to (44);
        \draw (30) to (46);
        \draw (31) to (48);
        \draw (32) to (50);
        \draw (33) to (52);
        \draw (34) to (54);
        \draw (35) to (56);
        \draw (36) to (43);
        \draw (37) to (45);
        \draw (38) to (47);
        \draw (39) to (49);
        \draw (40) to (51);
        \draw (41) to (53);
        \draw (42) to (55);
        \draw (43) to (57);
        \draw (43) to (44);
        \draw (45) to (58);
        \draw (45) to (46);
        \draw (47) to (59);
        \draw (47) to (48);
        \draw (49) to (60);
        \draw (49) to (50);
        \draw (51) to (61);
        \draw (51) to (52);
        \draw (53) to (62);
        \draw (53) to (54);
        \draw (55) to (63);
        \draw (55) to (56);
    \end{pgfonlayer}

\begin{pgfonlayer}{edgelayer}
\draw[red, line width=2.5pt, opacity=0.4] (10) -- ($(10)!0.5!(11)$);
\draw[red, line width=2.5pt, opacity=0.4] (11) -- ($(11)!0.5!(10)$);
\draw[red, line width=2.5pt, opacity=0.4] (10) -- ($(10)!0.5!(13)$);
\draw[red, line width=2.5pt, opacity=0.4] (13) -- ($(13)!0.5!(10)$);
\draw[red, line width=2.5pt, opacity=0.4] (22) -- ($(22)!0.5!(13)$);
\draw[red, line width=2.5pt, opacity=0.4] (13) -- ($(13)!0.5!(22)$);
\draw[red, line width=2.5pt, opacity=0.4] (22) -- ($(22)!0.5!(23)$);
\draw[red, line width=2.5pt, opacity=0.4] (23) -- ($(23)!0.5!(22)$);
\draw[red, line width=2.5pt, opacity=0.4] (26) -- ($(26)!0.5!(15)$);
\draw[red, line width=2.5pt, opacity=0.4] (15) -- ($(15)!0.5!(26)$);
\draw[red, line width=2.5pt, opacity=0.4] (26) -- ($(26)!0.5!(27)$);
\draw[red, line width=2.5pt, opacity=0.4] (27) -- ($(27)!0.5!(26)$);
\draw[red, line width=2.5pt, opacity=0.4] (37) -- ($(37)!0.5!(45)$);
\draw[red, line width=2.5pt, opacity=0.4] (45) -- ($(45)!0.5!(37)$);
\draw[red, line width=2.5pt, opacity=0.4] (38) -- ($(38)!0.5!(47)$);
\draw[red, line width=2.5pt, opacity=0.4] (47) -- ($(47)!0.5!(38)$);
\draw[red, line width=2.5pt, opacity=0.4] (41) -- ($(41)!0.5!(53)$);
\draw[red, line width=2.5pt, opacity=0.4] (53) -- ($(53)!0.5!(41)$);
\draw[red, line width=2.5pt, opacity=0.4] (42) -- ($(42)!0.5!(55)$);
\draw[red, line width=2.5pt, opacity=0.4] (55) -- ($(55)!0.5!(42)$);
\draw[red, line width=2.5pt, opacity=0.4] (46) -- ($(46)!0.5!(45)$);
\draw[red, line width=2.5pt, opacity=0.4] (45) -- ($(45)!0.5!(46)$);
\draw[red, line width=2.5pt, opacity=0.4] (48) -- ($(48)!0.5!(23)$);
\draw[red, line width=2.5pt, opacity=0.4] (23) -- ($(23)!0.5!(48)$);
\draw[red, line width=2.5pt, opacity=0.4] (48) -- ($(48)!0.5!(47)$);
\draw[red, line width=2.5pt, opacity=0.4] (47) -- ($(47)!0.5!(48)$);
\draw[red, line width=2.5pt, opacity=0.4] (54) -- ($(54)!0.5!(53)$);
\draw[red, line width=2.5pt, opacity=0.4] (53) -- ($(53)!0.5!(54)$);
\draw[red, line width=2.5pt, opacity=0.4] (56) -- ($(56)!0.5!(27)$);
\draw[red, line width=2.5pt, opacity=0.4] (27) -- ($(27)!0.5!(56)$);
\draw[red, line width=2.5pt, opacity=0.4] (56) -- ($(56)!0.5!(55)$);
\draw[red, line width=2.5pt, opacity=0.4] (55) -- ($(55)!0.5!(56)$);
\draw[red, line width=2.5pt, opacity=0.4] (2) -- ($(2)!0.5!(10)$);
\draw[red, line width=2.5pt, opacity=0.4] (10) -- ($(10)!0.5!(2)$);
\draw[red, line width=2.5pt, opacity=0.4] (11) -- ($(11)!0.5!(15)$);
\draw[red, line width=2.5pt, opacity=0.4] (15) -- ($(15)!0.5!(11)$);
\draw[red, line width=2.5pt, opacity=0.4] (22) -- ($(22)!0.5!(46)$);
\draw[red, line width=2.5pt, opacity=0.4] (46) -- ($(46)!0.5!(22)$);
\draw[red, line width=2.5pt, opacity=0.4] (26) -- ($(26)!0.5!(54)$);
\draw[red, line width=2.5pt, opacity=0.4] (54) -- ($(54)!0.5!(26)$);
\draw[red, line width=2.5pt, opacity=0.4] (30) -- ($(30)!0.5!(46)$);
\draw[red, line width=2.5pt, opacity=0.4] (46) -- ($(46)!0.5!(30)$);
\draw[red, line width=2.5pt, opacity=0.4] (31) -- ($(31)!0.5!(48)$);
\draw[red, line width=2.5pt, opacity=0.4] (48) -- ($(48)!0.5!(31)$);
\draw[red, line width=2.5pt, opacity=0.4] (34) -- ($(34)!0.5!(54)$);
\draw[red, line width=2.5pt, opacity=0.4] (54) -- ($(54)!0.5!(34)$);
\draw[red, line width=2.5pt, opacity=0.4] (35) -- ($(35)!0.5!(56)$);
\draw[red, line width=2.5pt, opacity=0.4] (56) -- ($(56)!0.5!(35)$);
\end{pgfonlayer}
\end{tikzpicture}} &
\scalebox{.65}{\begin{tikzpicture}
    \begin{pgfonlayer}{nodelayer}
        \node [style=Z dot] (0) at (0.00, 0.00) {};
        \node [style=Z dot] (1) at (0.00, -2.00) {};
        \node [style=Z dot] (2) at (0.00, -4.00) {};
        \node [style=X dot] (3) at (0.00, -6.00) {};
        \node [style=Z dot] (4) at (0.00, -8.00) {};
        \node [style=X dot] (5) at (0.00, -10.00) {};
        \node [style=X dot] (6) at (0.00, -12.00) {};
        \node [style=X dot] (7) at (0.00, -14.00) {};
        \node [style=Z dot] (8) at (1.00, -2.00) {};
        \node [style=X dot] (9) at (1.00, -10.00) {};
        \node [style=Z dot] (10) at (2.00, -4.00) {};
        \node [style=X dot] (11) at (2.00, -12.00) {};
        \node [style=Z dot] (12) at (3.00, 0.00) {};
        \node [style=X dot] (13) at (3.00, -4.00) {};
        \node [style=Z dot] (14) at (3.00, -8.00) {};
        \node [style=X dot] (15) at (3.00, -12.00) {};
        \node [style=Z dot] (16) at (4.00, -2.00) {};
        \node [style=X dot] (17) at (4.00, -6.00) {};
        \node [style=Z dot] (18) at (4.00, -10.00) {};
        \node [style=X dot] (19) at (4.00, -14.00) {};
        \node [style=Z dot] (20) at (5.00, 0.00) {};
        \node [style=X dot] (21) at (5.00, -2.00) {};
        \node [style=Z dot] (22) at (5.00, -4.00) {};
        \node [style=X dot] (23) at (5.00, -6.00) {};
        \node [style=Z dot] (24) at (5.00, -8.00) {};
        \node [style=X dot] (25) at (5.00, -10.00) {};
        \node [style=Z dot] (26) at (5.00, -12.00) {};
        \node [style=X dot] (27) at (5.00, -14.00) {};
        \node [style=none] (28) at (11.00, 0.00) {};
        \node [style=Z phase dot] (29) at (10.00, -2.00) {$m_1\pi$};
        \node [style=Z phase dot] (30) at (10.00, -4.00) {$m_2\pi$};
        \node [style=Z phase dot] (31) at (10.00, -6.00) {$m_3\pi$};
        \node [style=Z phase dot] (32) at (10.00, -8.00) {$m_4\pi$};
        \node [style=Z phase dot] (33) at (10.00, -10.00) {$m_5\pi$};
        \node [style=Z phase dot] (34) at (10.00, -12.00) {$m_6\pi$};
        \node [style=Z phase dot] (35) at (10.00, -14.00) {$m_7\pi$};
        \node [style=Z phase dot] (36) at (6.00, -1.00) {$\pi$};
        \node [style=Z phase dot] (37) at (6.00, -3.00) {$\pi$};
        \node [style=Z phase dot] (38) at (6.00, -5.00) {$\pi$};
        \node [style=Z phase dot] (39) at (6.00, -7.00) {$\pi$};
        \node [style=Z phase dot] (40) at (6.00, -9.00) {$\pi$};
        \node [style=Z phase dot] (41) at (6.00, -11.00) {$\pi$};
        \node [style=Z phase dot] (42) at (6.00, -13.00) {$\pi$};
        \node [style=X dot] (43) at (7.00, -1.00) {};
        \node [style=Z dot] (44) at (7.00, -2.00) {};
        \node [style=X dot] (45) at (7.00, -3.00) {};
        \node [style=Z dot] (46) at (7.00, -4.00) {};
        \node [style=X dot] (47) at (7.00, -5.00) {};
        \node [style=Z dot] (48) at (7.00, -6.00) {};
        \node [style=X dot] (49) at (7.00, -7.00) {};
        \node [style=Z dot] (50) at (7.00, -8.00) {};
        \node [style=X dot] (51) at (7.00, -9.00) {};
        \node [style=Z dot] (52) at (7.00, -10.00) {};
        \node [style=X dot] (53) at (7.00, -11.00) {};
        \node [style=Z dot] (54) at (7.00, -12.00) {};
        \node [style=X dot] (55) at (7.00, -13.00) {};
        \node [style=Z dot] (56) at (7.00, -14.00) {};
        \node [style=X phase dot] (57) at (8.00, -1.00) {$n_1\pi$};
        \node [style=X phase dot] (58) at (8.00, -3.00) {$n_2\pi$};
        \node [style=X phase dot] (59) at (8.00, -5.00) {$n_3\pi$};
        \node [style=X phase dot] (60) at (8.00, -7.00) {$n_4\pi$};
        \node [style=X phase dot] (61) at (8.00, -9.00) {$n_5\pi$};
        \node [style=X phase dot] (62) at (8.00, -11.00) {$n_6\pi$};
        \node [style=X phase dot] (63) at (8.00, -13.00) {$n_7\pi$};
    \end{pgfonlayer}
    \begin{pgfonlayer}{edgelayer}
        \draw (0) to (12);
        \draw (1) to (8);
        \draw (2) to (10);
        \draw (3) to (17);
        \draw (4) to (14);
        \draw (5) to (9);
        \draw (6) to (11);
        \draw (7) to (19);
        \draw (8) to (9);
        \draw (8) to (16);
        \draw (9) to (18);
        \draw (10) to (11);
        \draw (10) to (13);
        \draw (11) to (15);
        \draw (12) to (13);
        \draw (12) to (20);
        \draw (13) to (22);
        \draw (14) to (15);
        \draw (14) to (24);
        \draw (15) to (26);
        \draw (16) to (17);
        \draw (16) to (21);
        \draw (17) to (23);
        \draw (18) to (19);
        \draw (18) to (25);
        \draw (19) to (27);
        \draw (20) to (21);
        \draw (20) to (28);
        \draw (21) to (44);
        \draw (22) to (23);
        \draw (22) to (46);
        \draw (23) to (48);
        \draw (24) to (25);
        \draw (24) to (50);
        \draw (25) to (52);
        \draw (26) to (27);
        \draw (26) to (54);
        \draw (27) to (56);
        \draw (29) to (44);
        \draw (30) to (46);
        \draw (31) to (48);
        \draw (32) to (50);
        \draw (33) to (52);
        \draw (34) to (54);
        \draw (35) to (56);
        \draw (36) to (43);
        \draw (37) to (45);
        \draw (38) to (47);
        \draw (39) to (49);
        \draw (40) to (51);
        \draw (41) to (53);
        \draw (42) to (55);
        \draw (43) to (57);
        \draw (43) to (44);
        \draw (45) to (58);
        \draw (45) to (46);
        \draw (47) to (59);
        \draw (47) to (48);
        \draw (49) to (60);
        \draw (49) to (50);
        \draw (51) to (61);
        \draw (51) to (52);
        \draw (53) to (62);
        \draw (53) to (54);
        \draw (55) to (63);
        \draw (55) to (56);
    \end{pgfonlayer}

\begin{pgfonlayer}{edgelayer}
\draw[red, line width=2.5pt, opacity=0.4] (14) -- ($(14)!0.5!(15)$);
\draw[red, line width=2.5pt, opacity=0.4] (15) -- ($(15)!0.5!(14)$);
\draw[red, line width=2.5pt, opacity=0.4] (24) -- ($(24)!0.5!(25)$);
\draw[red, line width=2.5pt, opacity=0.4] (25) -- ($(25)!0.5!(24)$);
\draw[red, line width=2.5pt, opacity=0.4] (26) -- ($(26)!0.5!(15)$);
\draw[red, line width=2.5pt, opacity=0.4] (15) -- ($(15)!0.5!(26)$);
\draw[red, line width=2.5pt, opacity=0.4] (26) -- ($(26)!0.5!(27)$);
\draw[red, line width=2.5pt, opacity=0.4] (27) -- ($(27)!0.5!(26)$);
\draw[red, line width=2.5pt, opacity=0.4] (39) -- ($(39)!0.5!(49)$);
\draw[red, line width=2.5pt, opacity=0.4] (49) -- ($(49)!0.5!(39)$);
\draw[red, line width=2.5pt, opacity=0.4] (40) -- ($(40)!0.5!(51)$);
\draw[red, line width=2.5pt, opacity=0.4] (51) -- ($(51)!0.5!(40)$);
\draw[red, line width=2.5pt, opacity=0.4] (41) -- ($(41)!0.5!(53)$);
\draw[red, line width=2.5pt, opacity=0.4] (53) -- ($(53)!0.5!(41)$);
\draw[red, line width=2.5pt, opacity=0.4] (42) -- ($(42)!0.5!(55)$);
\draw[red, line width=2.5pt, opacity=0.4] (55) -- ($(55)!0.5!(42)$);
\draw[red, line width=2.5pt, opacity=0.4] (50) -- ($(50)!0.5!(49)$);
\draw[red, line width=2.5pt, opacity=0.4] (49) -- ($(49)!0.5!(50)$);
\draw[red, line width=2.5pt, opacity=0.4] (52) -- ($(52)!0.5!(25)$);
\draw[red, line width=2.5pt, opacity=0.4] (25) -- ($(25)!0.5!(52)$);
\draw[red, line width=2.5pt, opacity=0.4] (52) -- ($(52)!0.5!(51)$);
\draw[red, line width=2.5pt, opacity=0.4] (51) -- ($(51)!0.5!(52)$);
\draw[red, line width=2.5pt, opacity=0.4] (54) -- ($(54)!0.5!(53)$);
\draw[red, line width=2.5pt, opacity=0.4] (53) -- ($(53)!0.5!(54)$);
\draw[red, line width=2.5pt, opacity=0.4] (56) -- ($(56)!0.5!(27)$);
\draw[red, line width=2.5pt, opacity=0.4] (27) -- ($(27)!0.5!(56)$);
\draw[red, line width=2.5pt, opacity=0.4] (56) -- ($(56)!0.5!(55)$);
\draw[red, line width=2.5pt, opacity=0.4] (55) -- ($(55)!0.5!(56)$);
\draw[red, line width=2.5pt, opacity=0.4] (4) -- ($(4)!0.5!(14)$);
\draw[red, line width=2.5pt, opacity=0.4] (14) -- ($(14)!0.5!(4)$);
\draw[red, line width=2.5pt, opacity=0.4] (14) -- ($(14)!0.5!(24)$);
\draw[red, line width=2.5pt, opacity=0.4] (24) -- ($(24)!0.5!(14)$);
\draw[red, line width=2.5pt, opacity=0.4] (24) -- ($(24)!0.5!(50)$);
\draw[red, line width=2.5pt, opacity=0.4] (50) -- ($(50)!0.5!(24)$);
\draw[red, line width=2.5pt, opacity=0.4] (26) -- ($(26)!0.5!(54)$);
\draw[red, line width=2.5pt, opacity=0.4] (54) -- ($(54)!0.5!(26)$);
\draw[red, line width=2.5pt, opacity=0.4] (32) -- ($(32)!0.5!(50)$);
\draw[red, line width=2.5pt, opacity=0.4] (50) -- ($(50)!0.5!(32)$);
\draw[red, line width=2.5pt, opacity=0.4] (33) -- ($(33)!0.5!(52)$);
\draw[red, line width=2.5pt, opacity=0.4] (52) -- ($(52)!0.5!(33)$);
\draw[red, line width=2.5pt, opacity=0.4] (34) -- ($(34)!0.5!(54)$);
\draw[red, line width=2.5pt, opacity=0.4] (54) -- ($(54)!0.5!(34)$);
\draw[red, line width=2.5pt, opacity=0.4] (35) -- ($(35)!0.5!(56)$);
\draw[red, line width=2.5pt, opacity=0.4] (56) -- ($(56)!0.5!(35)$);
\end{pgfonlayer}
\end{tikzpicture}} 
\\
&
&
\\
 X_1X_3X_5X_7 &
 X_2X_3X_6X_7 &
 X_4X_5X_6X_7
\end{array}
\label{eq:neg_proxy_7} \ .
\end{equation}
And for the 15-to-1 protocol, post-selecting on the following logical check webs:
\begin{equation}
\setlength{\arraycolsep}{4pt}
\begin{array}{cccc}
\scalebox{.45}{\input{neg_state_proxy15_check0.tikz}} &
\scalebox{.45}{\input{neg_state_proxy15_check1.tikz}} &
\scalebox{.45}{\input{neg_state_proxy15_check2.tikz}} &
\scalebox{.45}{\input{neg_state_proxy15_check3.tikz}} \\
&&&\\[-2pt]
\resizebox{.15\linewidth}{!}{$X_{1} X_{3} X_{5} X_{7} X_{9} X_{11} X_{13} X_{15}$} &
\resizebox{.15\linewidth}{!}{$X_{2} X_{3} X_{6} X_{7} X_{10} X_{11} X_{14} X_{15}$} &
\resizebox{.15\linewidth}{!}{$X_{4} X_{5} X_{6} X_{7} X_{12} X_{13} X_{14} X_{15}$} &
\resizebox{.15\linewidth}{!}{$X_{8} X_{9} X_{10} X_{11} X_{12} X_{13} X_{14} X_{15}$}
\end{array}\,  \ .
\label{eq:neg_proxy_15}
\end{equation}
For either distillation circuits, these stabiliser proxies access only the logical $X$ initialisation and measurement outcomes.

\clearpage

\onecolumngrid
\section{Noisy and noiseless inplace \texorpdfstring{$\ket{Y}$}{Y} access}
\label{app:inplace_y_noise}

We compare physical noise on and off throughout the inplace $\ket{Y}$
preparation and output read-out routines. Both settings retain the same
operations and $\lfloor d/2\rfloor+2$ additional syndrome rounds per access.
The independently injected logical $Z$ errors of probability $p_{\rm inj}$
remain enabled. The remaining circuit noise is $p_{\rm circuit}=10^{-3}$,
with noiseless transversal CNOTs and three syndrome rounds at each time step.
The noiseless inplace comparison reduces the output error, but a nonzero
floor remains.

\begin{center}
\renewcommand{\arraystretch}{1.12}
\begin{tabular}{lcc}
\hline\hline
Noise location & Noisy inplace access & Noiseless inplace access\\
\hline
Entire inplace preparation and read-out & $p_{\rm circuit}$ & $0$\\
Remaining circuit operations & $p_{\rm circuit}$ & $p_{\rm circuit}$\\
Additional injected logical $Z$ & $p_{\rm inj}$ & $p_{\rm inj}$\\
\hline\hline
\end{tabular}
\end{center}

\begin{figure}[h!]
\centering
\includegraphics[width=.88\textwidth]{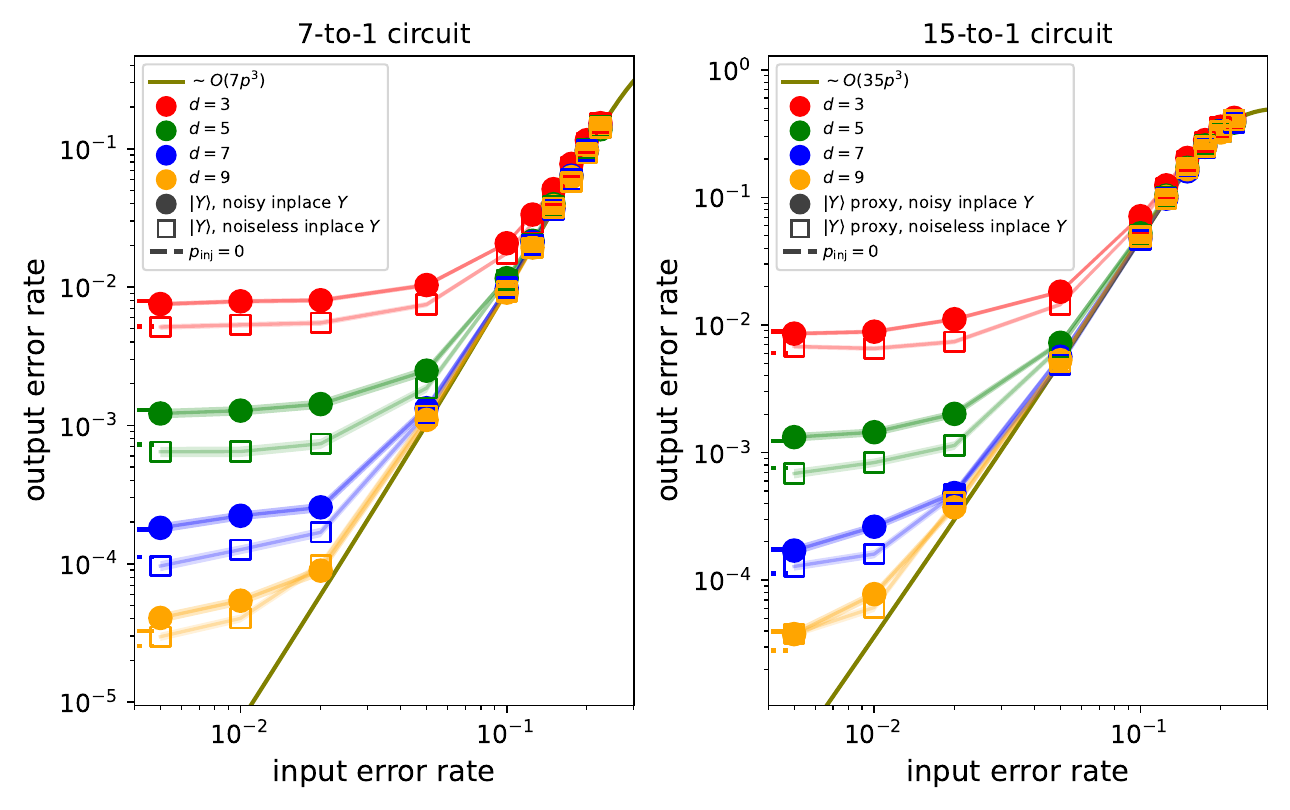}
\caption{Noisy and noiseless inplace $\ket{Y}$ access for $d=3,5,7,9$.
Filled circles include physical noise throughout the inplace routines;
open squares turn it off throughout those routines. Both retain
$p_{\rm inj}$. The olive curves are the exact ideal-protocol output errors.
}
\label{fig:inplace_y_noise}
\end{figure}

\end{document}